\documentclass[12pt]{article}
\usepackage[latin1]{inputenc}
\usepackage{amsmath}
\usepackage{color,soul}
\usepackage{xcolor}
\usepackage{amsfonts}
\usepackage{amssymb}
\usepackage{graphicx}
\usepackage{geometry}
\usepackage{hyperref}
\usepackage{multirow}
\usepackage[english]{babel}
\usepackage[bottom]{footmisc}
\usepackage{graphicx,algorithmic}
\usepackage[linesnumbered,ruled,vlined]{algorithm2e}
\SetCommentSty{mycommfont}

\SetKwInput{KwInput}{Input}                
\SetKwInput{KwOutput}{Output}              
\title{Unraveling heterogeneity of ADNI's time-to-event data using conditional entropy Part-I: Cross-sectional study.}
\author{Liao Shuting\footnotemark[1], Hsieh  Fushing \footnotemark[1] \footnotemark[2] \footnotemark[3], for the Alzheimer's Disease Neuroimaging Initiative \footnotemark[4].}
\date{}

\usepackage{diagbox}

\begin{document}
\maketitle
\footnotetext[1]{Graduate Group in Biostatistics, University of California, Davis, CA, 95616.}
\footnotetext[2]{Department of Statistics, University of California, Davis, CA, 95616.}
\footnotetext[3]{Correspondence: Fushing H. Department of Statistics, University of California at Davis, email:fhsieh@ucdavis.edu}
\footnotetext[4]{Data used in preparation of this article were obtained from the ADNI database(adni.loni.usc.edu). As such, the investigators within the ADNI contributed to the design and implementation of ADNI and/or provided data but did not participatein analysis or writing of this report. A complete listing of ADNI investiga-tors can be found at:\hyperlink{ http://adni.loni.usc.edu/wp-content/uploads/how to apply/ADNI Acknowledgement List.pdf.}{ http://adni.loni.usc.edu/wp-content/uploads/how to apply/ADNI Acknowledgement List.pdf.}}

\begin{abstract}

Through Alzheimer's Disease Neuroimaging Initiative (ADNI), time-to-event data: from the pre-dementia state of mild cognitive impairment (MCI) to the diagnosis of Alzheimer's disease (AD), is collected and analyzed by explicitly unraveling prognostic heterogeneity among 346 uncensored and 557 right censored subjects under structural dependency among covariate features. The non-informative censoring mechanism is tested and confirmed based on conditional-vs-marginal entropies evaluated upon contingency tables built by the Redistribute-to-the-right algorithm. The Categorical Exploratory Data Analysis (CEDA) paradigm is applied to evaluate conditional entropy-based associative patterns between the categorized response variable against 16 categorized covariable variables all having 4 categories. Two order-1 global major factors: V9 (MEM-mean) and V8 (ADAS13.bl), are selected for sharing the highest amounts of mutual information with the response variable.  This heavily censored data set is also analyzed using Cox's proportional hazard (PH) modeling and partial likelihood-based approach. Comparisons of PH and CEDA results on a global scale are complicated under the structural dependency of covariate features. To alleviate such complications, V9 and V8 are taken as two potential perspectives of heterogeneity and the entire 903 collections of subjects are respectively divided into two sets of four sub-collections according to their four categories. Censoring rates among sub-collections vary drastically from light to extreme. CEDA major factor selection protocol is again applied to all sub-collections to achieve the goal of local scale: To figure out which features provide extra information beyond V9 and V8. Graphic displays are developed to explicitly unravel conditional entropy expansions upon perspectives of heterogeneity in ADNI data. On the local scale, PH analysis is again carried out and results are compared with CEDA's. We conclude that, when facing structural dependency among covariates and heterogeneity in data, CEDA and its major factor selection provide significant merits for manifesting data's multiscale information content.
\end{abstract}

\vspace{1mm}
\clearpage

\section{Introduction.}
\paragraph{}Besides the pressing global climate issues, our humanity is facing another pressing issue: human aging. On October 1st, 2022 World Health Organization (WHO) released news regarding populations of elderly in the world quoted as:\\
`` By 2030, 1 in 6 people in the world will be aged 60 years or over. At this time the share of the population aged 60 years and over will increase from 1 billion in 2020 to 1.4 billion. By 2050, the world's population of people aged 60 years and older will double (2.1 billion). The number of persons aged 80 years or older is expected to triple between 2020 and 2050 to reach 426 million.'' To be more specific, a mentioned example in the news release is:``Japan 30\% of the population is already over 60 years old.'' That is, any common health issue related to aging would be one on a global scale.

In particular, as population aging is coming fast, so is human brain aging \cite{Craik2000,riddle2007}. Here, brain aging primarily means changes in cognitive functions \cite{Glisky2007}. When such changes go toward the dad directions, such as Alzheimer's disease as the focus of this study among many other diseases, the pressing health issue of human aging zooms really big.

As quoted from the home page of the Alzheimer's Disease Neuroimaging Initiative (ADNI) website: (\hyperlink{https://adni.loni.usc.edu}{https://adni.loni.usc.edu}),\\
``Alzheimer's disease (AD) is an irreversible neurodegenerative disease that results in a loss of mental function caused by the deterioration of brain tissue. It is the most common cause of dementia among people over the age of 65, affecting an estimated 5.5 million Americans, yet no prevention methods or cures have been discovered. For more information about Alzheimer's disease, visit the Alzheimer's Association.''

The Alzheimer's Disease Neuroimaging Initiative (ADNI) is a longitudinal multicenter study designed to develop clinical, imaging, genetic, and biochemical biomarkers for the early detection and tracking of Alzheimer's disease (AD). Since its launch more than a decade ago, the landmark public-private partnership has made major contributions to AD research, enabling the sharing of data between researchers around the world.

The goal of this study is in particular aligning with ADNI's first overarching goal: To detect AD at the earliest possible stage (pre-dementia) and identify ways to track the disease's progression with biomarkers. See the detailed Goals and Study design of the ADNI website at (\hyperlink{http://www.adni-info.org.}{http://www.adni-info.org.}).

So far studies using ADNI cross-sectional and longitudinal data from multiple modalities have reported two particularly relevant pieces of information: 1) AD pathology is already present in people with no outward sign of memory loss and these cognitively normal people may already have subtle brain atrophy; 2) Both the Cognitively Normal (CN) and Mild Cognitive Impairment (MCI) groups are pathologically heterogeneous. Some people show no signs of AD, some show signs of progressing to AD quickly, and others show signs of progressing to dementias other than AD.

Many cross-sectional studies on tracking progressions of AD with biomarkers using data collected from the ADNI database have reported applying the well-known Cox Proportional Hazard Regression model for the time-to-event data: from CN to MCI or from MCI to AD, and its well-studied partial likelihood approach as studies' chief inferential apparatus for selecting so-called significant biomarkers. However, this popular methodology of Survival analysis in Statistics has to face a series of fundamental questions when dealing with real-world data. When analyzing data from the ADNI database, not surprisingly, these essential questions, as listed below, are left unanswered in all studies found in literature \cite{Khajehpiri2022,Kueper2018,Li2018,Mubeen2017}:
\begin{description}
\item[Q1:] Does the Cox PH modeling assumptions on non-informative censoring mechanism violate the pattern-information embedded within data?
\item[Q2:] Does the data support the linearity-based additive effects of covariate features assumed by the Cox PH modeling structure?
\item[Q3:] Given that heterogeneity among subjects is intrinsic in ADNI data, could Cox PH provide reliable results, in particular when facing heavy censoring rates on the global and local scales and potentially complex interacting relations and effects among covariate features?
\item[Q4:] Is the partial likelihood based inferential approach valid?
\item[Q5:] If both the Cox PH model and the partial likelihood approach fail fundamentally, what are potential resolutions to achieve the overreaching goal of ADNI in this complex disease AD?
\item[Q6:] After all, how do we compute and identify pertinent perspectives of heterogeneity in cross-sectional ADNI data?
\item[Q7:]With the computed presence of heterogeneity, how to display the information contained in full?
\end{description}
We address Q1 through Q5 to a great extent, but only partially touch on Q6 and Q7 in this paper. Detailed and full discussions of the last two questions are deferred to Part-II. Here, we briefly explain why this series of questions is critical from a scientific perspective.

Testing whether the censoring scheme is non-informative or not in Q1 is the starter of data analysis applying methodologies in Survival Analysis, such as Cox PH modeling. However, this testing has not been carried out in the ADNI-associated literature. The reason might be that there is no simple testing statistic available in the Survival Analysis literature. As for Q2, the linearity-based additivity of covariate effects obviously doesn't work for categorical covariate variables, which is a common data type, such as sex and education. Even for quantitative variables, the sum of measurements of very distinct metric units is rather hard to explain literately and convincingly. With the presence of heterogeneity among subjects, Cox PH becomes rather limited with respect to fronts: censoring rates and unknown functional forms of interaction. The overall censoring rate is already high. It is more than $60\%$ in this cross-sectional study. The censoring rates can reach more than $90\%$ in some sub-collections of subjects defined by the potential perspective of heterogeneity. On the other front, as multiple reports depicting interacting effects in AD literature \cite{Altmann2014,Ardekani2020,Duarte2021}, such real-world interacting relations among covariate variables might be known prior, but their effects are hardly known functionally. Thus, any functional forms of interacting effects among covariate variables deem unrealistic and dangerous to subject matter science because of misinformation.

For Q4: the partial likelihood approach is strictly based on the correlation for evaluating associative relations between the time-to-event response variable and covariate features. This fact can be easily seen from the score equations derived from the partial likelihood, which involves $T_{(i)}$ minus its conditional expectation within the risk set. This format of partial likelihood score equations points to one strict and fundamental associative concept: correlation. It is known that the correlation concept is strictly designed for one quantitative variable against another quantitative variable, not for categorical ones. It is not valid for evaluating 1-to-2 or 2-to-k associative relations. This limit is consequential for the following reason. When facing an informative censoring scheme, the response and censoring variables are dependent. Then, the proper response variable must be 2D bivariate $(T_i, C_i)$. Under such a setting, the partial likelihood approach would not work. This fact again points to the critical role of Q1.

For Q5 and partially for Q6 and Q7, this paper proposes the Categorical Exploratory Data Analysis (CEDA) paradigm to resolve all issues raised from Q1 through Q4. In CEDA, we apply Shannon conditional entropy to evaluate associative relation, and mutual information to select so-called major factors underlying the dynamics of the response variable in relation to all covariate feature variables. When using two key Theoretical Information measurements, the CEDA naturally adopts the contingency table as its basic computational platform. Consequently, CEDA not only works for all data types, including the categorical one but also is capable of evaluating $k-to-k' $ associative relations for all integers $k$ and $k'$. Since $k$ categorized or categorical variable can be fused into one categorical variable via their contingency $k$-dim hypercubes. When facing censored data disregarding the censoring rate, we apply the re-distribution-to-the-right algorithm \cite{Efron1967} to build all sorts of contingency tables. For this simple reason, CEDA is capable of handling even $100\%$ censoring rates in any locality. We also can build the contingency table for categorized $T_i$ and $C_i$ to resolve the Q1 formally. This is a new resolution. And because CEDA is able to identify major factors of varying orders, the task of identifying pertinent perspectives of heterogeneity in Q6 is resolved to a great extent. We also develop a display called conditional-entropy-expansion to unravel the effects of all chosen perspectives of heterogeneity.

The organization of this paper is given as follows. In Section~\ref{Sec:data_descript}, we describe the cross-sectional ADNI data by providing brief descriptions for all variables: response, censoring, and 16 covariate features. In Section~\ref{Sec:CEDA}, we first review Theoretical Information measurements and rationales for major factor selection under independent and dependent covariate settings and illustrate how the Redistribution-to-the-right algorithm precisely works. In Section~\ref{Sec:sim}, we conduct three simulated experiments with varying censoring rates to showcase information relevant to Q1-Q5. The ADNI data is thoroughly analyzed in the lengthy Section~\ref{Sec:adni_analysis} with two perspectives of heterogeneity being studied and presented. In the last section of the Conclusion, we discuss the contributions of this Part-I paper, and briefly lay out computational tasks in Part-II. We hope this paper would help scientists to advance their research on Alzheimer's Disease and we are thankful to ADNI for making such an important database available to us.

\section{ADNI Data Description}\label{Sec:data_descript}

\paragraph{}Data used in the preparation of this article were obtained from the Alzheimer's Disease Neuroimaging Initiative (ADNI) database (\hyperlink{https://adni.loni.usc.edu}{https://adni.loni.usc.edu}). The ADNI was launched in 2003 as a public-private partnership, led by Principal Investigator Michael W. Weiner, MD. The primary goal of ADNI has been to test whether serial magnetic resonance imaging (MRI), positron emission tomography (PET), other biological markers, and clinical and neuropsychological assessment can be combined to measure the progression of mild cognitive impairment (MCI) and early Alzheimer's disease (AD).

We accessed the data file on June 9, 2022. It included 15,941 records from 1,094 participants. By removing any subjects with missing data, we ended up with the records of 903 subjects with a baseline diagnosis of Mild Cognitive Impairment (MCI) and at least one follow-up diagnosis. There are 346 subjects (211 males; 135 females) whose diagnosis progressed to Dementia (AD)  and their time-to-even $T_i$'s observed. During the follow-up, 557 individuals (328 males; 229 females) remained MCI, and their last scheduled exam times are observed as censoring time $C_i$s. Therefore, we observe $Y_i=(T_i \wedge C_i)$, the minimum of $T_i$ and $C_i$, and the censoring status $\delta_i=1_{[T_i \leq C_i]}$ for each of these 903 subjects.

Data feature descriptions are given as follows. The 16 covariate features used here are coming from three categories:  4 demographic features ($V1-V4$), 1 MRI-related feature ($V5$), and 11 clinical features ($V6-V16$). The features' abbreviation and definition are provided in Table~\ref{covtab} followed by detailed descriptions of 11 clinical features.
\begin{table}[ht]
\begin{center}
\resizebox{\columnwidth}{!}{
\begin{tabular}{ccc}\hline
 \textbf{Index} & \textbf{features} & \textbf{Definition} \\
  \hline
  $V1$ & age & Age at baseline \\
  $V2$ & gender  & Gender \\
  $V3$ &Education & Education Level \\
  $V4$ &APOE $\epsilon4$ & Apolipoprotein E $\epsilon$4 allele\\
  $V5$ & FLDSTRENG-bl  & MRI Scanner's Field Strength (1.5T or 3 T) used at baseline\\
  $V6$ & CDR-SB-bl &Clinical Dementia Rating at baseline (the sum of boxes)\\
  $V7$ &FAQ& Functional Activities Questionnaire\\
  $V8$ & ADAS13-bl & 13-item AD Assessment Scale-Cognitive Subscale at baseline\\
  $V9$ &MEM-mean & Mean of the Composite Cognitive Score for Memory\\
  $V10$ & MEM-std &  ``std'' of the Composite Cognitive Score for Memory\\
  $V11$ & EXF-mean & Mean of the Composite Cognitive Score for Executive Functioning\\
  $V12$ &EXF-std & ``std'' of the Composite Cognitive Score for Executive Functioning\\
  $V13$ &LAN-mean & Mean of the Composite Cognitive Score for Language\\
  $V14$ & LAN-std & ``std'' of the Composite Cognitive Score for Language\\
  $V15$ &VSP-mean & Mean of the Composite Cognitive Score for Visuospatial Functioning \\
  $V16$ &VSP-std & ``std'' of the Composite Cognitive Score for Visuospatial Functioning \\
   \hline
\end{tabular}}
\caption{Covariate features as biomarkers in the ADNI data analysis. ``std'': the standard deviation.}
\label{covtab}
\end{center}
\end{table}

Some further pieces of information regarding three clinical features: CDR-SB (V6) \cite{Morris1993} indicates the sum of scores for the following six domains of functioning: memory, orientation, judgment and problem solving, community affairs, home, and hobbies, and personal care. The CDR-SB ranges from 0 (no impairment) to 18 (severe impairment in all six domains); FAQ (V7) measures activities of daily living and ranges from 0 to 30 with higher scores reflecting greater cognitive impairment\cite{Pfeffer1982}; The modified ADAS-Cog 13-item scale (ADAS13(V8))\cite{Mohs1997} is a modified evaluation by adding a number cancellation task and a delayed free recall task to the 11-item ADAS-Cog (ADAS11). The higher scores suggest greater impairment.

In addition to the clinical features aforementioned, we consider four composite cognitive scores of 4 domains, including memory, executive functioning, language, and visuospatial functioning\cite{Mukherjee2022}. The scores are developed and calibrated by using the ADNI Neuropsychological batteries. During each available follow-up, the four composite cognitive scores are updated accordingly and thus they are time-varying. To extract information and characteristics for consideration, we make use of the mean and standard deviation of the composite cognitive score for each category until the event (i.e., AD or the last time point if censored) as two new features to describe the overall characteristics of each score. Therefore, we have eight total new features to represent the four composite cognitive scores which are shown in Table~\ref{covtab} ($V9-V16$). With such construction of time-varying scores of these 4 domains, we expect each subject has the mean and standard deviation of the composite cognitive scores at every exam time. Missing values of the composite cognitive scores at partial exam time points are found due to the intrinsic incomplete data (e.g., participant refusal to complete the study or the item was not administered due to the time limit) and this can be fixed by taking the average of available scores correspondingly. We remove subjects with no scores available across all exam time points as there is not enough useful information provided.

\section{CEDA Methodologies}\label{Sec:CEDA}
\paragraph{}
In this section, we first briefly review computational developments for CEDA's major factor selection protocol. This protocol is entirely based on Theoretical Information Measurements: marginal and conditional entropies and mutual information. Since we work only with categorical or categorized variables in CEDA. Therefore, the only version of entropy used here is Shannon entropy. Secondly, we illustrate the ``Redistribution-to-the right'' algorithm for building a contingency table with the presence of the right censored data. Since all CEDA computing is performed upon the contingency table platform. Thus, this algorithm indeed plays a critical role in this paper.

\subsection{CEDA's major factor selection protocol}
\paragraph{}In this subsection, we briefly review the concept and computing of conditional entropy and mutual information as two key Theoretical Information Measurements used throughout this paper. Detailed derivations of related formulas of these two measurements are referred to previous works in \cite{Chen2021,Chou2022,Hsieh2022,Chen2022}. We employ these two entropy-based measurements to evaluate potentially nonlinear directional association from a generic covariate feature variable denoted as $X$ to a generic response variable denoted as ${\cal Y}$, and the nondirectional association between two covariate feature variables, say $X_1$ and $X_2$.

Since we exclusively work on categorical or categorized variables in this paper. That is, any variable of continuous or discrete measurements is categorized with respect to a chosen version of the variable's histogram. For instance, the time-to-event $T$, the mean and standard deviation of test scores, and age are to be categorized. Further, any set of multiple categorical variables can be fused into a new categorical variable by redefining each distinct multiple-dimensional categorical vector as a category of the newly defined categorical variable. For instance, a bivariate vector $(X_1, X_2)$ would be taken and treated as a categorical variable. That is, the categorical variable, such as gender (V2), and the categorized variable, such as MEM-mean (V9), can be fused into a new categorical variable. Therefore, any set of covariate feature categorical or categorized variables, say $A$, is also taken as a categorical variable with the same name $A$. So $A$'s directional association to ${\cal Y}$ is evaluated in the same way as evaluating the directional association of any member of $A$ to a categorical response variable ${\cal Y}$.

Consider $A$ or $B$ as two different categorical covariate variables standing for two sets of covariate features. We evaluate the directional association of $A$ to ${\cal Y}$ upon their contingency table $C[A-vs-{\cal Y}]$ with categories of $A$ and ${\cal Y}$ being arranged along the row- and column-axes, respectively, as a conventional format used throughout this paper. Along the row-axis, each row of cell counts in $C[{\cal Y}-vs-A]$ is taken to define a conditional multinomial random variable, which is specified by its row-sum and the row-vector of proportions. For instance, a conditional (Shannon) entropy (CE) of ${\cal Y}$ at the $a$-th row of $C[A-vs-{\cal Y}]$ is calculated and denoted as:
\[
H[{\cal Y}|A=a]=(-1)\sum_{y\in \{y_1,.., y_r\}}\hat{p}[{\cal Y}=y|A=a] \log{\hat{p}[{\cal Y}=y|A=a]},
\]
with $(\hat{p}[{\cal Y}=y_1|A=a],.., \hat{p}[{\cal Y}=y_r|A=a] )$ as the $a$-th row's vector of proportions. This quantity of $H[{\cal Y}|A=a]$ indicates the amount of uncertainty about ${\cal Y}$ given the information of $A=a$ being known.

In contrast, the overall amount of uncertainty about ${\cal Y}$ given the information of $A$ with $a\in \{a_1, .., a_h\}$ is evaluated as the weighted average and denoted as:
\[
H[{\cal Y}|A]=\sum_{a\in \{a_1, .., a_h\}}\frac{n_a}{n}H[{\cal Y}|A=a],
\]
with $n_a$ being $a$-th row sum and the total sample size $n=\sum_{a\in \{a_1, .., a_h\}}n_a$. Further, the entropies of marginal column-wise vector of proportions $(\frac{n_{y_1}}{n},..., \frac{n_{y_r}}{n})$ and row-wise vector of proportions $(\frac{n_{a_1}}{n},..., \frac{n_{a_h}}{n})$ are denoted as $H[{\cal Y}]$ and $H[A]$, respectively.

It is known that the conditional entropy (CE) $H[{\cal Y}|A]$ conveys the expected amount of remaining uncertainty in ${\cal Y}$ after knowing $A$. Likewise, $H[A|{\cal Y}]$ conveys the expected amount of remaining uncertainty in $A$ after seeing ${\cal Y}$. The two conditional entropy drops, i.e. differences $H[Y]-H[{\cal Y}|A]$ and $H[A]-H[A|{\cal Y}]$, indicate the shared amount information between $A$ and ${\cal Y}$:
\begin{eqnarray*}
H[{\cal Y}]-H[{\cal Y}|A]&=&H[A]-H[A|{\cal Y}]\\
&=&H[A]+H[Y]-H[A,{\cal Y}]\\
&=&I[{\cal Y};A].
\end{eqnarray*}
where $I[{\cal Y};A]$ denotes the mutual information between ${\cal Y}$ and $A$.

Further, the conditional mutual information between the bivariate variable $(A,B)$ given ${\cal Y}$ is evaluated as:
\[
I[A; B|{\cal Y}]=H[A|{\cal Y}] +H[B|{\cal Y}]-H[(A,B)|{\cal Y}].
\]
Therefore, the mutual information $I[{\cal Y};(A,B)]$ can be estimated and decomposed as follows:
\begin{eqnarray*}
H[{\cal Y}]-H[{\cal Y}|(A, B)]&=&H[(A, B)]-H[(A,B)|{\cal Y}];\\
&=&H[A]+H[B]-I[A;B] - \{H[A|{\cal Y}] + H[B|{\cal Y}]-I[A;B|{\cal Y}]\};\\
&=&\{H[{\cal Y}]-H[{\cal Y}|A] + H[{\cal Y}]-H[{\cal Y}|B]\}+\{I[A;B|{\cal Y}]-I[A;B]\},
\end{eqnarray*}
where the first two terms are individual CE-drops attributed to $A$ and $B$ and the third term is the difference of conditional and marginal mutual information of $A$ and $B$. In particular, we term $A$ and $B$ achieve their ecological effect if this term: $\{I[A;B|{\cal Y}]-I[A;B]\}$, is positive. This positiveness indicates the potential for $A$ and $B$ being concurrently present within the dynamics underlying ${\cal Y}$. The essence of achieving the ecological effect is that $A$ and $B$ have the potential of being conditional dependent under ${\cal Y}$ disregarding whether they are marginal dependent or not.

However, the above decomposition precisely conveys the interpretable meaning of conditional mutual information when $I[A;B]=0$ as the two involving feature sets $A$ and $B$ are marginally independent. Thus, if $I[A;B|{\cal Y}]$ is significantly larger than $\min\{H[{\cal Y}]-H[{\cal Y}|A], H[{\cal Y}]-H[{\cal Y}|B]\}$, then we are certain that $A$ and $B$ achieve a significant interacting effect in reducing the uncertainty of ${\cal Y}$. Therefore, we particularly evaluate the so-called successive conditional entropy (SCE) drop as:
\[
H[{\cal Y}]-H[{\cal Y}|(A, B)]-\max\{H[{\cal Y}]-H[{\cal Y}|A], H[{\cal Y}]-H[{\cal Y}|B]\}.
\]
The task of identifying any realistic interacting effect is always essential in any real-world data analysis because such effects could provide a critical understanding of the system under study. Thus, this SCE-drop would be reported in all tables. Its merit also includes checking whether $A$ and $B$ achieve their ecological effect, which is required before considering whether they have an interacting effect or not.

On the other hand, if $A$ and $B$ are indeed highly associated in the marginal sense via certain unknown dependency, then $I[A;B] >0$. That is, the term $\{I[A;B|{\cal Y}]-I[A;B]\}$, could be negative. Hence, when $A$ and $B$ are associated, we face two chief difficulties. The first difficulty is that it is hard to determine whether the minimum CE-drop: $\min\{H[{\cal Y}]-H[{\cal Y}|A], H[{\cal Y}]-H[{\cal Y}|B]\}$, due to either $A$ or $B$ is indeed significant or not. That is since their ecological effect is failed to be seen, we can not be sure whether $A$ and $B$ are concurrently present within the dynamics underlying ${\cal Y}$. The second difficulty is that, even $\{I[A;B|{\cal Y}]-I[A;B]\}$ is positive with a moderate, not large enough size, then it becomes difficult to assess whether the $A$ and $B$ have a significant interacting effect or not.

In order to resolve these two difficulties, a de-associating procedure is proposed in \cite{Hsieh2022} by simply subdividing the entire data set with respect to a target covariate variable's categories. For instance, $A$ is the target covariate variable, and the entire data set is divided into $h$ sub-collections with respect to $\{a_1, .., a_h\}$. That is, $A$ is a constant within each sub-collection. Hence, the association between $B$ and $A$ within each of these $ h$ sub-collection disappears. Overall speaking, all covariate variables are less associated within each sub-collection. The merit of the de-associating procedure is evidently seen in Section~\ref{Sec:adni_analysis} of ADNI data analysis. In fact, the most critical merit of this de-associating procedure indeed rests on the fact that $A$'s perspective of heterogeneity embedded within the dynamics underlying ${\cal Y}$ can be much more easily discovered. As such we can discover two versions of heterogeneity from the perspectives of $A$ and $B$ and compare these two versions of heterogeneity. By doing so, we discover authentic heterogeneity-based pattern information hidden in data. Ideally, if we could identify all relevant perspectives of heterogeneity, then collectively we can have the data's full information content. This is termed the ideal scenario in data analysis.

By summarizing all aforementioned developments in this subsection, we arrive at the CEDA's Major Factor Selection (MFS) protocol under two settings. The first setting is developed in \cite{Chen2021} to deal with independent or slightly dependent covariate feature variables only, while the second set is developed in \cite{Hsieh2022} to deal with the heavily dependent covariate setting. Nonetheless, by adopting the de-associating procedure within any heavily dependent setting, the MFS protocol originally built under an independent setting is indeed applied within the sub-collection levels, at which all heavily dependent covariate feature indeed become much less dependent. As such, we discover which covariate features or feature-sets can indeed provide extra amounts of information beyond which targeted covariate features or feature-sets. This computational capability is essential in any real data analysis. Since the goal of data analysis is aimed at the ideal scenario: data's full information content.

It is worth reiterating that, with the potential presence of heterogeneity, the goal of MFS protocol is not set to create one ultimate collection of major factors of various orders that can collectively and concurrently reduce the uncertainty of ${\cal Y}$ to the lowest level. The goal of the MFS protocol is to precisely explore and extract pattern information pertaining to perspective-specific heterogeneity. By exploring many perspectives of heterogeneity, we hope we can get much closer to the data's full information content.

\subsection{Redistribution-to-the-right}
\paragraph{}Evaluations of association between two categorical or categorized variables are performed on the contingency table platform. Without involving censored data points, the construction of a contingency table is straightforward by counting the number of data points falling into each cell defined by one category from each of the two variables. Nonetheless, when censoring is involved with one variable such as the response survival time variable $T$, then constructing a contingency table of any covariate variable $X$ against $T$ is not a straightforward computing task. We illustrate how to achieve such a task in this subsection.

Denote the contingency table to be built as $C[X-vs-T]$ with $T$ being censored by $C$. In this subsection, we first consider the variables $T$ and $C$ being measured at their original time scale before being categorized. Recall that the possibly right-censored survival time data set is denoted as
$\{(Y_i, \delta_i, {\cal X}_i|i=1,.., n\}$ with $Y_i= (T_i \wedge C_i)$ as the minimum of $T_i$ and $C_i$ and $\delta_i=1_{[T_i \leq C_i]}$ indicating the binary censoring status: 0 for being censored and 1 for being uncensored. Let  $n_c=n-\sum^n_{i=1}\delta_i=n-n_{u}$ denote the number of censored data points and $n_c/n $ the censoring rate of this data set. For simplicity, we assume all $\{Y_i\}$ are distinct and its order statistics is denoted as $\{Y_{(i)}\}$. The $K$-dim covariate vector ${\cal X}_i=(X_{1i}, X_{2i},.., X_{Ki})$ records the measurements of $K$ feature variables $\{X_k|k=1,.., K\}$.

The Kaplan-Meier estimation \cite{Kaplan1958} of survival function $S(t)=Pr[T_i >t]$ of $T_i$ based only on $\{(Y_i, \delta_i)|i=1,.., n\}$, as if without knowledge of  $\{{\cal X}_i|i=1,..,n\}$, is built as \cite{Miller1981}:
\[
\hat{S}(t)=\prod_{i: Y_{(i)}\leq t}(1-\frac{1}{n-i+1})^{\delta_{(i)}}.
\]
It is evident that this empirical distribution $\hat{S}(t)$ has all its jumps at uncensored time points
$\{(Y_i,\delta_i)|\delta_i=1\}$ with possibly unequal jump-sizes. This phenomenon is characterized as so-called ``redistribution-to-the-right'' in \cite{Efron1967}, that is, the empirical weight $\frac{1}{n}$ of any censored data point is equally redistributed to all data points: uncensored as well as censored, found on its right-hand side. A weight received by any censored data point is likewise re-distributed to all data points on its right. In Table~\ref{Redist0}, we specifically illustrate this redistributing algorithm for the 3rd ordered survival time $Y_{(3)}$, which is censored ($\delta_{(3)}=0$) among $10(=n)$ data points with the 6th and 8th being censored as well. For expositional simplicity, we denote and mark these three ordered censored time points as $Y^c_{(3)}$, $Y^c_{(6)}$, and $Y^c_{(8)}$.

It is noted that not only $Y^c_{(3)}$ is subject to the such redistribution-the-right algorithm in constructing the empirical survival distribution $\hat{S}(t)$, but also all covariate measurements $\{X_{k(3)}|k=1, .., K\}$ are subject to the same redistribution when we construct contingency table $C[X_{k}-vs-T]$. It is because that we only observe $(X_{k(3)}, Y^c_{(3)} )$, not the unobserved $ (X_{k(3)}, T^c_{(3)})$. And the unobserved $T^c_{(3)}$ is indeed larger than $Y^c_{(3)}$. Since there are two more censored survival time points beyond $Y^c_{(3)}$. The entire redistribution algorithm takes three steps to finish. In Table~\ref{Redist1}, we show the results of the redistribution of empirical weights pertaining to the three censored data points.

\begin{table}[h!]
\centering
\begin{tabular}{ccccccccccc}\hline
$X/Y$ &$Y_{(1)}$ &$Y_{(2)}$&$Y^c_{(3)}$&$Y_{(4)}$&$Y_{(5)}$&$Y^c_{(6)}$&$Y_{(7)}$&$Y^c_{(8)}$&$Y_{(9)}$&$Y_{(10)}$\\ \hline
$X_{k3}$ &0.0&0.0&0.0&$\frac{1}{7}$&$\frac{1}{7}$&$\frac{1}{7}$&$\frac{1}{7}$&$\frac{1}{7}$&$\frac{1}{7}$&$\frac{1}{7}$\\
$X_{k3}$&0.0&0.0&0.0&$\frac{1}{7}$&$\frac{1}{7}$&0.0&$\frac{1}{7}+\frac{1}{28}$&$\frac{1}{7}+\frac{1}{28}$&$\frac{1}{7}+\frac{1}{28}$ &$\frac{1}{7}+\frac{1}{28}$\\
$X_{k3}$&0.0&0.0&0.0&$\frac{1}{7}$&$\frac{1}{7}$&0.0&$\frac{1}{7}+\frac{1}{28}$&0.0&$\frac{1}{7}+\frac{1}{28}+\frac{5}{56}$ &$\frac{1}{7}+\frac{1}{28}+\frac{5}{56}$\\
\hline
\end{tabular}%
\caption{A censored data point's ($X_3$) step-by-step redistribution-to-the-right at three ordered censored time points $Y^c_{(3)}$, $Y^c_{(6)}$ and $Y^c_{(8)}$ along the axis. All row-sums of weights are equal to 1.}
\label{Redist0}
\end{table}

\begin{table}[h!]
\centering
\begin{tabular}{ccccccccccc}\hline
$X/Y$ &$Y_{(1)}$ &$Y_{(2)}$&$Y^c_{(3)}$&$Y_{(4)}$&$Y_{(5)}$&$Y^c_{(6)}$&$Y_{(7)}$&$Y^c_{(8)}$&$Y_{(9)}$&$Y_{(10)}$\\ \hline
$X_{k(1)}$ &1 &0 &0 &0 &0 &0 &0 &0 &0 &0\\
$X_{k(2)}$ &0 &1 &0 &0 &0 &0 &0 &0 &0 &0\\
$X_{k(3)}$&0&0&0&$\frac{1}{7}$&$\frac{1}{7}$&0&$\frac{5}{28}$&0&$\frac{15}{56}$ &$\frac{15}{56}$\\
$X_{k(4)}$ &0 &0 &0 &1 &0 &0 &0 &0 &0 &0\\
$X_{k(5)}$ &0 &0 &0 &0 &1 &0 &0 &0 &0 &0\\
$X_{k(6)}$ &0 &0 &0 &0 &0 &0 &$\frac{1}{4}$ &0 &$\frac{3}{8}$ &$\frac{3}{8}$ \\
$X_{k(7)}$ &0 &0 &0 &0 &0 &0 &1 &0 &0 &0\\
$X_{k(8)}$ &0 &0 &0 &0 &0 &0 &0 &0 &$\frac{1}{2}$ &$\frac{1}{2}$\\
$X_{k(9)}$ &0 &0 &0 &0 &0 &0 &0 &0 &1 &0\\
$X_{k(10)}$ &0 &0 &0 &0 &0 &0 &0 &0 &0 &1\\\hline
\end{tabular}%
\caption{Redistribution-to-the-right of three censored data data points among 7 uncensored ones along the ordered survival time axis. All row-sums of weights are equal to 1.}
\label{Redist1}
\end{table}

Upon the entire data set $\{(Y_i, \delta_i, {\cal X}_i|i=1,.., n\}$, a $n\times n$ weight-redistribution matrix is constructed and fixed according to the layout of $n$ order statistics' $\{(Y_{(i)}, \delta_{(i)})\}$ censoring statuses. Denote this weight-redistribution matrix as ${\cal W}$ as shown in Table~\ref{Redist1}. It is noted that all columns according to all censored survival times $\{Y_{(i)}|\delta_{(i)}=0\}$ are $n$-dim zero-vectors in ${\cal W}$. Thus, when the uncensored survival times are categorized by a specific way of grouping on $\{T_{(i')}|\delta_{(i')}=1, i'=1,.., n_u\}$, then ${\cal W}$ will be subject to the same grouping along its column axis. Consequently, the weights contributed to each bin by any $k$-th covariate measurement of $i$-th individual is specified. In this fashion, we are able to construct contingency tables $C[X_{k}-vs-T]$ for evaluating associations between $X_k$ and $T$. We can likewise construct a contingency table $C[(X_{k_1},.., X_{k_l})-vs-T]$ for a feature-set $\{X_{k_1},.., X_{k_l}\}$ and $T$.

\section{Computer experiments with increasing right censoring rates.}\label{Sec:sim}
\paragraph{} In this section, we report a simple computational study by applying methodologies proposed in the previous section on three simulated right-censored survival time data sets. These three data sets are generated by the same functional structures but have distinct censoring rates: $10\%$, $20\%$, and $30\%$. We hope to demonstrate the stably evolving conditional entropy evaluations and at the same time to show correct selections of major factors across these three different censoring rates.

We employ an integral equation, which has been proposed and studied in \cite{Hsieh2012}, to generate survival time $T$ as the time of using up the unobserved reserve value $U$ with respect to an exhausting rate specified by $e^{\{V1+sin(2\pi(V2+V3))+V7^2\}} \lambda_0(t)$. The term $\lambda_0(t)$ is taken as the baseline hazard rate. The integral equation is given as follows:
\[
U = \int^T_0 e^{\{V1+sin(2\pi(V2+V3))+V7^2\}} \lambda_0(t) dt.
\]
Here the term $sin(2\pi(V2+V3))$ in the exponent of the integrand is designed to have an interacting relational effect of variables V2 and V3 through a sine function. This simple functional form signals the nonlinearity, on one hand, and the departure from the classic product format of interacting effect, on the other.

Denote the hazard rates of $U$ and $T$ as  $\lambda_{U}(\cdot)=\Lambda'_U (\cdot)$ and $\lambda_{T}(\cdot)=\Lambda'_T (\cdot)$, respectively. Their relationships are characterized as follows, see more details in \cite{Hsieh2012}:
\begin{eqnarray*}
&&e^{-\Lambda_T(t)}=Pr[T >t]\\
&=&Pr[\int^T_0 e^{\{V1+sin(2\pi(V2+V3))+V7^2\}} \lambda_0(s) dts > \int^t_0 e^{\{V1+sin(2\pi(V2+V3))+V7^2\}} \lambda_0(s) ds]\\
&=&Pr[U>\int^t_0 e^{\{V1+sin(2\pi(V2+V3))+V7^2\}} \lambda_0(s) ds];\\
&=&e^{-\Lambda_U(\int^t_0 e^{\{V1+sin(2\pi(V2+V3))+V7^2\}}\lambda_0(s) ds)}.
\end{eqnarray*}
Therefore, we have the cumulative hazard rate and hazard rate of $T$ being specified as follows:
\begin{eqnarray*}
\Lambda_T(t)&=&\Lambda_U(\int^t_0 e^{\{V1+sin(2\pi(V2+V3))+V7^2\}} \lambda_0(s)ds);\\
\lambda_T(t)&=&\lambda_U(\int^t_0 e^{\{V1+sin(2\pi(V2+V3))+V7^2\}}\lambda_0(s)ds)\\
&\cdot & e^{\{V1+sin(2\pi(V2+V3))+V7^2\}} \lambda_0(t).
\end{eqnarray*}
That is, if $\lambda_{U}(\cdot)$ is a constant function, that is, $U ~ Exp(.)$, then we have:
\[
\lambda_T(t)=e^{\{V1+sin(2\pi(V2+V3))+V7^2\}} \lambda_0(t),
\]
which is in a format of Cox's proportional hazard setting.

To simulate three experimental data sets, for simplicity, we use the Weibull baseline hazard function $\lambda_0(t)=kt^{k-1}$ with $k=1.5$ and Exponential distributed reserve function, $U \sim Exp(1.5)$. There are 10 mutually independent covariate variables $\{V1, .., V10\}$. They are all randomly sampled from Uniform$[0,1]$ distribution. The three right censoring variables are also Exponentially distributed with three different chosen rates to create preset censoring rates. For CEs calculation, all covariates are categorized into 10 uniformed bins, and the response variable $T$ is also categorized into 10 bins as well based on their Kaplan-Meier estimates. Each simulated data set is 10,000 $(=n)$ in size.

\paragraph{[Experiment-: $10\%$ censoring rate]}We report our CE evaluations in Table~\ref{CEexample1}. The row-wise CEs are ranked from the top-to-bottom in the three feature settings, respectively. In the 1-feature setting, we see only V1 and V7 having significant SCE-drops. So they are individual order-1 major factor candidates. The CEs of V2 and V3 are even as low as that of those random noise features.

The interacting effect of V2 and V3 are visible in the 2-feature setting by having a SCE-drop being many times of CEs of either V2 or V3. That is, feature-pair (V2, V3) is an order-2 major factor candidate. The SCE-drop of the feature-pair (V1, V7) is larger than V7's SCE-drop. This ecological effect indicates that V1 and V7 are currently present in the dynamics of response. So they are likely to be the order-1 major factors together. It is noted that CEs and SCE-drops of all the feature pairs of random noise provide the baseline of comparison. In the 3-feature setting, the feature triplets (V1, V2, V3) and (V2, V3, V7) achieve the lowest CEs and their SCE-drops indicate the ecological effects being achieved. That is, these three major factors: (V2, V3), V1, and V7, are concurrently present within the dynamics of $T$. Despite increasing uncertainty along with increasing censoring rates, exactly the same conclusions of major factors can be drawn from Table~\ref{CEexample2} and Table~\ref{CEexample3} based on the two simulated data sets with $20\%$ and $30\%$ censoring rates, respectively.

As for the Cox PH results, V1 and V7 are significant across the three experiments, while in contrast the two features: V2 and V3, are never seen as simultaneously significant in the three experiments. We only observe that V3 is somehow significant in experiment-1, V2 in experiment-2. But both features are insignificant in experiment-3. These experimental results converge to the fact that any interacting effect with non-product format is likely ignored by Cox PH results, especially when the censoring is high.

\begin{table}[h!]
\centering
\resizebox{\columnwidth}{!}{
\begin{tabular}{cccc|ccc|ccc}
\hline
1-feature & CE     & SCE-dp & $p$-value(PH)      & 2-feature & CE     & SCE-dp & 3-feature   & CE     & SCE-dp \\ \hline
V7        & 1.0103 & 0.0158 & \textless{}2e-16 & V2\_V3    & 0.9292 & 0.0955 & V1\_V2\_V3  & 0.8114 & 0.1178 \\
V1        & 1.0156 & 0.0105 & \textless{}2e-16 & V1\_V7    & 0.9898 & 0.0205 & V2\_V3\_V7  & 0.812  & 0.1172 \\
V9        & 1.0246 & 0.0014 & 0.7942           & V2\_V7    & 0.9973 & 0.013  & V2\_V3\_V10 & 0.8217 & 0.1075 \\
V3        & 1.0247 & 0.0013 & 0.0515           & V7\_V8    & 0.9982 & 0.0121 & V2\_V3\_V5  & 0.8221 & 0.1072 \\
V2        & 1.0247 & 0.0013 & 0.3575           & V3\_V7    & 0.9984 & 0.0118 & V2\_V3\_V4  & 0.8238 & 0.1054 \\
V4        & 1.0248 & 0.0013 & 0.5773           & V7\_V9    & 0.9993 & 0.011  & V2\_V3\_V8  & 0.8267 & 0.1026 \\
V10       & 1.0248 & 0.0012 & 0.7622           & V5\_V7    & 0.9994 & 0.0108 & V2\_V3\_V9  & 0.8274 & 0.1018 \\
V6        & 1.0249 & 0.0012 & 0.654            & V6\_V7    & 0.9996 & 0.0106 & V2\_V3\_V6  & 0.8282 & 0.101  \\
V8        & 1.0251 & 0.001  & 0.9028           & V7\_V10   & 0.9999 & 0.0104 & V1\_V7\_V8  & 0.8728 & 0.117  \\
V5        & 1.0251 & 0.001  & 0.4253           & V4\_V7    & 1.0004 & 0.0099 & V1\_V2\_V7  & 0.8743 & 0.1155 \\ \hline
\end{tabular}}
\caption{Experiment-1 (censoring rate is $10\%$): Ranked conditional entropies (CE) and successive CE-drop for selected feature-sets; ``SCE-dp'' short for ``SCE-drop''; ``$p$-value(PH)'' for the fitted Cox PH model. }
\label{CEexample1}
\end{table}

\begin{table}[h!]
\centering
\resizebox{\columnwidth}{!}{
\begin{tabular}{cccc|ccc|ccc}
\hline
1-feature & CE     & SCE-dp & $p$-value(PH)      & 2-feature & CE     & SCE-dp & 3-feature   & CE     & SCE-dp \\ \hline
V7        & 1.0759 & 0.0116 & \textless{}2e-16 & V2\_V3    & 1.0034 & 0.0827 & V2\_V3\_V7  & 0.8997 & 0.1037 \\
V1        & 1.0773 & 0.0102 & \textless{}2e-16 & V1\_V7    & 1.0563 & 0.0196 & V1\_V2\_V3  & 0.9032 & 0.1002 \\
V10       & 1.0861 & 0.0014 & 0.5769           & V1\_V10   & 1.0649 & 0.0124 & V2\_V3\_V6  & 0.9056 & 0.0978 \\
V3        & 1.0862 & 0.0013 & 0.4753           & V1\_V9    & 1.0652 & 0.0121 & V2\_V3\_V10 & 0.9058 & 0.0977 \\
V8        & 1.0866 & 9e-04  & 0.7523           & V7\_V8    & 1.0652 & 0.0107 & V2\_V3\_V8  & 0.9108 & 0.0927 \\
V2        & 1.0866 & 8e-04  & 0.0866           & V3\_V7    & 1.0652 & 0.0107 & V2\_V3\_V4  & 0.9122 & 0.0913 \\
V6        & 1.0867 & 8e-04  & 0.9617           & V7\_V10   & 1.0657 & 0.0102 & V2\_V3\_V5  & 0.9136 & 0.0898 \\
V9        & 1.0868 & 7e-04  & 0.5642           & V4\_V7    & 1.0658 & 0.0101 & V2\_V3\_V9  & 0.9174 & 0.0861 \\
V4        & 1.0868 & 7e-04  & 0.7306           & V6\_V7    & 1.0658 & 0.0101 & V1\_V7\_V10 & 0.9545 & 0.1017 \\
V5        & 1.0869 & 6e-04  & 0.986            & V1\_V5    & 1.0665 & 0.0108 & V1\_V7\_V9  & 0.9553 & 0.101  \\ \hline
\end{tabular}}
\caption{Experiment-2 (censoring rate is $20\%$): Ranked conditional entropies (CE) and successive CE-drop for selected feature-sets; ``SCE-dp'' short for ``SCE-drop''; ``$p$-value(PH)'' for the fitted Cox PH model. }
\label{CEexample2}
\end{table}

\begin{table}[h!]
\centering
\resizebox{\columnwidth}{!}{
\begin{tabular}{cccc|ccc|ccc}
\hline
1-feature & CE     & SCE-dp & $p$-value(PH)      & 2-feature & CE     & SCE-dp & 3-feature   & CE     & SCE-dp \\ \hline
V1        & 1.0996 & 0.0096 & \textless{}2e-16 & V2\_V3    & 1.0466 & 0.0616 & V1\_V2\_V3  & 0.9561 & 0.0906 \\
V7        & 1.1007 & 0.0085 & \textless{}2e-16 & V1\_V7    & 1.0851 & 0.0146 & V2\_V3\_V7  & 0.9609 & 0.0858 \\
V9        & 1.1078 & 0.0014 & 0.697            & V1\_V8    & 1.0894 & 0.0102 & V2\_V3\_V8  & 0.9647 & 0.0819 \\
V6        & 1.1081 & 0.0011 & 0.436            & V1\_V2    & 1.0899 & 0.0097 & V2\_V3\_V6  & 0.965  & 0.0816 \\
V10       & 1.1082 & 0.001  & 0.871            & V1\_V9    & 1.0904 & 0.0093 & V2\_V3\_V5  & 0.9667 & 0.08   \\
V2        & 1.1083 & 0.001  & 0.988            & V1\_V4    & 1.0906 & 0.009  & V2\_V3\_V10 & 0.9676 & 0.079  \\
V8        & 1.1084 & 8e-04  & 0.562            & V7\_V9    & 1.0908 & 0.0099 & V2\_V3\_V4  & 0.9679 & 0.0788 \\
V5        & 1.1084 & 8e-04  & 0.877            & V1\_V10   & 1.0908 & 0.0088 & V2\_V3\_V9  & 0.9683 & 0.0784 \\
V4        & 1.1087 & 6e-04  & 0.808            & V1\_V5    & 1.0912 & 0.0084 & V1\_V3\_V7  & 0.9981 & 0.0869 \\
V3        & 1.1089 & 3e-04  & 0.323            & V6\_V7    & 1.0912 & 0.0095 & V1\_V7\_V8  & 0.9996 & 0.0854 \\ \hline
\end{tabular}}
\caption{Experiment-3 (censoring rate is $30\%$): Ranked conditional entropies (CE) and successive CE-drop for selected feature-sets; ``SCE-dp'' short for ``SCE-drop''; ``$p$-value(PH)'' for the fitted Cox PH model. }
\label{CEexample3}
\end{table}

\section{ADNI data analysis}\label{Sec:adni_analysis}

\paragraph{}For the CEDA paradigm, the scheme of data categorization for all quantitative variables is given as follows. Measurements of each quantitative covariate feature are grouped into 4 equal-spaced bins. As we only have 903 subjects in total, 4 bins would be an appropriate choice to conduct the CEDA analysis to avoid the effect of the curse of dimensionality. There are 2 binary categorical features, GENDER ($V2$) and FLDSTRENG-bl ($V5$). As for the $T_i$ and $C_i$ which have their ranges within $[6,162]$ (month). It is noted that the maximum observed $T_i$ is $138$ and the observed $C_i$ is 162. We opt to obtain time bins by the following scheme: dividing $[6,162]$ into 4 bins: $[6,46), [46,85), [85,139), [139,163)$ so that all observed $T_i$ are included in the first three bins. In this way, we are able to clearly learn the structure and characteristics between the observed time ($T_i$) and censoring time ($C_i$).

Denote the right censoring ADNI data set as $\{(Y_i, {\cal V}_i, \delta_i)|i=1,.., n\}$ with $Y_i=(T_i \wedge C_i)$, ${\cal V}_i=\{V1_{i}, ..,V16_{i}\}$, $\delta_i$ the censoring status and sample size $n=903$. If the $i$th subject is uncensored, $\delta_i=1$, then  $Y_i=T_i  (< C_i)$ is its observed survival time, while if the $i$th subject is censored, $\delta_i=0$, then $Y_i=C_i (<T_i) $ is the censoring time defined as the exam date of first no-show. The total number of uncensored data points is $n_o=\sum^{903}_{i=1} \delta_i=346$, while the total number of censored data points is $n_c=n-n_o=557.$ So, this data set's censoring rate is over $61\%$.

\begin{figure}[h!]
 \centering
   \includegraphics[width=0.7\textwidth]{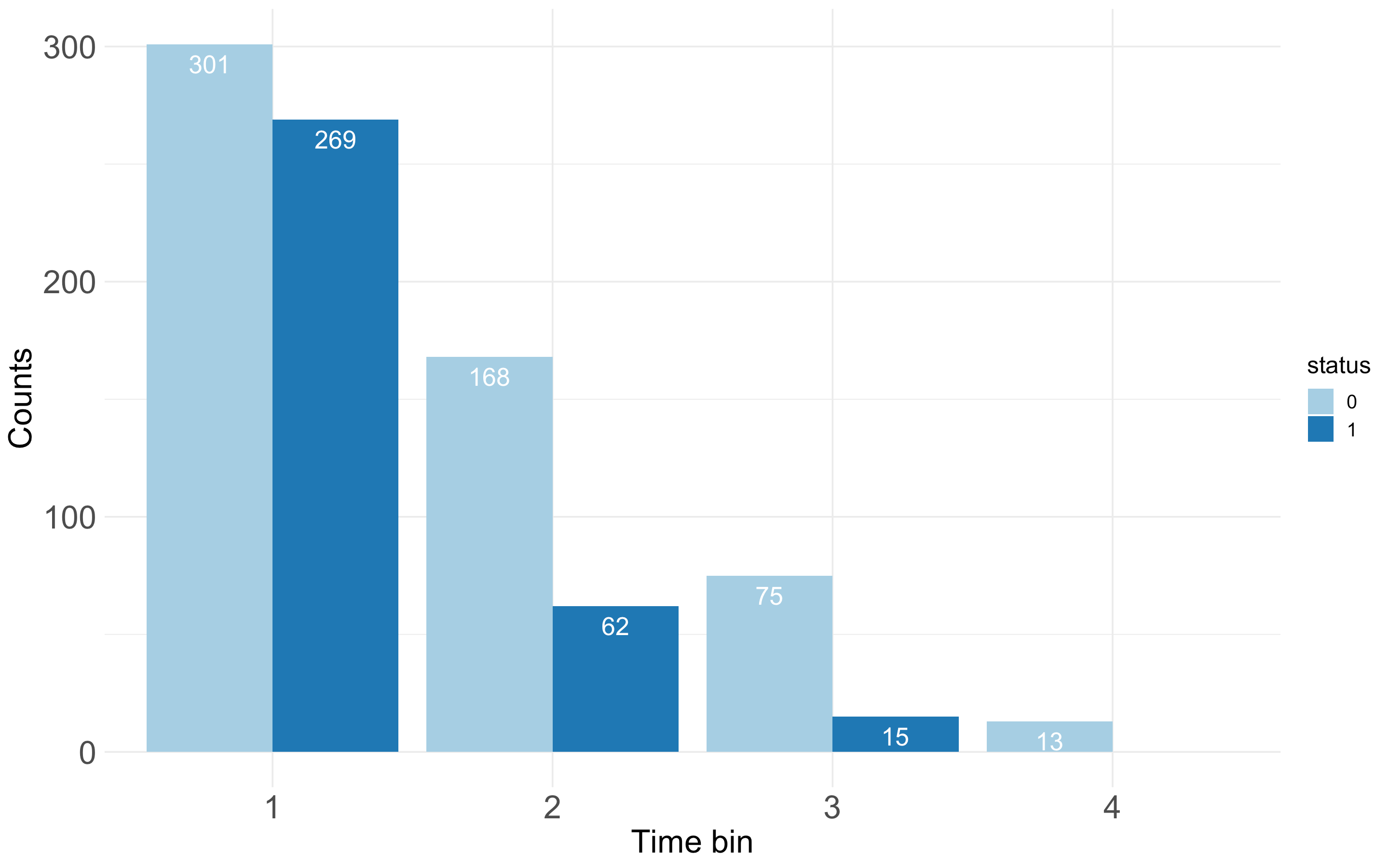}
 \caption{Histograms of 557 $C_i$ (censoring time) (in dark blue color bars) and 346 observed $T_i$ (in light-blue bars).}
 \label{histogram}
 \end{figure}

Consider the two ensembles of observed censoring and survival times: $\{C_i|\delta_i=0, i=1,..,557\}$ and $\{T_i|\delta_i=1, i=1,..,346\}$.
It is noted the largest $C_i$ among the censored data point is $\max\{C_i| \delta_i=0, i=1,...,557\}=162$, while the largest $T_i$ among the uncensored data point is $\max\{T_i| \delta_i=0, i=1,...,346\}=138$. Therefore, we have to take the convention that the largest censored data is taken as an uncensored one as usually done within many computational operations in Survival Analysis, for instance, in constructing Kaplan-Meier estimation of the survival function\cite{Kaplan1958}. We make use of this convention because the redistribution-to-the-right algorithm developed by B. Efron in \cite{Efron1967} is heavily applied in this paper. With respect to these two ensembles, we report their histograms with respect to four bins: $[6,46)$, $[46,85)$,  $[85,139)$, $[139,163)$, in Figure~\ref{histogram}.

Since each individual subject's potential $T_i$ and $C_i$ are to be realized and observed at a common setting: a scheduled examination date, it becomes not at all obvious whether the $T_i$ is indeed stochastically independent of $C_i$. Needed rigorous testing is constructed and conducted in the next subsection. The efforts invested in confirming this answer are essential because either positive or negative answers to this required fundamental assumption are expected to have significant impacts on the validity of any applications of the Cox Proportional Hazard model, which would be briefly reviewed below.

The Cox Proportional Hazard model, which was proposed by D.R. Cox in his 1972 landmark paper \cite{Cox1972}, is the most fundamental and the most popular modeling structure employed in the statistical topic of Survival Analysis. It is also widely used in analyzing data derived from ADNI as well. The classic version of the Cox proportional hazard (PH) model on the right censored data set is described as follows. Given  $\{(Y_i, {\cal V}_i, \delta_i)|i=1,.., n\}$ and assuming $T_i$ being stochastically independent of $C_i$, the hazard rate function $\lambda_{\tilde{T}_i}(t)$ of $\tilde{T}_i$ is specified as:
\[
\lambda_{T_i}(t)=\lambda_0(t) e^{\sum^{16}_{k=1}\beta_k Vk_i},
\]
for all $i=1, .., n$. The assumed structural linearity and additivity are designed to accommodate all effects of the 16 covariate variables. The exponent of $\lambda_{T_i}(t)$ certainly can involve interacting effects among the 16 covariate features effects. Nevertheless, given no prior knowledge of which forms of interacting effects pertain to which pairs or triplets of features, the inclusion of interacting effects would result in an unrealistic and complex model.

Under the PH model structures and non-informative censoring assumption, the partial likelihood approach proposed in \cite{Cox1972} is still the most widely used inference methodology in Survival Analysis. Nevertheless, it is worth reiterating that the global structure embraced by this PH model indeed is built upon an assumed homogeneity across the entire population contained in ADNI. From a rigorous standpoint, this homogeneous assumption is neither natural nor scientific. Though it might be practical, it is just parsimonious at best. Since this data set from ADNI is subject to two characteristics that could significantly impact results derived from Cox proportional hazard model structure. These two characteristics are: 1) the presence of heavy censoring; 2) the hidden heterogeneity among subjects. Both characteristics likely violate the validity regarding the global structures. In the last three subsections of this section, the partial likelihood results of the Cox proportional hazard model are compared with results derived based on our major factor selections on two scales: global and local.

\subsection{Redistribute-to-the-right weight matrix.}
\paragraph{}In this data analysis, all computations are primarily performed on the platform of the contingency table. This platform explicitly and correctly facilitates all calculations of measurements of conditional entropy, as such, we evaluate directed or indirect associations between the survival time $T_i$ and any other covariate variables. When building a contingency table for such association involving variable $T_i$, we need to adopt a convention in Survival Analysis: the censored subject having its censoring time being equal to $\max\{C_j| \delta_j=0, j=1,...,557\}=162$ is converted into an uncensored subject because this censoring time is beyond $\max\{T_i| \delta_i=1, i=1,...,346\}=138$. With this convention, we illustrate how to build a $903\times 347$ weight-matrix by applying the [Redistribution-to-the-right] algorithm discussed in Section~\ref{Sec:CEDA} in this subsection. This weight matrix would the basis for building a contingency table pertaining to any covariate features or feature-sets against variable $T_i$.

For illustrative purposes, we build a $903\times 347$ matrix lattice for distributed weights pertaining to the variable $\{Y_i\}$ against the variable $\{T_i\}$. By breaking all ties, we make all values of $\{Y_i|i=1,.., 903\}$ distinct, and then ordered and arranged them from bottom-to-top along this matrix's row-axis, while all observed $\{T_{i'}|\delta_{i'}=1, i'=1,.., 347\}$ are made distinct, ordered and arranged from left-to-right along this matrix's column-axis. It is noted again here that the largest uncensored $T_{(347)}=162$, which is originally censored.

\begin{figure}[h!]
 \centering
   \includegraphics[width=0.7\textwidth]{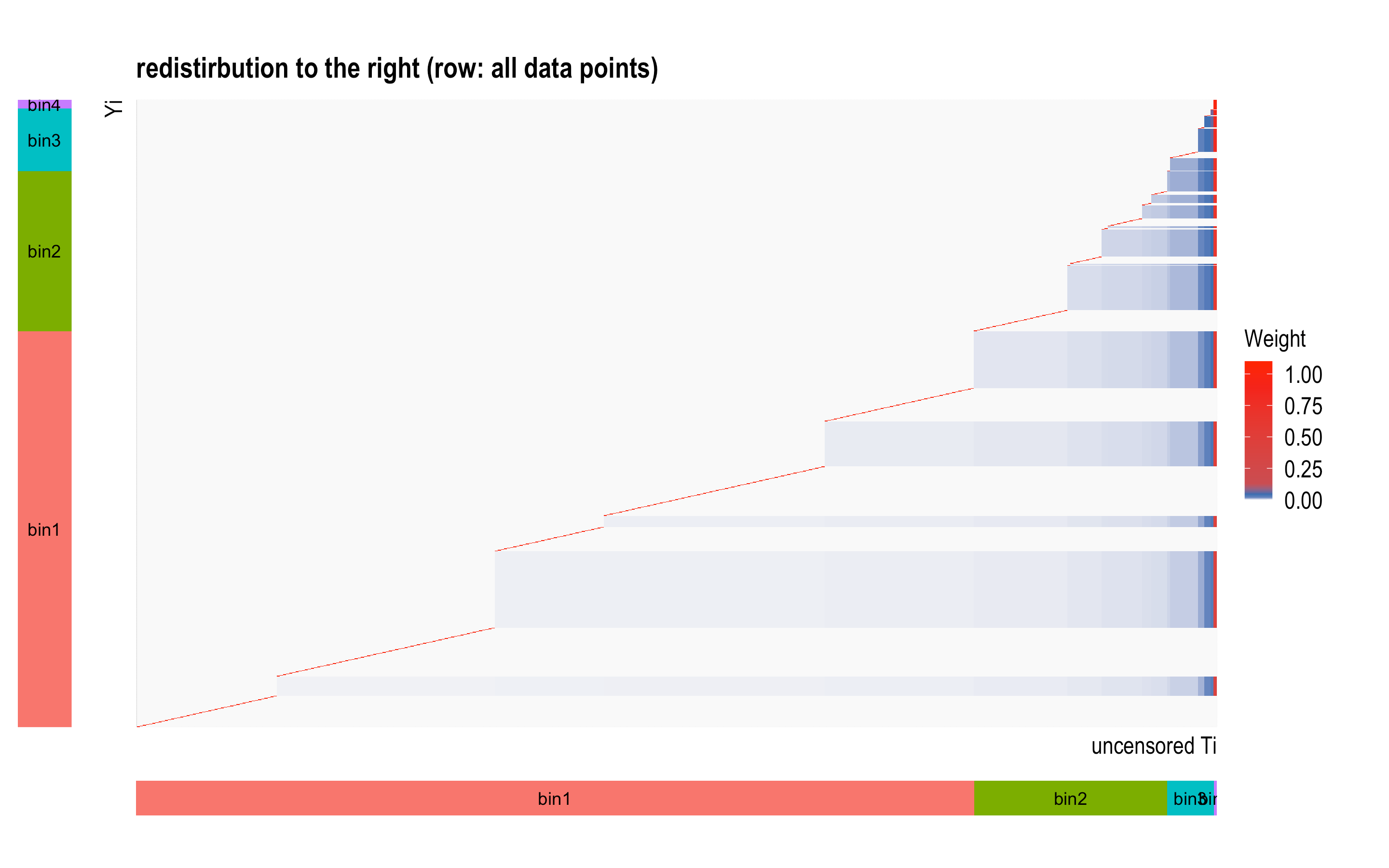}
 \caption{Matrix of distributed weights of individual 903 ranked $Y_i$ against 347 (=346+1) observed survival time $T_i$; two color bars indicate the corresponding 4 divided time bins.}
 \label{weightmatrix}
 \end{figure}

Upon this $903\times 347$ matrix lattice, each of its rows is constructed in the following fashion:
\begin{description}
\item[[$\delta_{(i)}=1$]:] If the $i$-th ranked $Y_{(i)}$ is uncensored ($\delta_{(i)}=1$), then the weight 1 goes to the $i'$-th column of $T_{(i')}=Y_{(i)}$, which is color-marked as a red-dot in Figure~\ref{weightmatrix};
\item[[$\delta_{(i)}=0$]:] If the $i$-th ranked $Y_{(i)}$ is censored ($\delta_{(i)}=0$), then the weight 1 would be re-distributed to all columns of $T_{(i')}>Y_{(i)}$ according to the [Redistribution-to-the right] algorithm, are represented by in changing-color-segment in Figure~\ref{weightmatrix}.
\end{description}
Let this matrix of distributed weights be denoted by ${\cal W}[Y, T]$, which is displayed in Figure~\ref{weightmatrix}.

This weight matrix ${\cal W}[Y, T]$ would play a very fundamental role in all our applications of major factor selection protocol throughout this data analysis from here on. Since we can replace this feature of time variable $Y_i$ with any one of the 16 features or feature-sets. That is, each row of ${\cal W}[Y, T]$ is identified via the subject's ID and its censoring status $\delta_{i}$. As such, a weight matrix ${\cal W}[Vk, T]$ is obtained by permuting the rows according to the increasing orders with respect to ordered values of $Vk$.

For any $k=1, .., 16$, on the column-axis of ${\cal W}[Vk, T]$, the values $\{T_{(i')}|\delta_{(i')}=1, i'=1,.., 347\}$ are grouped into four bins: $[6,46)$, $[46,85)$, $[85,139)$, $[139,163)$. It is noted that the bin $[139, 163]$ contains only censored data points. Likewise, upon the row-axis of ${\cal W}[Vk, T]$, the values $\{V_{k_{(i)}}| i=1,.., 903\}$ are also grouped into the four bins, which can be determined with respect to a histogram of $Vk$.

\subsection{Testing non-informative censoring assumption.}
\begin{figure}[h!]
 \centering
   \includegraphics[width=1\textwidth]{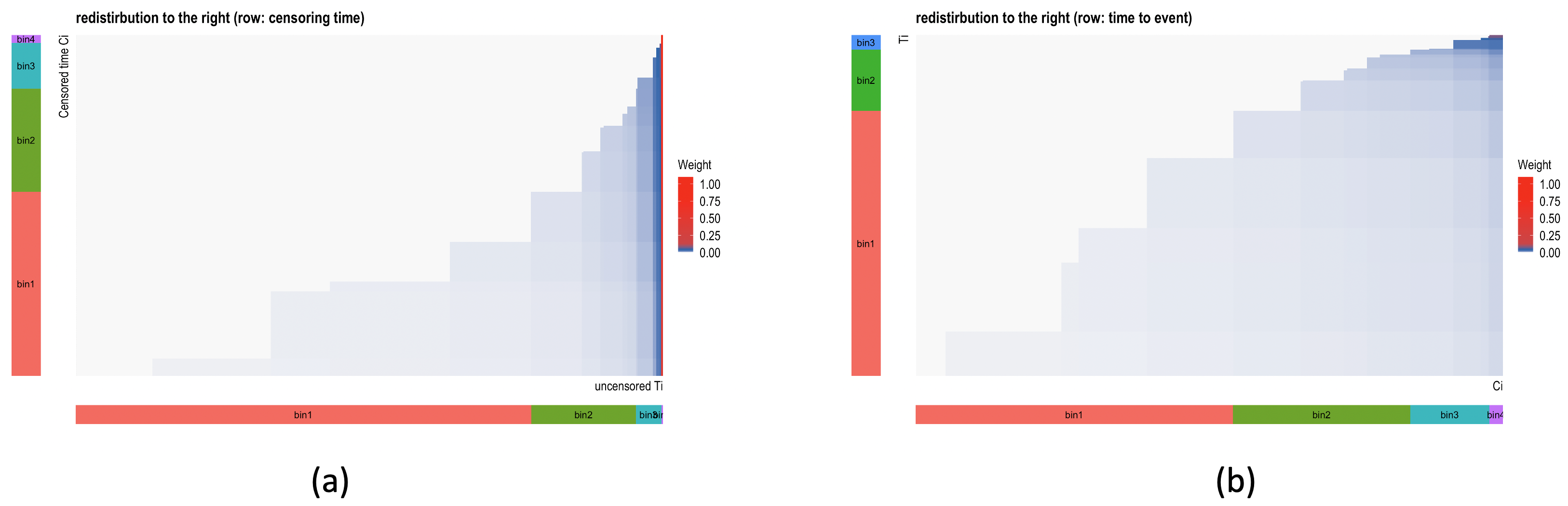}
 \caption{(a) Matrix of distributed weights of individual 557 ranked $C_i$ against 347 (=346+1) observed survival time $T_i $ and (b) matrix of distributed weights of individual 346 ranked $T_i$ against 557 observed censored time $C_i $; color bars indicate the corresponding 4 divided time bins.}
 \label{2weightmatrix}
 \end{figure}

For testing the non-informative censoring assumption, we need to explore the associative relation between survival time ($T_i$) and censoring time ($C_i$). However, we only observed $Y_i==(T_i \wedge C_i)$ as the minimum $T_i$ and $C_i$. That is, when $\delta_{i}=0$, $T_i$ is censored by $C_i$. But when $\delta_{i}=1$, $C_i$ is censored by $T_i$. The dual roles of $T_i$ and $C_i$ make this missing-data mechanism symmetric. Therefore, when $\delta_{i}=0$ and observed $C_i$, we figure out where the missing bivariate$ (C_i, T_i)$ could potentially located by using the [Redistribution-to-the-right] algorithm. The $557\times 347$ matrix of distributed weights, denoted as ${\cal W}[C, T]_{\delta_{i}=0}$, is reported in panel (a) of Figure~\ref{2weightmatrix}. This weight matrix is indeed obtained simply by deleting the 346 rows of ${\cal W}[Y, T]$ corresponding to $\delta_{i}=1$.

In contrast, when $\delta_{i}=1$ and observed $T_i$, we again figure out where the missing bivariate$ (C_i, T_i)$ could potentially be located by using the [Redistribution-to-the-right] algorithm. The $346\times 557$ matrix of distributed weights, denoted as ${\cal W}[T, C|]_{\delta_{i}=1}$, is reported in panel (b) of Figure~\ref{2weightmatrix}.

By applying the binning-scheme with respect to the four interval regions: $[6,46)$, $[46,85)$,  $[85,139)$, $[139,163)$, on both axes of ${\cal W}[C, T]_{\delta_{i}=0}$ and transposed matrix ${\cal W}^T[T, C|]_{\delta_{i}=1}$, we obtain two $4\times 4$ contingency tables, as reported in {4x4tableCT}. The contingency table ${\cal C}[C, T]$ denotes the sum of these two $4\times 4$ contingency tables.

\begin{table}[h!]
\centering
\begin{tabular}{c|cccc}\hline
\diagbox[width=5.5em]{bin-C}{bin-T} & 1 & 2  & 3  & 4 \\ \hline
1 &  46.70  & 60.58 & 54.15 & 139.56 \\
2 & 0.00  & 17.33 & 42.12 & 108.55 \\
3 &0.00 & 0.00 & 28.79 & 61.21 \\
4 & 0.00 & 0.00 &  0.00 &13.00\\
\hline
1 &  58.88 & 0.00& 0.00 &0.00\\
2 & 122.22& 22.53 &0.00& 0.00\\
3 &72.09& 31.65 & 7.60& 0.00\\
4 &17.81& 7.82 & 7.40  &0.00\\\hline
\end{tabular}%
\caption{Contingency table ${\cal C}[C, T]$: the sum of the $4\times 4$ contingency table derived from $557\times 347$ weight matrix ${\cal W}[C, T]_{\delta_{i}=0}$ (upper half) and the $4\times 4$ contingency table derived from $346\times 557$ weight matrix ${\cal W}^T[T, C|]_{\delta_{i}=1}$ (lower half).}
\label{4x4tableCT}
\end{table}

This contingency table ${\cal C}[C, T]$ reported in Table~\ref{4x4tableCT} indeed manifests the multiple aspects of associative relations between the censoring time $C_i$ and survival time $T_i$. From the row- and column-wise aspects, we calculate the row-wise and column-wise Shannon conditional entropies and re-scale them by Shannon entropies of vectors of proportions of column-sums and row-sums, accordingly. Hypothetically, if a row-wise re-scaled CE is close to 1, then we know the information about $C_i$ not helping us in predicting $T_i$, while a column-wise re-scaled CE is away from 1 and smaller than 1, then we know the information of $T_i$ indeed helping us in predicting $C_i$.

\begin{figure}[h!]
 \centering
\includegraphics[width=1\textwidth]{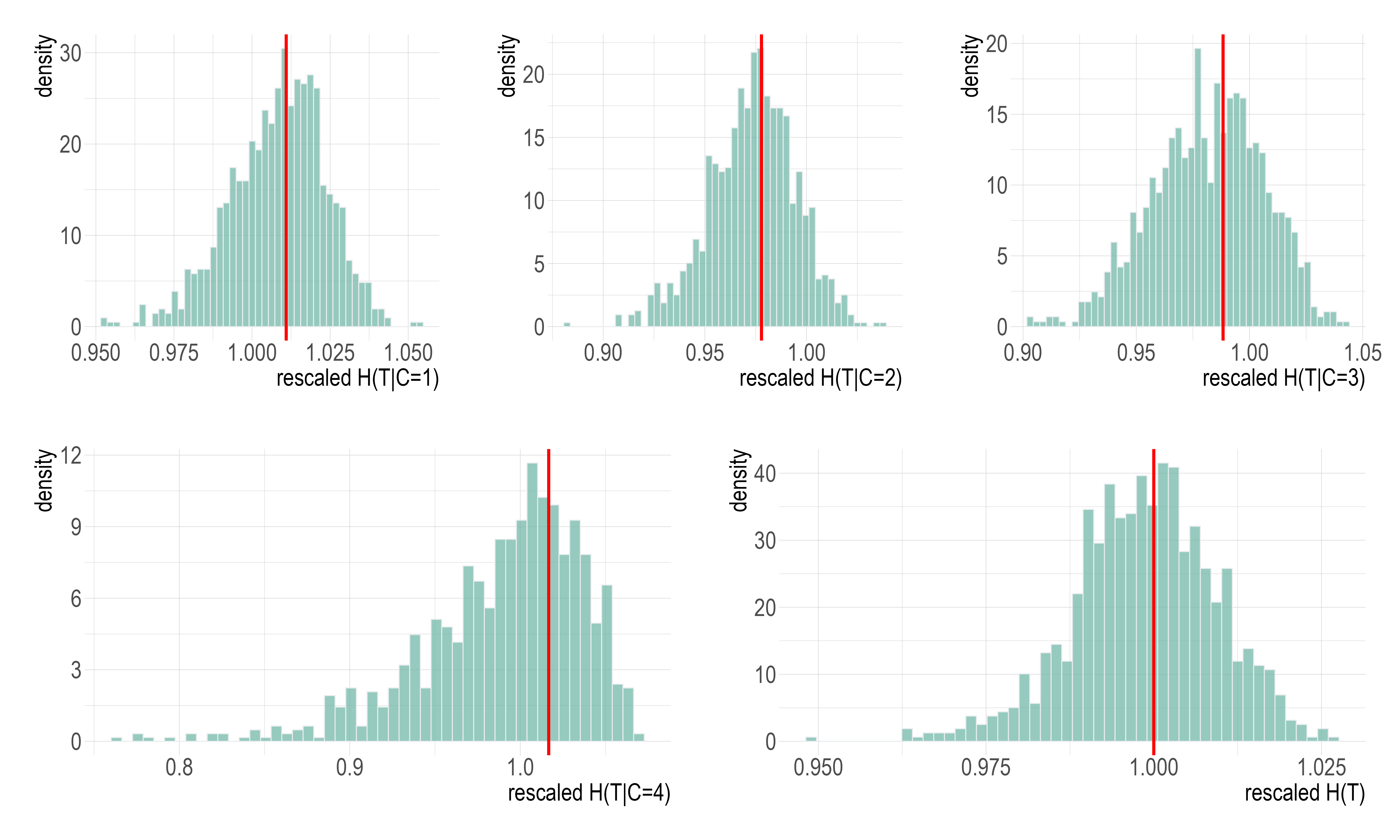}
 \caption{Four row-wise simulated alternative distributions marked with vertical lines at observed re-scaled CEs \{1.0109, 0.9778, 0.9883, 1.0167\} against one simulated null distribution marked with one vertical line at 1.00. The entropy of vector of column-sum proportions is 1.2979.}
 \label{rowCEtest}
 \end{figure}

\begin{figure}[h!]
 \centering
   \includegraphics[width=1\textwidth]{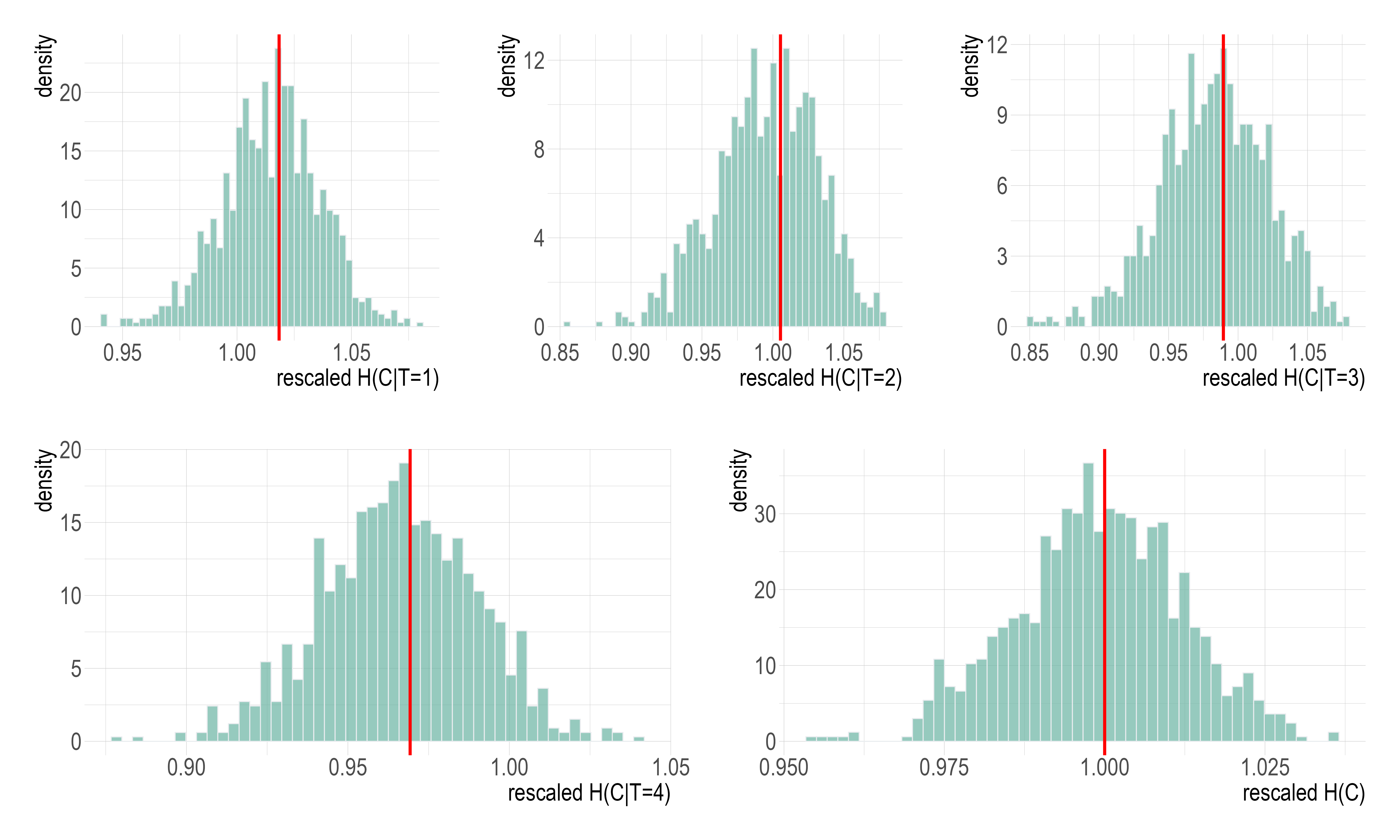}
 \caption{Four column-wise simulated alternative distributions marked with vertical lines at observed re-scaled CEs \{1.0184, 1.0053, 0.9894, 0.9693\} against one simulated null distribution marked with one vertical line at 1.00. The entropy of vector of row-sum proportions is 1.2111.}
 \label{colCEtest}
 \end{figure}

Our testing results from row-aspect are displayed in Figure~\ref{rowCEtest} and testing results from column-aspect are displayed in Figure~\ref{colCEtest}. Based on the Multinomial random mechanism, we simulate four row-wise alternative distributions against the null distribution based on the vector of column-sums as displayed in Figure~\ref{rowCEtest}. All five distributions heavily overlap. Therefore, the sum of Type-I and Type-II errors is large. Likewise, as displayed in Figure~\ref{colCEtest}, similar results are found through the four column-wise alternative distributions against the null distribution based on the vector of row-sums. Thus, we conclude that associative relations between the censoring time $C_i$ and survival time $T_i$ are not evident. The non-informative censoring assumption stands.

The critical implication derived from the confirmation of the noninformative censoring scheme is regarding which response variables, either $T_i$ or $(T_i, C_i)$, are legitimate and should be used in Survival analysis. Since $T_i$ is stochastically independent of $C_i$, we are not required to use $(T_i, C_i)$ as the legitimate 2D response variable, or to use the categorical variable defined by the 16 categories in Table~\ref{4x4tableCT}. That, we only need to use the categorized variable of $T_i$ defined by the 4D vector of column-sums in Table~\ref{4x4tableCT} throughout the categorical exploratory data analysis (CEDA) carried out in this paper.

\subsection{Using $\delta_i$ as a response variable.}\label{Sec:delta_i}
After confirming the fundamental assumption of non-informative censoring, we first look into the issue of whether the observed data of censoring status $\{\delta_i|i=1, .., n\}$ could indeed shed light on the dynamics of $T_i$. The rationale is that $\delta_i=1_{[T_i \leq C_i]}=1-1_{[T_i > C_i]}$ is categorical transformation of $T_i$. Since $T_i$ is stochastically independent of $C_i$, this transformed data logically should still carry relevant information. So, as the first step of our categorical exploratory data analysis (CEDA), we seek for any covariate variables $\{V1, ..,V16\}$ that is highly associated with $\delta_i$. We perform our CE computations by employing $\delta_i$ as a binary response variable and all covariate variables are categorized to have 4 bins based on their individual histograms. The relevant CEs are reported in Table~\ref{CEdropDel}.

\begin{table}[h!]
\centering
\begin{tabular}{cccc}\hline
1-feature & Feature name & CE&	SCE-drop \\ \hline
V9&	MEM-mean&0.5137&0.1518\\
V8&	ADAS13.bl&0.5731&0.0924\\
V7&	FAQ	&0.6154&0.0501\\
V6&	CDRSB.bl&0.6156&0.0500\\
V13&LAN-mean&0.6200&0.0455\\
V11&EXF-mean&0.6207&0.0448\\
V4&	APOE4	&0.6420&0.0236\\
V16&VSP-std&0.6530&0.0125\\
V15&VSP-mean&0.6560&0.0095\\
V5&	FLDSTRENG.bl&0.6571&0.0084\\
V1&	AGE&0.6575&	0.0080\\
V12&EXF-std&0.6595&0.0060\\
V10&MEM-std&	0.6600&0.0055\\
V14&LAN-std&0.6650&	0.0006\\
V3&	PTEDUCAT	&0.6650&0.0005\\
V2&	PTGENDER&0.6653&	0.0002\\\hline
\end{tabular}%
\caption{Conditional entropies (CE) of 16 features with censoring status $\delta_i$ as a response variable. }
\label{CEdropDel}
\end{table}

From Table~\ref{CEdropDel}, we see that $\delta_i$ is apparently associated with V9 (MEM-mean) and V8 (ADAS13.bl), both of which achieve about $23\%$ and $13\%$ reductions of $\delta_i$'s uncertainty, respectively. This result implies that the directed association from $T_i$ to the 16 covariate features surely would be much stronger than the results reported strongly.

\subsection{Computing major factors}
\begin{figure}[h!]
 \centering
   \includegraphics[width=1\textwidth]{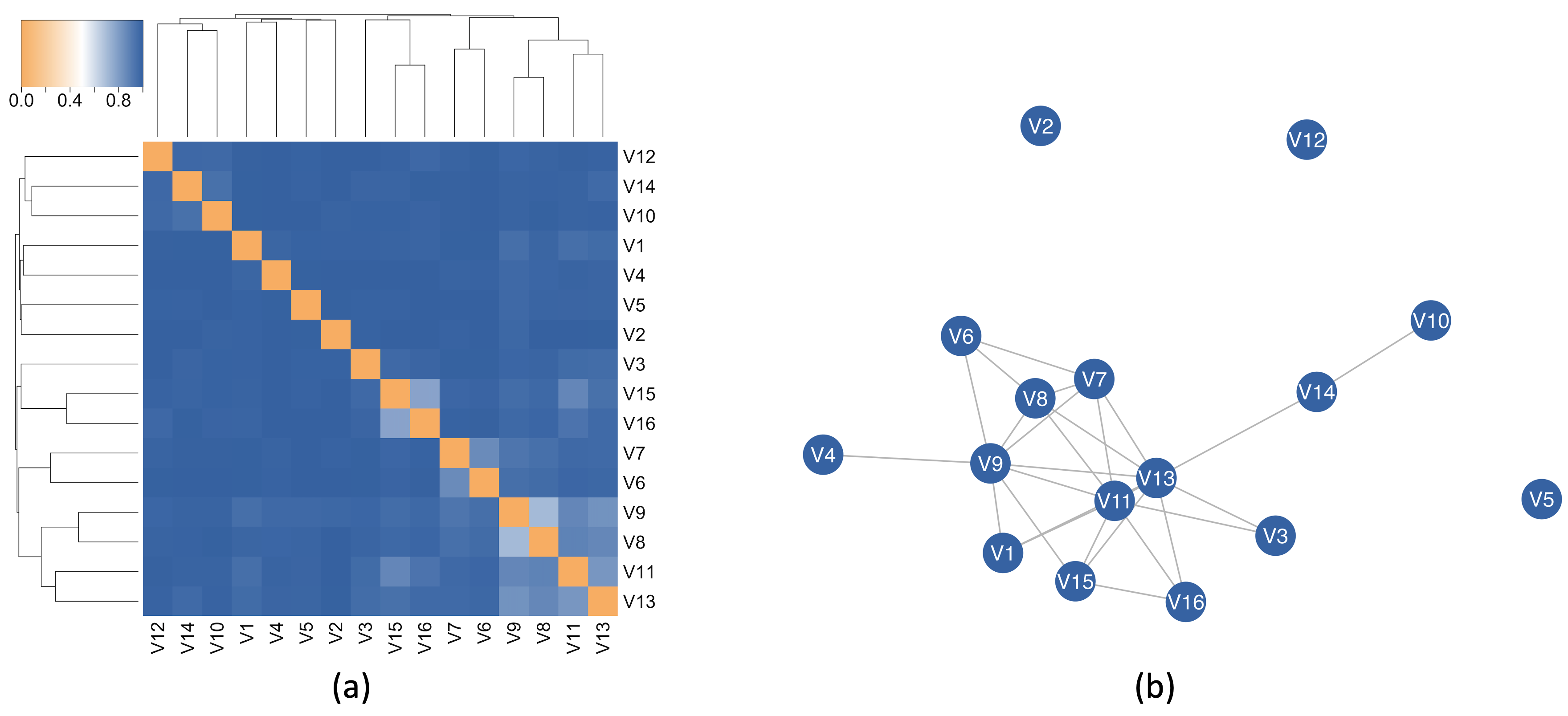}
 \caption{Heatmap (a) and network (b) of 16 covariate features in terms of Mutual Conditional Entropy (MCE). The threshold of the linkage is 0.97.}
 \label{heatnet}
 \end{figure}

Before exploring the directed association of all 16 covariate variables $\{V1, ..,V16\}$ to response variable $T_i$, we display mutual association among these 16 covariate features via a $16\times 16$ heatmap and a network of 16 nodes in the two panels of Figure~\ref{heatnet}. From panel (a), we see two highly associated feature-pairs: \{V9, V8\} and \{V15, V16\}, two moderately associated feature-pairs:\{V6, V7\}, \{V11, V13\}, and moderately associated feature-triplet:\{V10, V12, V14\}. These feature-sets are also mutually associated with varying degrees. In contrast, panel (b) reveals a global picture of the association among these 16 features with respect to a chosen threshold.

It is worth reiterating that, as mentioned in Section~\ref{Sec:sim} and detailed in \cite{Hsieh2022}, any highly associative relationships among features would require extra computational efforts for explicitly clarifying and evaluating their joint conditional dependency or interacting effects much harder. We demonstrate why such required efforts are needed in this subsection by implementing our major factor selection protocol facilitated by contingency-table-based CE computations within this ADNI data set. Nonetheless, as would be discussed in the two subsections after this one, such efforts are indeed needed anyway for exploring heterogeneity within data.

The first step of the major factor selection protocol is performed by calculating the conditional entropies (CEs) of all possible feature-sets of the collection of 16 covariate features from the 1-feature setting up to the 3-feature setting. Our CEs computations end at the 3-feature setting is a necessary choice for reflecting the uncensored sample size 346. All CEs of the 1-feature setting is reported in Table~\ref{CEdrop1F} coupled with results derived from Cox PH models on an individual feature basis. Specifically, we also report $p$-values of partial likelihood estimates of $\{\beta_1, ..,\beta_{16}\}$ parameters. As for CEs of 2-feature and 3-feature settings, we only select and report 12 primary top-ranked pairs and triplets in Table~\ref{CEdrop2n3F} due to the large numbers of feature-pairs and feature-triplets.

\begin{table}[h!]
\centering
\begin{tabular}{ccccc}\hline
1-feature &Feature name& CE&	SCE-drop & $p$-value(PH) \\ \hline
V9 &	MEM-mean	&1.1581 &0.1397& \textless{}2e-16\\
V8 &	ADAS13.bl	            &1.1954  &0.1023&0.5007\\
V11&	EXF-mean	   &1.2354  &0.0624&0.0007\\
V7 & FAQ	                       &1.2419  &0.0559&\textless{}2e-4\\
V6 &CDRSB.bl                  &1.2492   &0.0486&\textless{}2e-4\\
V13& LAN-mean   &1.2526  &0.0452&0.5007\\
V4 & APOE4	                   &1.2697  &0.0281&0.0013\\
V5 & FLDSTRENG.bl  	&1.2849    &0.0129&0.2386\\
V15 &	VSP-mean     &1.2859  &0.0118&0.2343\\
V14	& LAN-std    &1.2863   &0.0115&0.0002\\
V12	& EXF-std	     &1.2872   &0.0106&0.0100\\
V10&	MEM-std      &1.2879  &0.0099&0.7764\\
V1 & AGE	                     &1.2896   &0.0082&0.6314\\
V16 &	VSP-std	    &1.2900   &0.0078&0.0739\\
V3 &PTEDUCAT	        &1.2952    &0.0026&0.0025\\
V2&	PTGENDER	            &1.2973   &	0.0005&0.1110\\\hline
\end{tabular}%
\caption{Conditional entropies (CE) of 16 features with $T$ as response variable, along with $p$-values of features via Cox Partial likelihood approach.}
\label{CEdrop1F}
\end{table}

Before comparing CEs and Cox's PH results, it is necessary to keep in mind that each CE is pertaining to one individual feature, while a $p$-value of covariate feature is evaluated under a global PH modeling structure. Therefore, these 16 parameter estimates are correlated, so are their $p$-values. From Table~\ref{CEdrop1F}, we observe that feature: MEM-mean (V9), achieves the lowest CE, while ADAS13.bl (V8) is ranked 2nd. Both are potential order-1 major factor candidates, though they are not likely concurrently present when interpreting reduction of marginal uncertainty of $T_i$. That is, they could potentially join together in reducing uncertainty of $T_i$ under conditional settings, as would be seen in the last two subsections regarding heterogeneity from both V9 and V8 perspectives. This possibility is in sharp contrast with PH results, which indicate V9 is more essential by having a $p$-value nearly zero than V8 with a $p$-value 0.5007. In other words, the act of dismissing the importance of V8 could come with a cost of ignoring important information of heterogeneity via V8 perspective, as would be seen in the last subsection of this section.

The EXF-mean(V11), FAQ (V7), CDRSB.bl (V6), LAN-mean (V13) and APOE4 (V4) are ranked from the 3rd to the 7th. Though these four features' individual CE-drops are significantly less than the CE-drops of the top two, their CE-drops are still significantly larger than CE-drops of the rest of the 10 features. Therefore, it is still possible for the individual features of \{ V6, V7, V11, V13, V4 \} to be a candidate of stand-alone order-1 major factors. We also know the fact that \{V6, V7\} and \{V11, V13\} are moderately associated, while V4 is less associative with members of these two pairs. Thus, it is less likely that both V6 and V7 are concurrently present as two separate order-1 major factors, as are V11 and V13. On one hand, according to the PH results, V11 has a near zero $p$-value, but V13 has a $p$-value larger than 0.50. From the aspect of feature-pair \{V11, V13\}, the CEDA and PH results are coherent. On the other hand, both V6 and V7 have $p$-values near zeros. This PH result is not coherent with entropy-based CEDA results regarding the presence of order-1 major factors. The feature V4 has a $p$-value 0.0013 from the PH results. This computed PH result is not strongly incoherent from the CEDA perspective.

Behind these top 7 ranked features, the CE-drops of the rest of the 9 features are more or less falling into 2 tiers: \{MEM-std (V10), VSP-mean (V15), FLDSTRENG.bl (V5), VSP-std (V16), EXF-std (V12), LAN-std (V14), AGE (V1)\} for tier-1; \{PTEDUCAT (V3), PTGENDER (V2)\} for tier-2. From the CEDA perspective, feature members of these two tiers are increasingly less likely to be stand-alone order-1 major factors. But each one of them can couple with other features to become a potential candidate of order-2 major factor. In contrast, from the PH perspective, V14, V12, and V3 are highly significant by having $p$-values being equal to or less than 0.01. In Table~\ref{CEdrop1F}, from the CEDA perspective, these features only achieve $0.9\%$, $0.8\%$ and $0.2\%$ of uncertainty reduction on $T_i$. Thus, the degrees of incoherence between PH and CEDA results are somehow evident.

Next, we turn the comparison to CEDA results of 2-feature or 3-feature settings in Table~\ref{CEdrop2n3F} with a focus on interacting effects among feature-pairs and feature-triplets. Some of the above conflicting results between CEDA-based major factor selection and Cox PH could be reconciled, but not all.

Before specifically discussing potential interacting effects in the 2-feature and 3-feature settings in Table~\ref{CEdrop2n3F}, it is worth reiterating that an interacting effect of two or three features is referred to the confirmed presence of their conditional dependency given the response variable $Y$. Based on a relatively loose criterion in terms of SCE-drops, we identify and see how much extra uncertainty of response variable $T_i$ is reduced by including a less potential feature variable relative to 3 or more times this feature's individual CE-drop, see details in \cite{Chen2021,Hsieh2022}. A more strict criterion will be based on a criterion constructed based on the dominant feature. This strict version of the interacting effect is not used here. Further, the explicit form of interacting effect is left unknown when applying this criterion, it is somehow visible through the contingency table against the response variable $T$. In comparison, the task of confirming any conditional dependency of multiple orders is not at all simple under an assumed global structural model, such as Cox's PH model. In this paper, we explicitly demonstrate how to resolve this task.

Though the feature-set \{V7, V9\} achieves the lowest CE under the 2-feature setting, this pair's SCE-drop (0.0294) is less than the CE-drop (0.0559) of V7. This fact indicates that the conditional mutual information $I[V7, V9|T]$ is less than the marginal mutual information $I[V7, V9]$. That is, we can't confirm whether the feature-set \{V7, V9\} gives rise to a significant interacting effect or not. To confirm or dispute this fact. we either explicitly evaluate the mutual information $I[V7, V9]$, or evaluate how much extra information can V7 provide going beyond V9. We take the latter approach in the next subsection. Since we face the same issue for feature-sets \{V6, V9\}, \{V9, V11\}, \{V8, V9\} and \{V4, V9\}, which are the top-ranked 5 feature-pairs.

As for feature-set \{V9, V10\}, we know that V9 and V10 are marginally independent, so their marginal mutual information $I[V9, V10]=0$. And the SCE-drop (0.0148) of V10 is larger than its CE-drop (0.0099) under the 1-feature setting. Therefore, their conditional mutual information $I[V9, V10|Y]\approx 0.0049$ is barely positive. Thus, according to the above criterion for confirming the interacting effect, the feature-set \{V9, V10\} has a very slight chance of being conditionally dependent given $Y$. However, the V10 could still play the role of assisting the order-1 major factor V9 in facilitating the information content of this ADNI data set or any predictive decision-making locally, as discussed in the next subsection.

In contrast, the three feature-sets: \{V2, V9\}, \{V2, V8\} and \{V3, V8\} have their SCE-drops being almost three times of V2 and V3's individual CE-drops or more. All involving CE-drops are confirmed to pass the reliability check. Therefore, we confirm the interacting effects of V2 and V3 with V8 at least. In fact, V1, V2 and V3 apparently have interacting effects with many features: V6, V7, V10, V11, V12, V15 and V16 as well. It is worth mentioning that some of the aforementioned interacting effects are indeed clinically confirmed and reported in \cite{Altmann2014,Ardekani2020,Duarte2021}.

\begin{table}[h!]
\centering
\begin{tabular}{ccc|ccc}\hline
2-feature & CE&	SCE-drop  & 3-feature&CE&	SCE-drop\\ \hline
V7\_V9 (1)  & 1.1287 &0.0294  & V7\_V9\_V11 (1) &1.1002 &	0.0285\\
V6\_V9 (2) & 1.1346 & 0.0234 &V6\_V9\_V11 (2) &1.1035 &0.0311\\
V9\_V11 (3) &	1.1385 &0.0196 &V4\_V7\_V9 (3)&1.1039 &	0.0247\\
V8\_V9(4)& 1.1405  &0.0175 & V7\_V8\_V9 (4) &1.1076 & 0.0210\\
V4\_V9 (5) &1.1418  &0.0162 & V7\_V9\_V12 (5) &1.1078 &0.0209\\
V9\_V10 (6) &1.1432 &0.0148 & V6\_V7\_V9 (6) &1.1089 &	0.0197\\
V1\_V9(11) &1.1519 &0.0061   & V6\_V9\_V10 (7)&	1.1090 &0.0255\\
V3\_V9(14) &1.1533 &0.0047  &V7\_V9\_V10 (8) &1.1095 & 0.0191\\
V2\_V9(15) &1.1558 &0.0022  & V9\_V11\_V12 (9) &1.1099 &	0.0286\\
V8\_V16(22) &1.1832&0.0121 &V4\_V9\_V11 (10) &1.1108 & 0.0239\\
V3\_V8(26)  &1.1847 &0.0106 & V3\_V7\_V9 (16) &1.1129 &	0.0158\\
V2\_V8(30)  &1.1925 &0.0029  &V1\_V6\_V9 (20) &1.1144  &0.0202\\
\hline
\end{tabular}%
\caption{Ranked conditional entropies (CE) and successive CE-drop for selected feature-sets. }
\label{CEdrop2n3F}
\end{table}

In the 3-feature setting, we see that V9 is present in all the top-ranked feature-triplets. Its most often companion feature is either V7 or V11. There is a range of features for the third member of the triplet. Because of the marginal dependency of V9 on other features, the reliable pattern information becomes harder to be confirmed or disputed.

This phenomenon would be resolved to a great extent by subdividing the entire data collection with respect to categories of V9, which is called ``de-associating'' in \cite{Hsieh2022}. As reported in the four tables given in the next subsection in the next subsection, by applying this simple computational procedure, we can take away or significantly reduce all covariate features' associations with V9. Overall, the rest of the 15 covariate features become less associative within each sub-collection of study subjects defined by the 4 categories of V9, respectively. Another functional merit of de-associating is that this computing procedure allows us to figure out which feature variables can provide extra information beyond V9 while holding V9 constant. Therefore, this procedure is an effective way of discovering heterogeneity from the perspective of V9. We do the same heterogeneity exploration from the V8 perspective, as well as in the last subsection.

\subsection{Heterogeneity w.r.t V9 (MEM-mean)}
\paragraph{\{V9=1\}:$199(n_o)-vs-67(n_c)$.}
Upon the sub-collection: \{V9=1\}, of 266 subjects having the lowest ordinal category of MEM-mean (V9=1), we see the drastic ranking changes among the 14 covariate features in Table~\ref{CEdropV9eq1}. The highest-ranked feature is V7, which is ranked 4th on the overall CE-drop in Table~\ref{CEdrop1F}. The 2nd ranked one is V6, which is ranked 4th in Table~\ref{CEdrop1F}, while the originally ranked 11th V16 is now ranked 4th. The V8 is ranked 3rd here, while it was ranked 2nd in Table~\ref{CEdrop1F}. However, significant ranking drops are seen on features: V13, V4, and V5, which were previously ranked 6th to 8th. Now they are ranked 10th, 13th, and 15th here.

\begin{figure}[h!]
 \centering
   \includegraphics[width=0.9\textwidth]{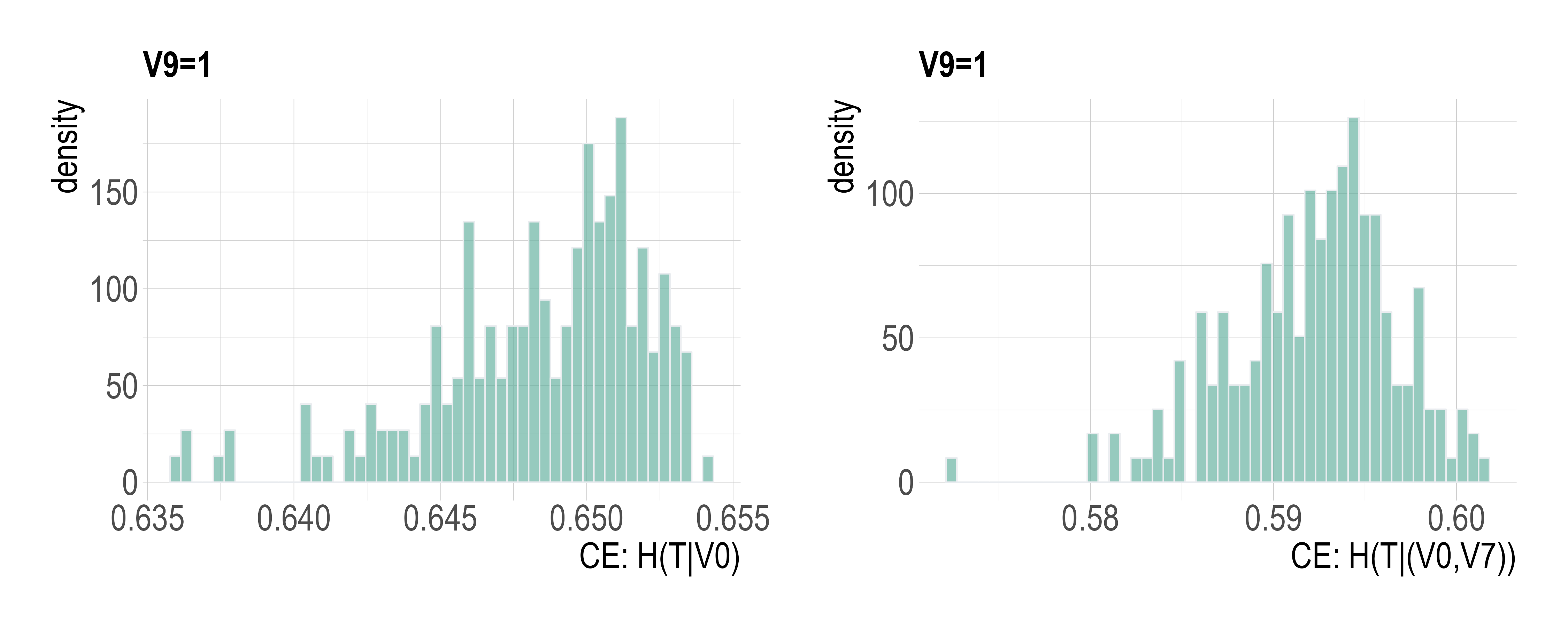}
 \caption{Null distributions for reliability checks for 1-feature and 2-feature settings: Left: simulated CEs of $H[T|V0]$; Right: simulated CEs of $H[T|(V0, V7)]$ in sub-collection \{V9=1\} based on 200 simulated $U[0,1]$ based features. }
 \label{Reliachec1}
 \end{figure}

For reliability check on CE calculations, we generate a random $U[0,1]$ noise feature, denoted as V0, repeat this 200 times and calculate CEs: $H[Y|V0]$. The simulated null distribution for the 1-feature setting is given in the left of Figure~\ref{Reliachec1}. The top 6 ranked 1-features all have very small p-values. That is, their CEs are significant and real. The V1 has the p-value on the borderline. As for the 2-feature setting, as shown in the right of Figure~\ref{Reliachec1}, we found that only the top 3 feature-pairs have only borderline p-values. Nonetheless, we still carry out the task of identifying potential interacting effects as follows.

The most significant result observed from CEs of 1-feature and 2-feature is that majority of feature-pairs achieve the ecological effect, except the feature-pair \{V6, V7\}. This is a strong indication that all involving features are much less associated due to the de-associating procedure. Further, as listed below, we see many feature-pairs just like these three pairs: \{V7, V15\}, \{V7, V13\}, \{V7, V3\} and \{V12, V14\}, achieve SCE-drops that are at least three times of V15, V13, and V3's individual CE-drops. These results collectively suggest the following collection of 16 candidates of order-2 major factors:
$$
\{\{V7, V3\}, \{V7, V4\}, \{V7, V13\}, \{V7, V15\}, \{V8, V3\},
$$
$$\{V8, V4\}, \{V11, V4\}, \{V16, V2\},  \{V14, V12\} \{V1, V4\},
$$
$$\{V10, V4\}, \{V12, V3\}, \{V12, V4\}, \{V12, V13\}, \{V2, V4\},  \{V13, V3\}\}.
$$

Such results strongly indicate that not only members of \{V7, V8, V11, V16, V1, V10\} are concurrent candidates of order-1 major factor, but also they are coupled with members of \{V4, V3, V2, V13, V15\} to be candidates of order-2 major factors. That is, these feature-pairs indeed provide extra information beyond what V9 can provide at least in this sub-collection. This fact is the strongest evidence of heterogeneity that goes far beyond 2-feature setting of Table~\ref{CEdropV9eq1}.

Next, it is also worth mentioning one sharp contrasting result: the feature-pair \{V7, V8\} achieves the ecological effect. So, V8 indeed can be concurrently present with V7 in reducing the uncertainty of $T_i$ under the sub-collection \{V9=1\}. This result is indeed rather significant because V8 is an important feature in AD literature. It is also interesting that the feature V16 (VSP-std), which doesn't demonstrate a significant role in the overall setting at all in Table~\ref{CEdrop1F} and Table~\ref{CEdrop2n3F}, surprisingly joins V7 and V8 in reducing the uncertainty of the response variable $Y$ within this sub-collection of \{V9=1\}. This is the second piece of evidence of heterogeneity embedded within ADNI data.

The third piece of evidence of heterogeneity is collectively provided by the ranking changes pertaining to the originally lowest-ranked three features in Table~\ref{CEdrop1F}: V1(Age), V3(PTEDUCAT), and V2(PTGENDER). They are now ranked 6th, 11th, and 12th, while V4's ranking changed from 7th to 10th.

When comparing CEDA and PH results, we clearly notice a few conflicting results observed from PH results reported in the 3rd column of Table~\ref{CEdropV9eq1}. Given that feature-pair \{V7, V6\} doesn't achieve the ecological effect, their concurrent presence is not confirmed in CEDA analysis. However, PH results strongly indicate both features are significant simultaneously. While \{V7, V8\} achieves the ecological effect in CEDA analysis, on the contrary, the PH results indicate V8 is insignificant by having a $p$-value 0.1209. Further, results based on the PH model indicate features: \{V11, V16, V1\}, have $p$-values less than 0.05 and V14 has a $p$-value between 0.05 and 0.1, while the rest of the features are not significant. That is, PH results have completely missed all potential interacting effects like the 16 feature-pairs identified by our major factor selection protocol. This is a chief difference between CEDA and PH results.

\begin{table}[h!]
\centering
\begin{tabular}{cccc|ccc}\hline
1-feature & CE&SCE-drop&$p$-value(PH) & 2-feature & CE&SCE-drop\\ \hline
V7&	0.6112 &0.0432 &0.0001&V7\_V11 &0.5877 &0.0235\\
V6&	0.6307 &0.0238 &0.0461&V7\_V16 &0.5879 &0.0233\\
V8&0.6347 &0.0198 &0.1209&V7\_V8	 &0.5886 &0.0226\\
V11&0.6364&0.0182 &0.0205&V1\_V7 &0.5906 &0.0206\\
V16&0.6375 &	0.0170 &0.0004&V4\_V7 &0.5919 &0.0193\\
V14&0.6390 &0.0156&0.0714&V7\_V14 &0.5925 &0.0187\\
V1&0.6442 &0.0104 &0.0002&V6\_V7&0.5930 &0.0182\\
V10&0.6461 &0.0084 &0.6311&V3\_V7 &0.5952 &0.0160\\
V12&0.6502 &	0.0043 &0.3741& V7\_V10&0.5967 &0.0145\\
V4&0.6503 &	0.0042 &0.5398&V7\_V15 &0.5971 &0.0141\\
V3&	0.6506 &0.0039 &0.0334&V7\_V12	&0.5984 &0.0128\\
V2&0.6506 &	0.0039 &0.1480& V2\_V7 &0.5994 & 0.0118\\
V13&	0.6511 &0.0034 &0.3634&V11\_V14&0.6005 & 0.0358\\
V15&0.6534 & 0.0011 &0.3666&V7\_V13 &0.6008 &0.0104\\
V5&	0.6542 &0.0003 &0.9179&V12\_V14 (18)&0.6056 &0.0333\\\hline
\end{tabular}%
\caption{\{V9=1\}:Top 15 ranked conditional entropies (CE) and successive CE-drop under 1-feature and 2-feature settings. $n=266$ with 199 uncensored data points.}
\label{CEdropV9eq1}
\end{table}

What are the potential consequences that could be derived from the identified collection of 16 candidates of order-2 major factors? How these consequences of many faced heterogeneity embedded within data would impact our understanding of the dynamics of progression from MCI to AD? The full discussions of these two essential questions would be deferred to the ongoing Part-II of this study. Here, we only briefly mention some clues leading toward the to-be-proposed resolutions.

\begin{figure}[h!]
 \centering
\includegraphics[width=0.7\textwidth]{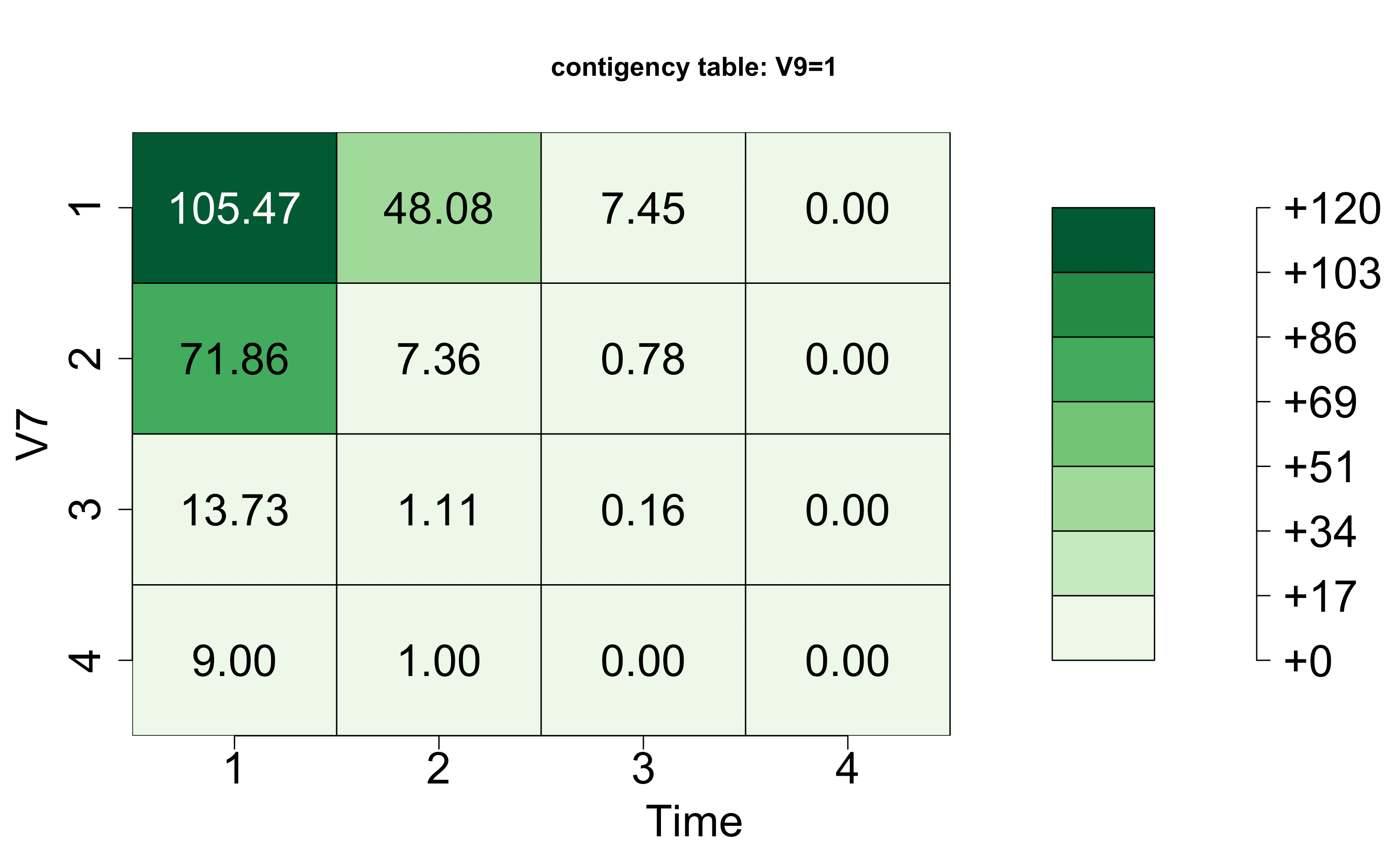}
 \caption{Contingency table of $C[V7-vs-T]$ within sub-collection $V9=1$. }
 \label{ConV9eq1toV7}
 \end{figure}

Consider the contingency table $C[V7-vs-T|V9=1]$ shown in Figure~\ref{ConV9eq1toV7} as a way of precisely revealing the first layer of heterogeneity with respect to V7 within the sub-collection \{V9=1\}. We see relatively distinct 4 rows. In particular on the 2nd, 3rd and 4th rows, we intuitively predict $T_i$ falling the category-1 of T if V7 is in categories \{2, 3, 4\}. This predictive decision-making is correct with varying correct rates and surely subject to some varying error rates across these three categories. In contrast, when V7 is seen in category \{1\}, then the correct rate goes down and the error rate goes up. Would the interacting effect of \{V7, V3\} help?

\begin{figure}[h!]
 \centering
\includegraphics[width=0.7\textwidth]{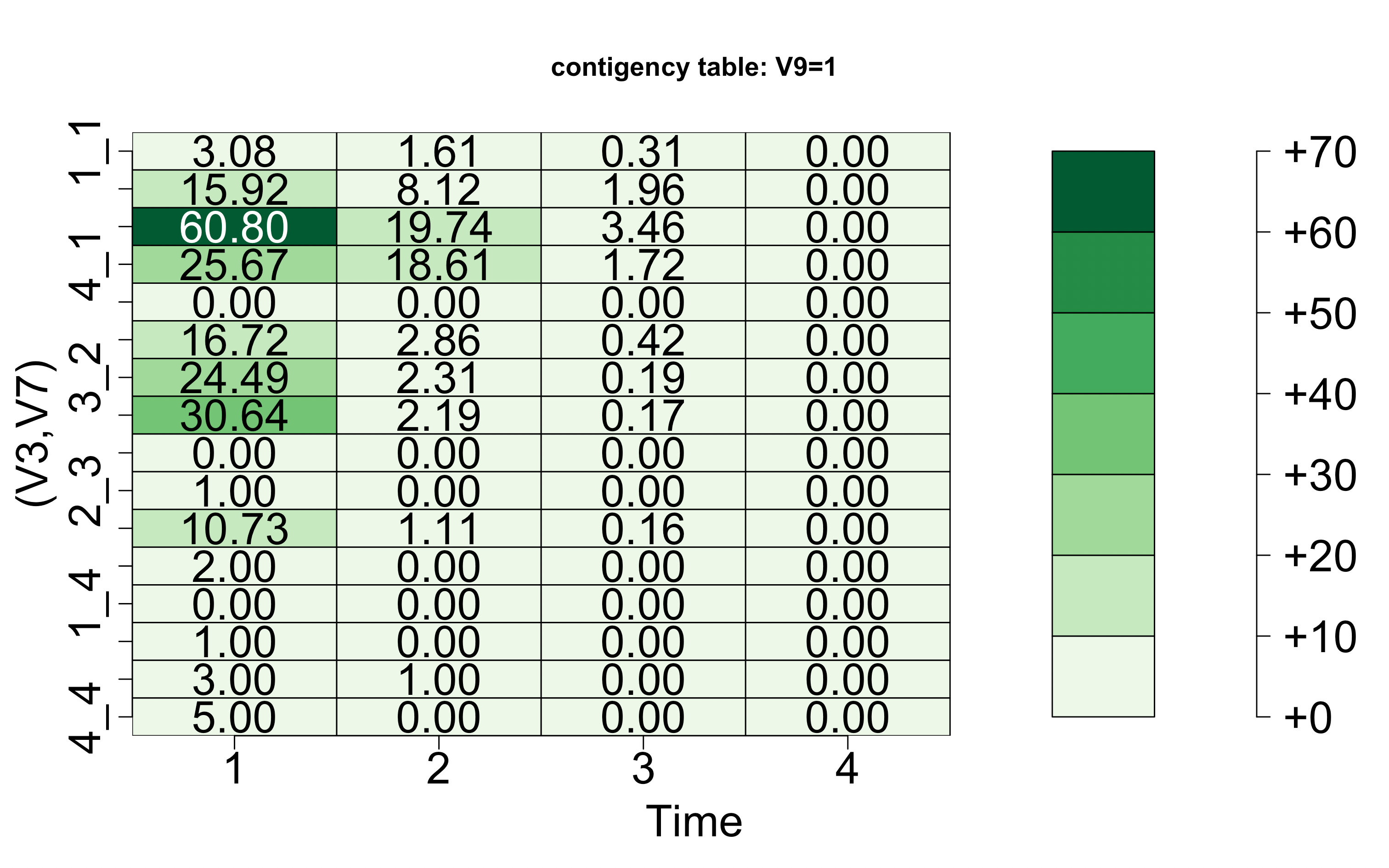}
 \caption{Contingency table of $C[(V3, V7)-vs-T]$ within the sub-collection $V9=1$. }
 \label{ConV9eq1toV3V7}
 \end{figure}

\begin{figure}[h!]
 \centering
\includegraphics[width=0.7\textwidth]{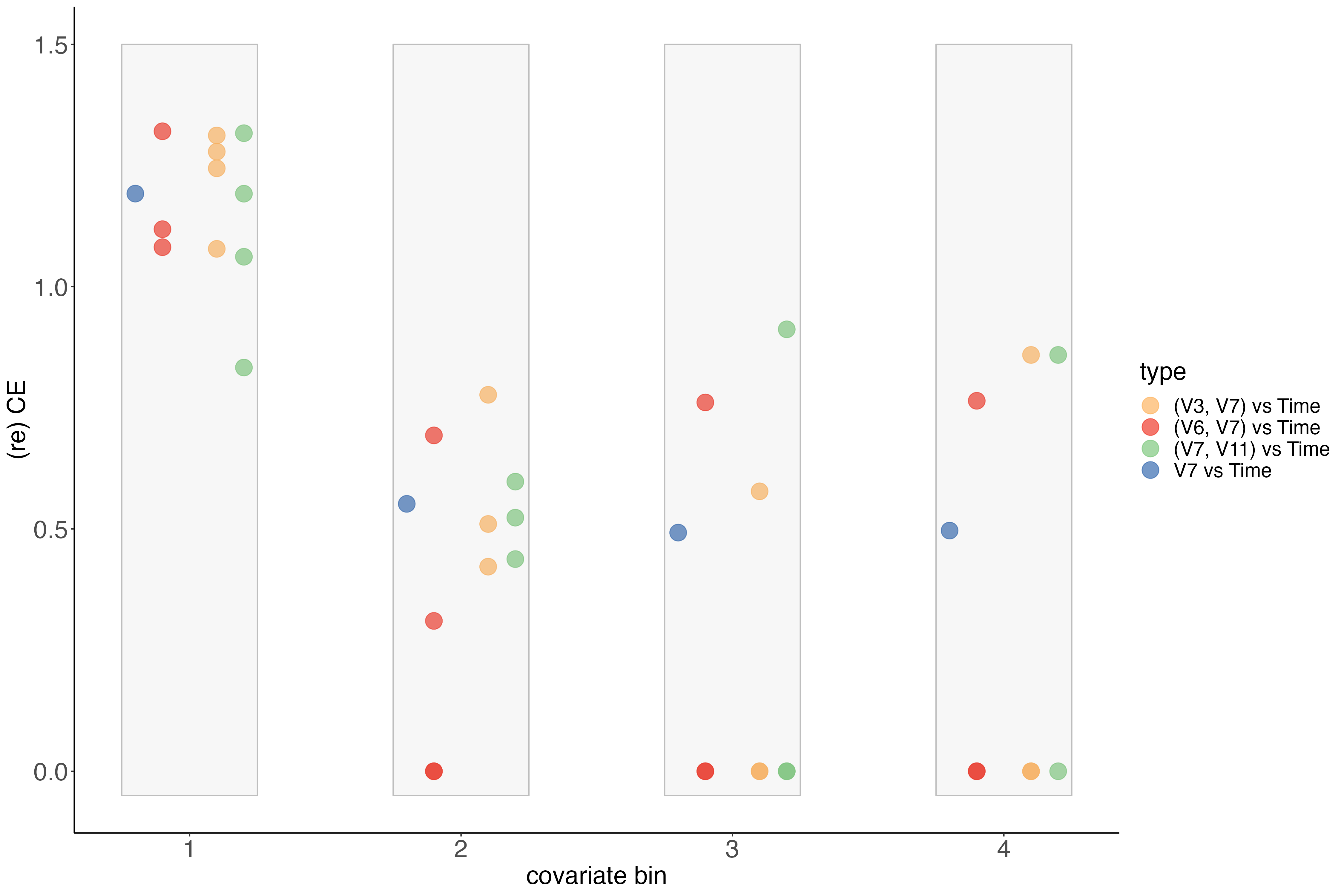}
 \caption{Heterogeneity of CE-expansions ($V9=1$): V7-vs-T (blue dots); (V7, V6)-vs-T (red dots); (V7, V3)-vs-T (orange dots); (V7, V11)-vs-T (green dots). The CE of $T$ in sub-collection \{V9=1\} is equal to 0.6545. There are overlapping dots at and near zero.}
 \label{legendV9eq1}
 \end{figure}

The contingency table $C[(V3,V7)-vs-T|V9=1]$ is shown in Figure~\ref{ConV9eq1toV3V7}. Each category of V7 is divided into 3 or 4 bivariate-categories with respect to the four categories of V3. To precisely see which rows of V7 are being improved by bivariate-categories, we present the re-scaled CEs with respect to marginal CE of $T$ within sub-collection \{V9=1\} in Figure~\ref{legendV9eq1}: re-scaled CEs pertaining to categories of V7 (in blue dots), re-scaled CEs pertaining to categories of (V3,V7) (in orange dots). It is noted that there are overlapping dots at zeros in bivariate-categories: (2, 3), (4, 3), (2, 4), and (4,4). We propose to encode those subjects falling into these bivariate categories as, for example, V9-1-V7-3-V3-2-T1, V9-1-V7-3-V3-4-T1, V9-1-V7-4-V3-2-T1, and V9-1-V7-4-V3-4-T1. Such new code-ID indicates when a subject can be identified without the uncertainty of categories of $T$. If we relax the coding criterion just slightly to allow a small positive CE, such as 0.05, then we will also include subjects falling into bivariate-categories (3,3). The 12 subjects will be encoded with code-ID: V9-1-V7-3-V3-3-T1.

\begin{figure}[h!]
 \centering
   \includegraphics[width=0.7\textwidth]{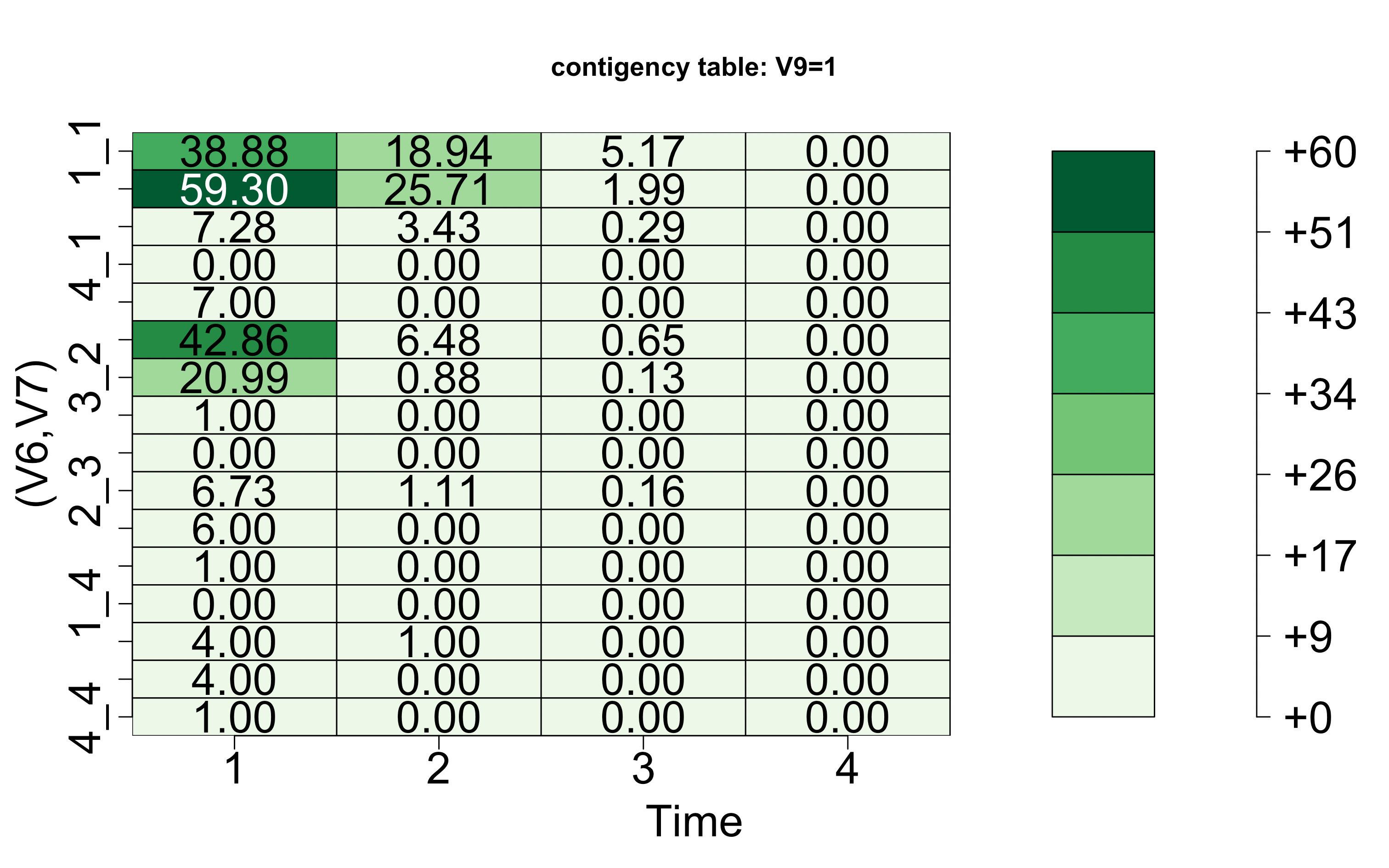}
 \caption{Contingency table of $C[(V6, V7)-vs-T]$ within the sub-collection $V9=1$.}
 \label{ConV9eq1toV6V7}
 \end{figure}

Further, even though feature-pair \{V7, V6\} doesn't achieve the ecological effects and just ranked above feature-pair \{V7, V3\}, we still can see many of their bivariate-categories achieving rather low or even zero CEs in Figure~\ref{ConV9eq1toV6V7}. Subjects falling into those bivariate-categories are qualified for short code-IDs. See also Figure~\ref{legendV9eq1}, where are many overlapping dots at or very near zero CEs.

\paragraph{\{V9=2\}:$141(n_o)-vs-332(n_c)$.}
The sub-collection \{V9=2\} involves a heavy censoring rate. It is more than $\frac{2}{3}$. Our CE computations reported in Table~\ref{CEdropV9eq2} reveal very distinct ranking among 1-feature and 2-feature settings from that found on the sub-collection: \{V9=1\}.
This distinction is an indication of a large-scale heterogeneity.

In the 1-feature setting, the lowest CE is surprisingly achieved by the feature: V4, and the 2nd and 3rd-ranked CEs are achieved by V6 and V7, respectively. Unlike in \{V9=1\}, V10 (MEM-std) rises to the 4th, which is ranked ahead of V11 and V8. V8 is ranked 3rd in in \{V9=1\}. We also had significant ranking drops for V16 and V1: 5th to 9th for V16; 7th to 14th for V1. These ranking drops would significantly impact the CEs of feature-pairs as seen in the 2-feature setting.

\begin{figure}[h!]
 \centering
   \includegraphics[width=0.99\textwidth, height=2in]{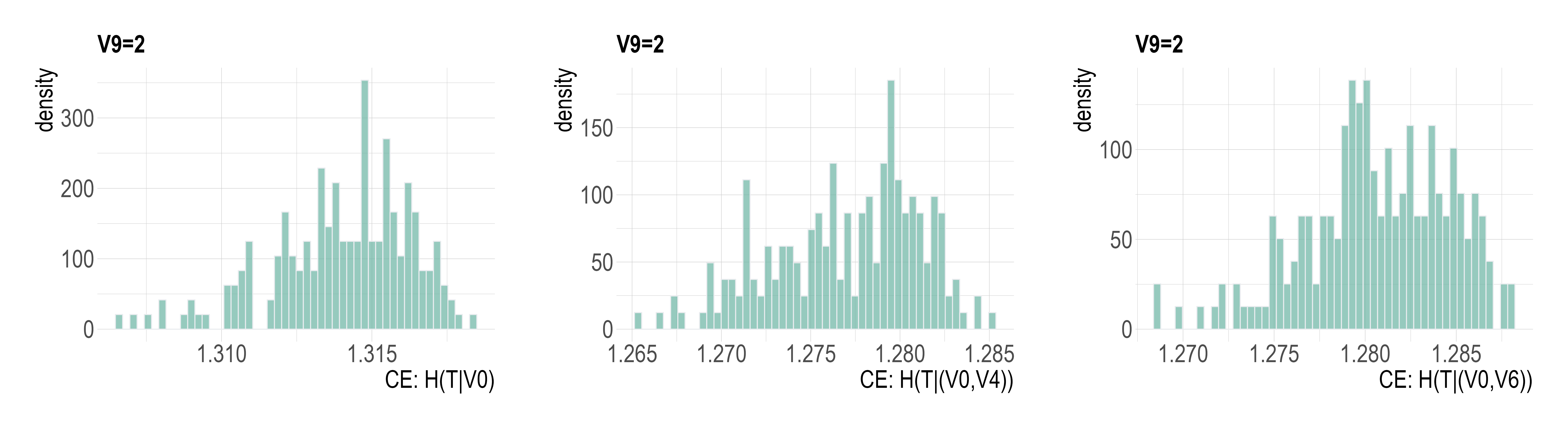}
 \caption{Null distributions for reliability checks at 2-feature settings: Left: simulated CEs of $H[T|(V0, V4)]$ ; Middle: simulated CEs of $H[T|(V0, V6)]$; Right: simulated CEs of $H[T|(V0, V6)]$, in sub-collection \{V9=2\} based 200 simulated $U[0,1]$ based features.}
 \label{Reliachec2}
 \end{figure}

Also, for reliability check on CE calculations, we simulate a random $U[0,1]$ noise feature, denoted as V0, repeat this 200 times and calculate CEs: $H[Y|V0]$ as shown in the left of Figure~\ref{Reliachec2}. We see that, due to having more uncensored data points in this sub-collection \{V9=2\},  many more single features have very small p-values. As for the 2-feature setting, the middle and right of Figure~\ref{Reliachec2} show the null distributions of simulated CEs of $H[T|(V0, V4)]$ and $H[T|(V0, V6)]$, respectively. Via panel (a), even the 14th ranked feature-pair (V4, V8) has a relatively smaller p-value, while via panel (b), the 15 ranked feature-pair (V7, V10) has a very small p-value. Again, we accordingly carry out the task of identifying potential interacting effects as follows.

For the 2-feature setting, the most striking pattern is that all $\binom{5}{2}(=10)$ feature-pairs of the top 5 ranked features: V4, V6, V7, V10, and V11, appear in the top 15 list and achieve the ecological effects. This computational fact indicates that the two members of each pair can be concurrently present as order-1 major factors. Such a phenomenon is made possible by the de-associating procedure, which makes the features less associated or even independent by taking off their association with V9. Further, the 17th ranked pair: \{V3, V11\}, demonstrates a strong interacting effect by having the SCE-drop of adding V3 being more than 5 times V3's CE-drop. Similar interacting effects with less strength are also seen, such as \{V1, V6\}, \{V7, V13\}, just to list a few.
\begin{table}[h!]
\centering
\begin{tabular}{cccc|ccc}\hline
1-feature & 	CE&	SCE-drop &$p$-value(PH)  & 2-feature& 	CE&	SCE-drop\\ \hline
V4&	1.2897 &0.0287 &0.0000 & V6\_V11 &1.2582& 0.0348\\
V6&	1.2931 &0.0253 &0.0002&V4\_V7&1.2589 & 0.0308\\
V7 &1.2948 &	0.0236&0.0000&V4\_V12 &1.2591 &	0.0306\\
V11&1.2981 &	0.0203 &0.0842&V4\_V6 &1.2595 & 0.0302\\
V10&1.2986&	0.0198& 0.4121&V7\_V11&1.2609 &0.0339\\
V8&	1.2998&0.0186 &0.0102&V6\_V10&1.2621&	0.0309\\
V12&1.3058 &	0.0126 &0.0885&V4\_V11&1.2635 &	0.0262\\
V15 &1.3086 &0.0099 &0.0298&V4\_V10&1.2671 &0.0226\\
V16	&1.3128 &0.0056 &0.1459&V10\_V11&1.2672 &0.0310\\
V5	&1.3129 &0.0056&0.8022&V7\_V12&1.2676 &0.0272\\
V14&1.3129 &	0.0056 &0.0713&V6\_V15&1.2700 &	0.0230\\
V3& 1.3144 &0.0040&0.0014&V11\_V12&1.2704 & 0.0277\\
V13&	1.3146 & 0.0039 &0.5264&V4\_V8&1.2712&0.0185\\
V1&	1.3163 &0.0022&0.4214& V7\_V10&1.2729 &0.0219\\
V2&	1.3117 &0.0013 &0.9507&V3\_V11 (17)	&1.2765 &0.0216\\\hline
\end{tabular}%
\caption{\{V9=2\}:Top 15 ranked conditional entropies (CE) and successive CE-drop under 1-feature and 2-feature settings. $n=473$ with 141 uncensored data points.}
\label{CEdropV9eq2}
\end{table}

\begin{figure}[h!]
 \centering
   \includegraphics[width=0.7\textwidth]{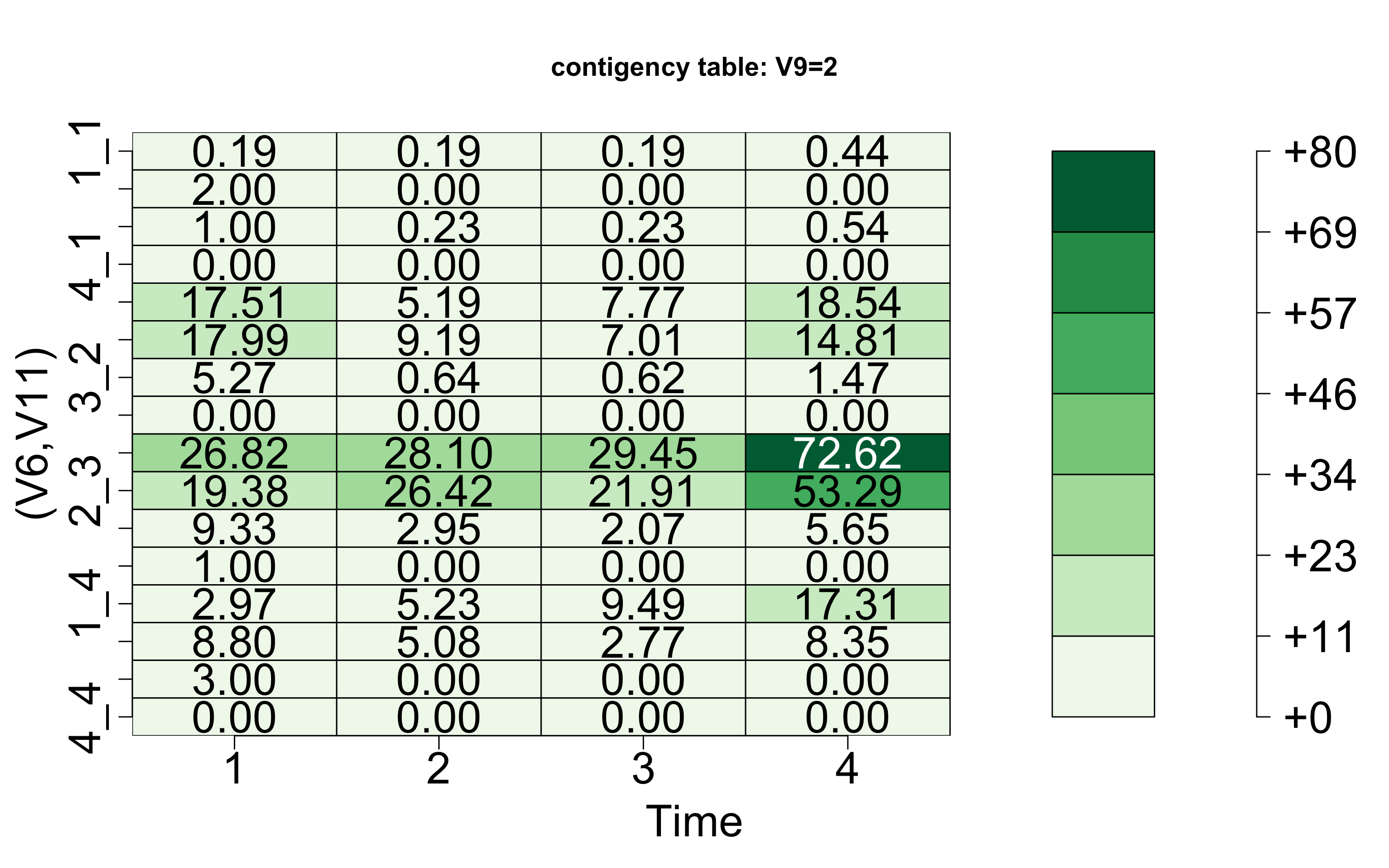}
 \caption{Contingency table of $C[(V6, V11)-vs-T]$ within the sub-collection $V9=2$. }
 \label{ConV9eq2toV6V11}
 \end{figure}

\begin{figure}[h!]
 \centering
   \includegraphics[width=0.7\textwidth]{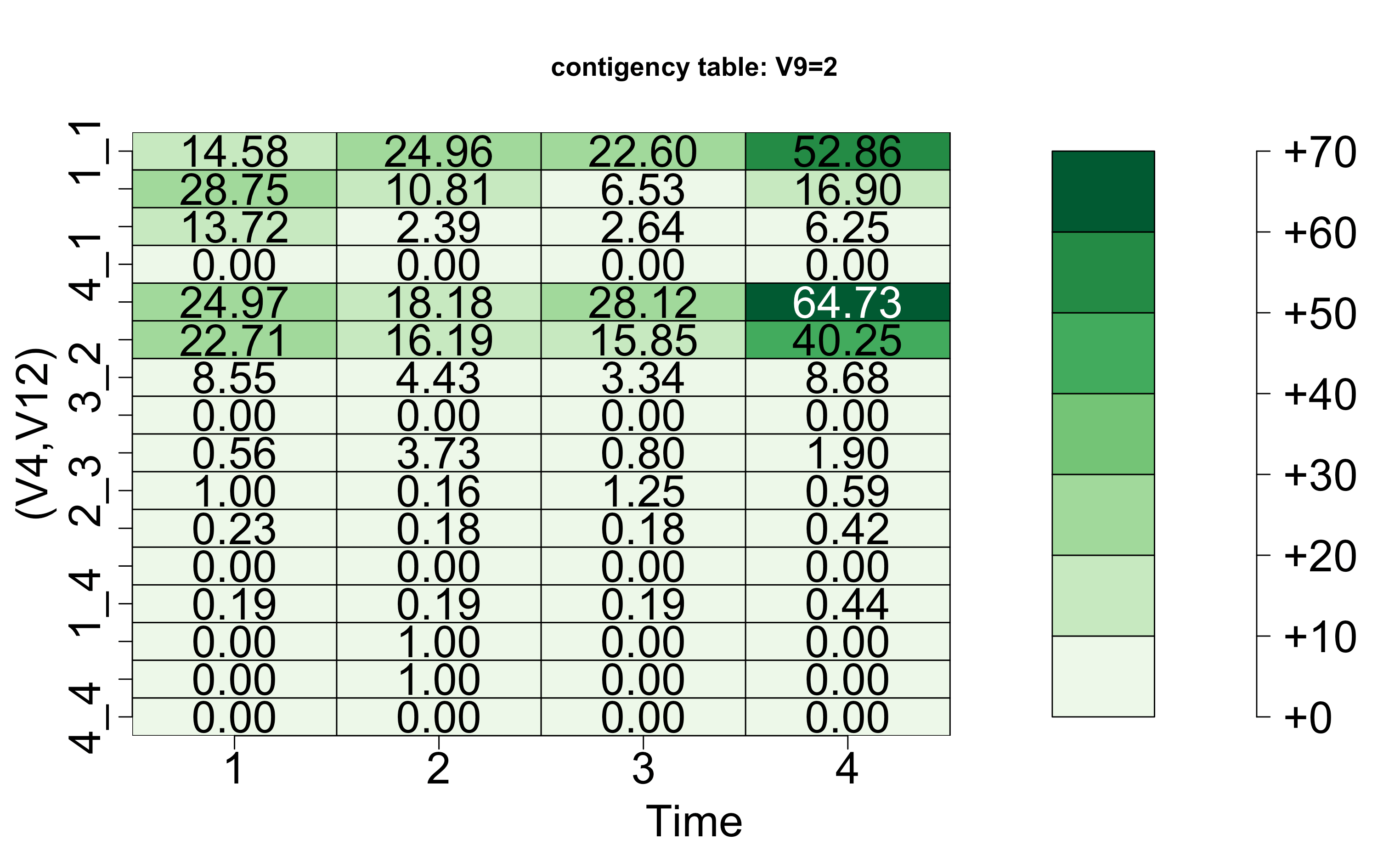}
 \caption{Contingency table of $C[(V4, V12)-vs-T]$ within the sub-collection $V9=2$.}
 \label{ConV9eq2toV4V12}
 \end{figure}

\begin{figure}[h!]
 \centering
   \includegraphics[width=0.7\textwidth]{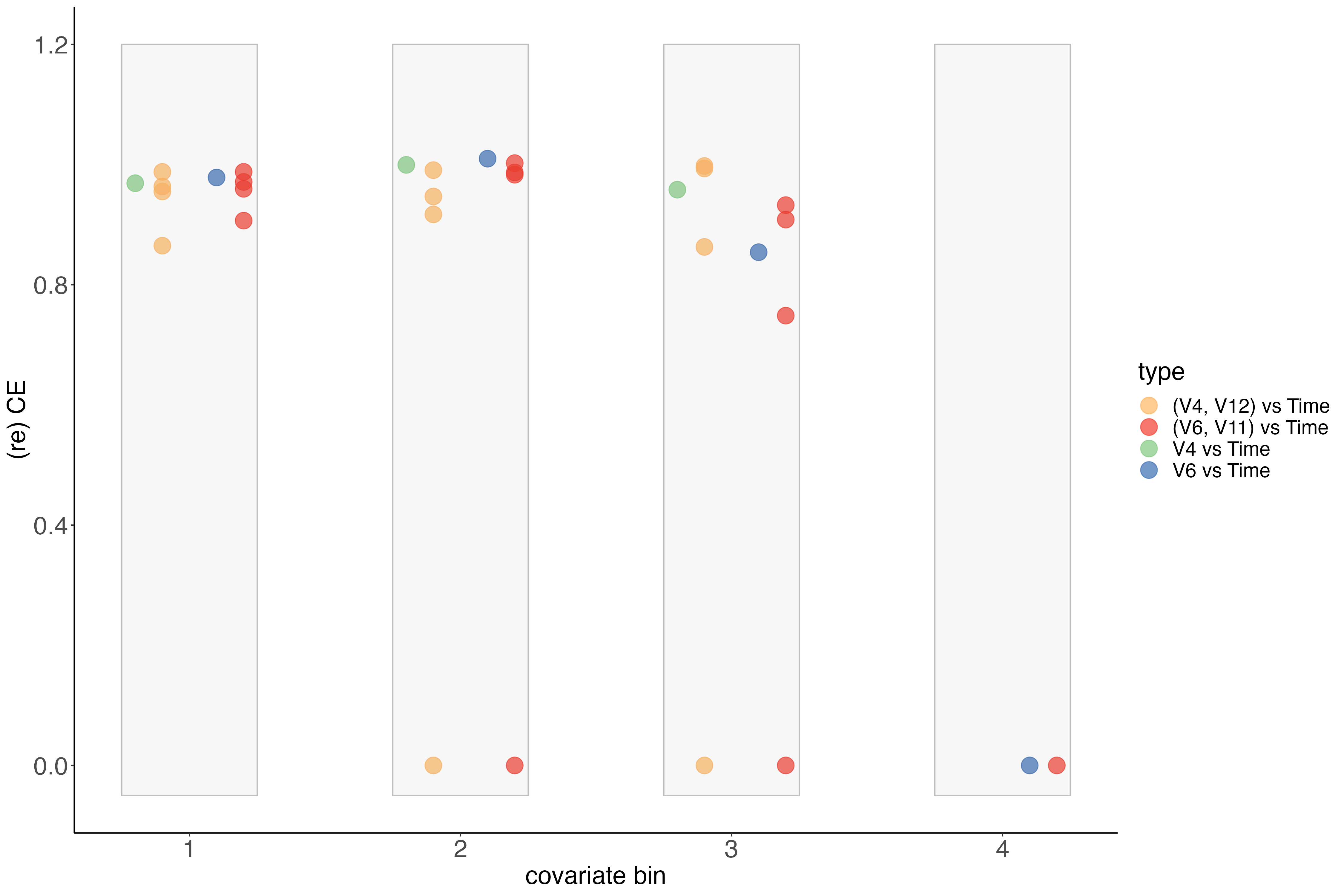}
 \caption{Heterogeneity of CE-expansion ($V9=2$) of V6-vs-T (blue dots) and V4-vs-T (green dots) and and (V6, V11)-vs-T (red dots) and (V4, V12)-vs-T (orange dots). The CE of $T$ in sub-collection \{V9=2\} is equal to 1.3183.}
 \label{legendV9eq2}
 \end{figure}

These aforementioned order-1 features and order-2 feature-pairs are potential candidates for building their contingency tables against $T$ in order to map out the heterogeneity within this sub-collection \{V9=2\}. However, this mapping out would not yield very decisive or precise predictions because these features and feature-sets can't achieve significant uncertainty reductions. For instance, even the top-ranked feature-pair \{V6, V11\} can only achieve less than $5\%$ reduction of uncertainty of $T$, which is calculated as $\frac{0.0348+0.0253}{0.0348+0.0253+1.2582}$. Details of this overall conclusion in this sub-collection \{V9=2\} can be seen through three figures: Figure~\ref{ConV9eq2toV6V11} for contingency table $C[(V6, V11)-vs-T]$; Figure~\ref{ConV9eq2toV4V12} for contingency table of $C[(V4, V12)-vs-T]$ and Figure~\ref{legendV9eq2} for CE-expansion plots pertaining to contingency tables $C[(V6, V11)-vs-T]$ and $C[(V4, V12)-vs-T]$. Based on these three figures, we see only a few bivariate-categories can receive short code-IDs. We can make the same conclusion through the plots of CE-expansions pertaining to feature-pairs \{V6, V11\} and \{V4, V12\} presented in Figure~\ref{legendV9eq2}.

Therefore, we can conclude that subjects in this sub-collection  \{V9=2\} need different perspectives to look through them and analyze their data. For instance, we need to use different features or feature-sets to define sub-collections. We perform a different set of sub-collections defined by V8 in the next subsection.

For comparing the above CEDA results with the PH results, which are reported in the third column of Table~\ref{CEdropV9eq2}, we see incoherent results from the aspects of \{V10, V12, V15, V14, V3\}. V10 is ranked 5th in CEDA results, but it is insignificant in PH results. While features: \{V12, V15, V14, V3\} are ranked 7th, 8th, 11th, and 12th, respectively, they are all significant in PH results. Further, PH results indicate that the 4th ranked V11 has a $p$-value much larger than the 12th ranked V3's, even though feature-pair \{V3, V11\} seems to have an interacting effect according to CEDA results. Nonetheless, the uncertainty reduction achieved by \{V3, V11\} is rather low. By summarizing these incoherent comparisons between CEDA and PH results, we suspect that PH results could be rather unreliable because of the presence of heavy censoring in this sub-collection  \{V9=2\}. We further see its unreliability turning into incapability in the next sub-collections: \{V9=3\} and \{V9=4\}.

\paragraph{\{V9=3\}: $4(n_o)-vs-147(n_c)$.}
In the sub-collection \{V9=3\}, there are only 4 uncensored and 147 censored data points. The PH hazard regression model becomes completely incapable of extracting any reliable information from data pertaining to this sub-collection. In sharp contrast, our contingency-table-based CEDA computations are not affected. Based on Table~\ref{CEdropV9eq3}, the CE of $T$ is about one-third of CE of $T$ in sub-collection \{V9=2\} and one-half of CE of $T$  in sub-collection \{V9=1\}. Further, some of the rest of the 15 features still offer reasonable amounts of information beyond the information of \{V9=3\}. This is one striking aspect of heterogeneity with respect to V9 and CEDA computations.

\begin{figure}[h!]
 \centering
\includegraphics[width=0.7\textwidth]{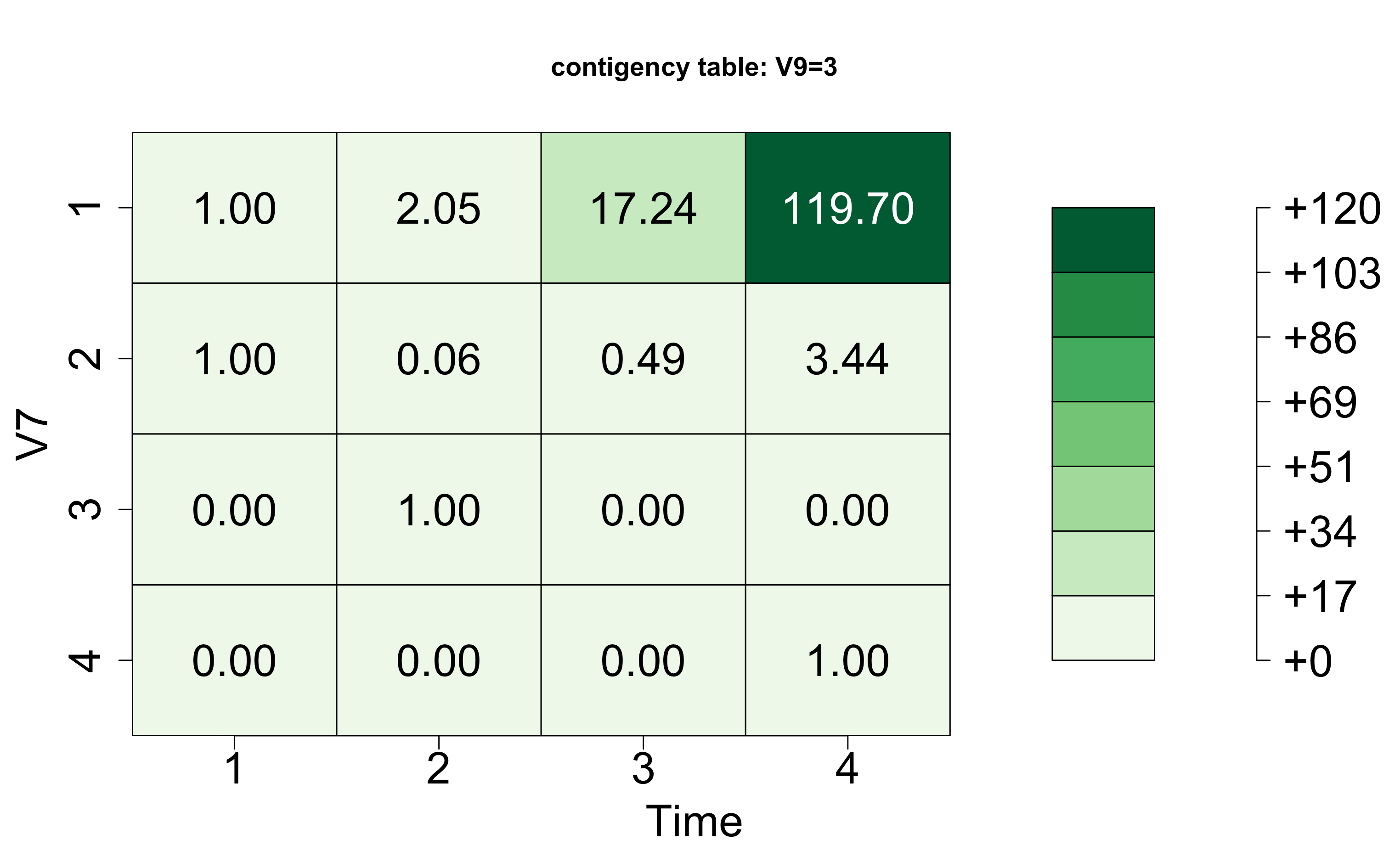}
 \caption{Contingency table of $C[V7-vs-T]$ within sub-collection $V9=3$. }
 \label{ConV9eq3toV7}
 \end{figure}

Based on Table~\ref{CEdropV9eq3}, the top two ranked features are V7 and V6, respectively. So, the presences of V7 and V6 are among the top three ranked features across sub-collections: from \{V9=1\} to \{V9=3\}. This result strongly indicates that either V7 or V6 could provide extra information beyond V9 at least within these three sub-collections. The evidence of V7 is seen in Figure~\ref{ConV9eq3toV7} having two rows with relatively low CEs: the 3rd and 4th. Furthermore, based on results of 1-feature setting in Table~\ref{CEdropV9eq3},  the fact that V10 is ranked 4th is unseen in the sub-collections:\{V9=1\} and \{V9=2\}.

\begin{table}[h!]
\centering
\begin{tabular}{cccc|ccc}\hline
1-feature &CE&	SCE-drop&$p$-value(PH) & 2-feature& CE&	SCE-drop\\ \hline
V7&	0.4951&0.0429 &0.0000&V7\_V11&0.4699&0.0252\\
V6&	0.5081&0.0299 &0.0000&V6\_V10&0.4734&0.0347\\
V11&0.5155 &	0.0225&0.0000&V7\_V10&0.4743 &0.0208\\
V10&0.5191 &0.0188 &0.0000&V7\_V15&0.4781&0.0170\\
V4&	0.5196&0.0183&0.9980&V1\_V7&	0.4801&0.0150\\
V8&0.5228&0.0152&0.0000&V6\_V7&0.4829&0.0122\\
V1&	0.5241&0.0139&0.0000&V4\_V7	&0.4841&0.0110\\
V3&	0.5260&0.0120&0.0000&V3\_V7&0.4856&0.0095\\
V15&0.5293&0.0087&0.0000&V6\_V11&0.4864&0.0216\\
V5&0.5308&0.0072&0.0000&V7\_V8&0.4865&0.0086\\
V13&0.5324&0.0056&0.0000&V7\_V12&0.4871&0.0080\\
V14&0.5339&0.0041&0.8120&V10\_V11&0.4874&0.0280\\
V16&0.5358&0.0022&0.0000&V2\_V7&0.4875&0.0076\\
V12&0.5370&0.0010&0.0000&V7\_V13&0.4887&0.0064\\
V2&	0.5378&0.0002&0.0000&V4\_V6 (17)&0.4898&0.0182\\\hline
\end{tabular}%
\caption{\{V9=3\}:Top 15 ranked conditional entropies (CE) and successive CE-drop under 1-feature and 2-feature settings. $n=147$ with 5 uncensored data points.}
\label{CEdropV9eq3}
\end{table}

\begin{figure}[h!]
 \centering
   \includegraphics[width=0.9\textwidth]{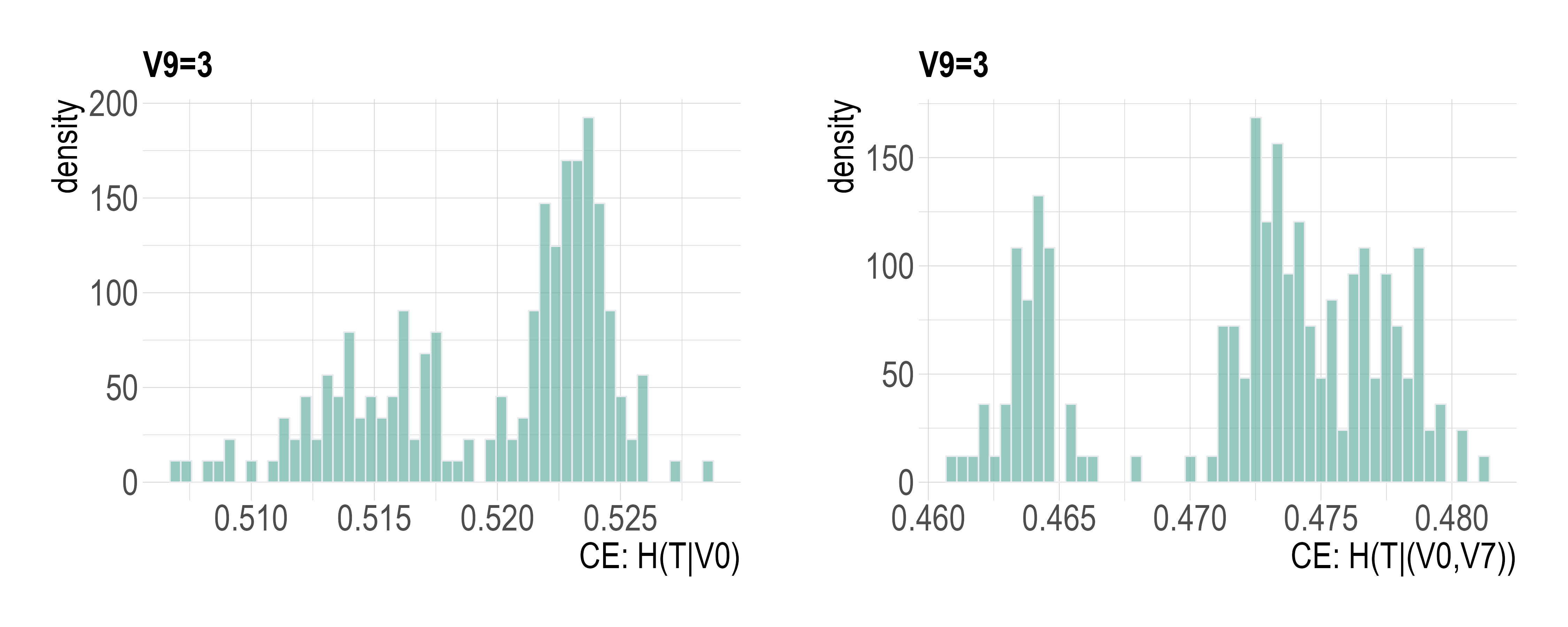}
 \caption{Null distributions for reliability checks for 1-feature and 2-feature settings: Left: simulated CEs of $H[T|V0]$; Right: simulated CEs of $H[T|(V0, V7)]$ in sub-collection \{V9=3\} based on 200 simulated $U[0,1]$ based features. }
 \label{Reliachec3}
 \end{figure}

For the reliability check on CE calculations, we use the same simulation plan and calculate CEs: $H[Y|V0]$ as shown in the left of Figure~\ref{Reliachec3}. In sharp contrast, due to having a very little number of uncensored data points in this sub-collection \{V9=3\}, we see only V7 and V6 have very small p-values. As for the 2-feature setting, we see that no feature-pairs have small p-values according to the right of Figure~\ref{Reliachec3} of the null distributions of simulated CEs of $H[T|(V0, V7)]$. However, we still accordingly carry out the task of identifying potential interacting effects for the continuum of our discussion across all sub-collection as follows.

\begin{figure}[h!]
 \centering
   \includegraphics[width=0.7\textwidth]{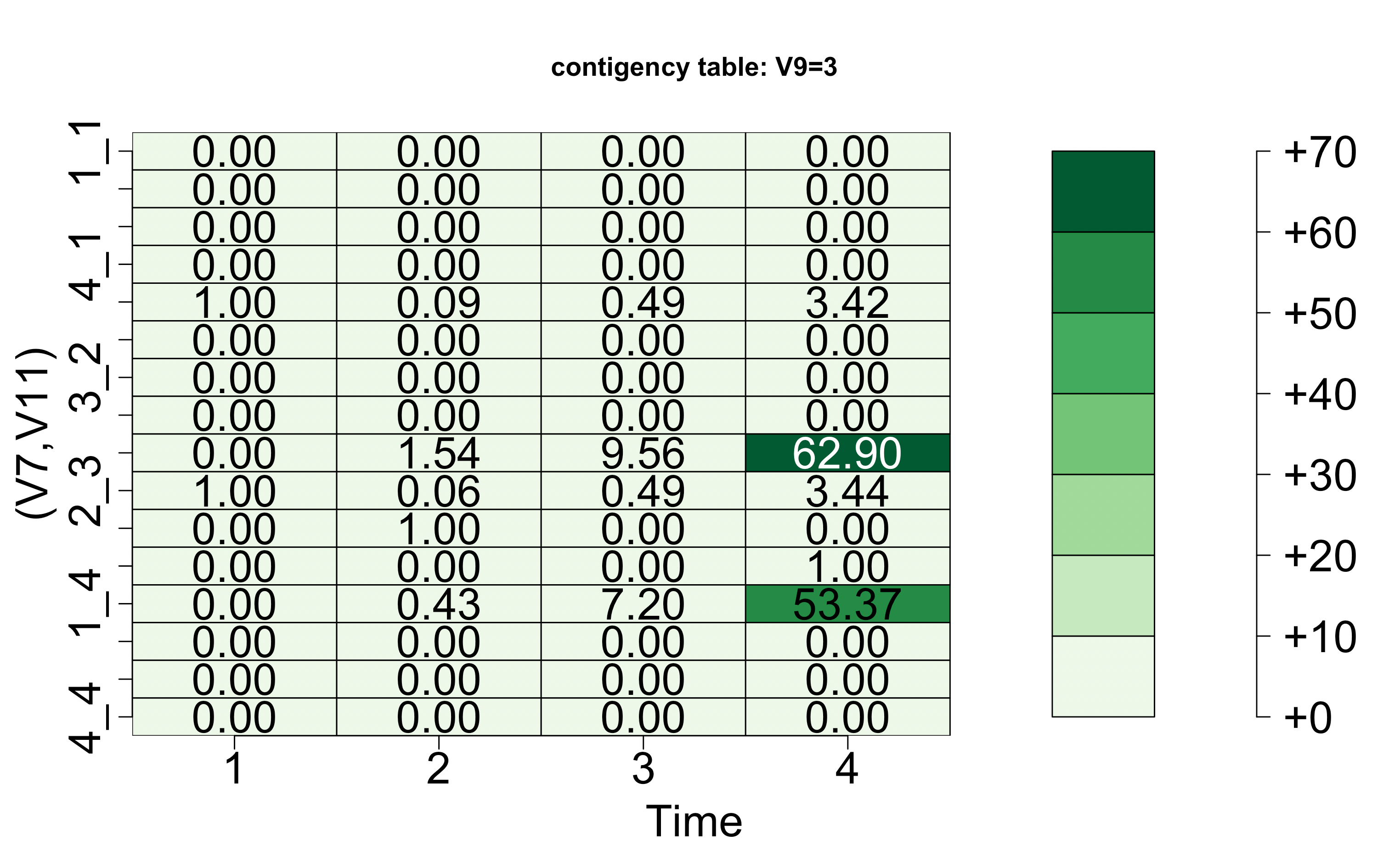}
 \caption{Contingency table of $C[(V7, V11)-vs-T]$ within the sub-collection $V9=3$. }
 \label{ConV9eq3toV7V11}
 \end{figure}

\begin{figure}[h!]
 \centering
\includegraphics[width=0.7\textwidth]{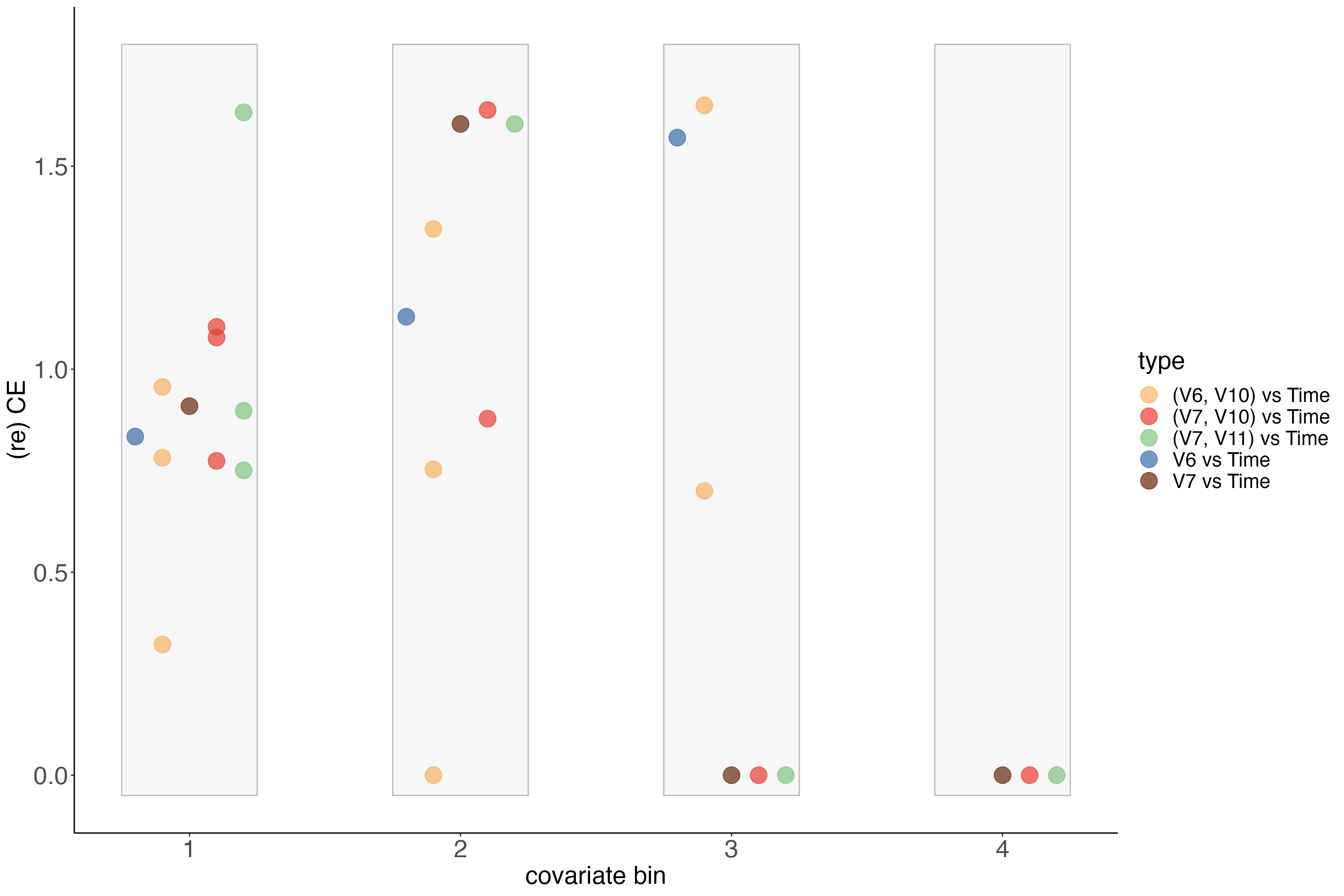}
 \caption{Heterogeneity of CE-expansion ($V9=3$): V7-vs-T (blown dots), V6-vs-T (blue dots),   (V6, V10)-vs-T (orange dots), (V7, V10)-vs-T (red dots) and (V7, V11)-vs-T (green dots). The CE of $T$ in sub-collection \{V9=3\} is equal to 0.5380.}
 \label{legendV9eq3}
 \end{figure}

As for results of 2-feature setting, feature-pairs: \{V6, V10\}, \{V7, V11\}, \{V10, V11\}, achieve ecological effects, but not \{V6, V7\}. Another interesting observation is that there are no evident interacting effects present in this sub-collection. Therefore, we can conclude a reasonable collection of order-1 major factors: \{V7, V11\}. The effect of coupling V11 with V7 is seen through the contingency table in Figure~\ref{ConV9eq3toV7V11}. Their CE-expansion together with  CE-expansions with respect to feature-pairs: \{V6, V10\}, \{V7, V11\} and \{V7, V10\}, are reported in Figure~\ref{legendV9eq3}. The degrees of expansions offered by V10 and V11 are relatively small.

\paragraph{\{V9=4\}: $1(n_o)-vs-16(n_c)$}
In the sub-collection \{V9=4\}, there is only one data point out of 17 uncensored. From Table~\ref{CEdropV9eq4}, the CE of $T$ is the smallest among the four sub-collection. Based on CEs of 1-feature and 2-feature settings reported in the table, the rest of the 15 features still offer some extra information beyond the information of \{V9=4\}. This is another striking aspect of heterogeneity with respect to V9. In sharp contrast, the PH regression model simply doesn't provide any meaningful result.

\begin{figure}[h!]
 \centering
\includegraphics[width=0.7\textwidth]{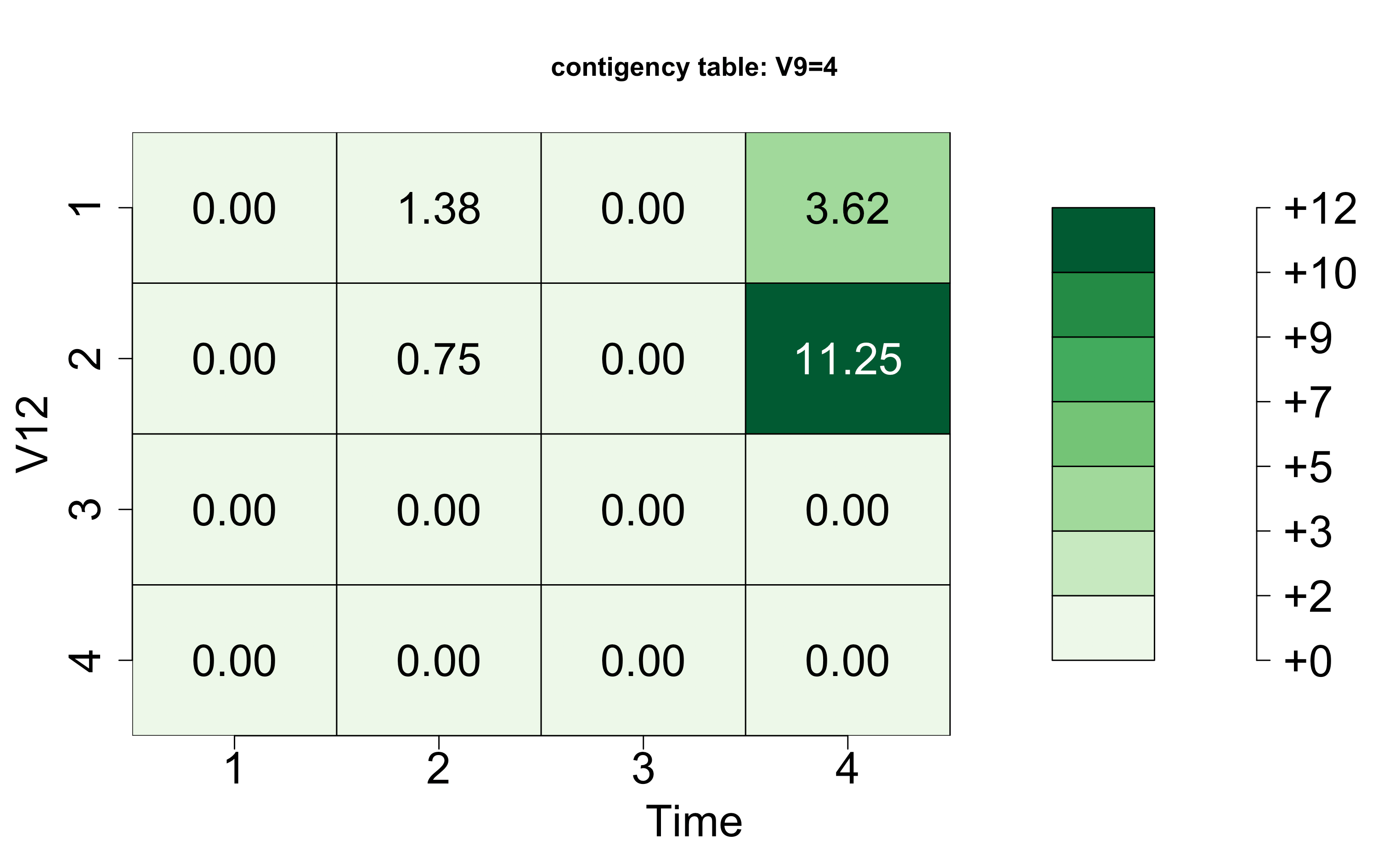}
 \caption{Contingency table of $C[V12-vs-T]$ within sub-collection $V9=4$. }
 \label{ConV9eq4toV12}
 \end{figure}

Based on the results of 1-feature in Table~\ref{CEdropV9eq4}, the two facts: 1) the top two ranked features are \{V12, V14\}; 2) V7 is ranked at the bottom, are totally new by being totally different from the previous three sub-collections. From Figure~\ref{ConV9eq4toV12}, the contingency table $C[V12-vs-T]$ shows the 2nd row with relatively low CEs.

\begin{figure}[h!]
\centering
\includegraphics[width=0.9\textwidth]{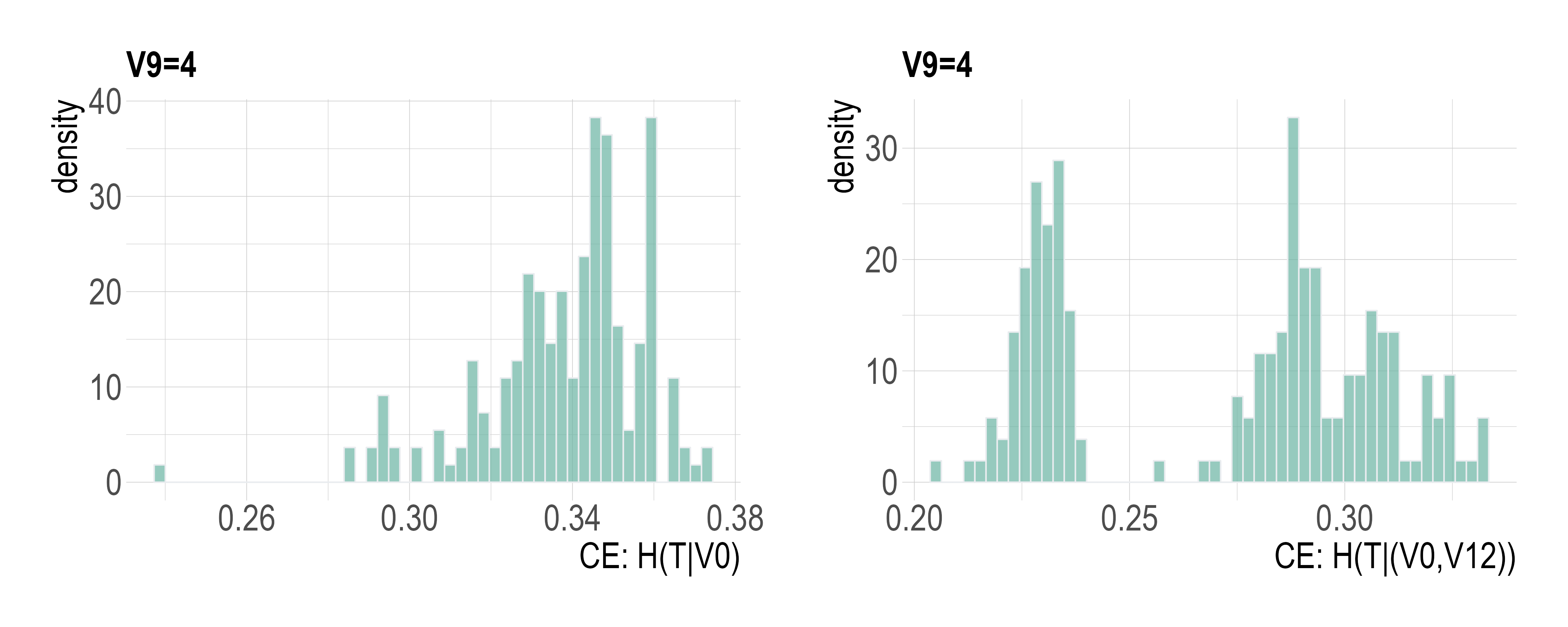}
\caption{Null distributions for reliability checks for 1-feature and 2-feature settings: Left: simulated CEs of $H[T|V0]$; Right: simulated CEs of $H[T|(V0, V12)]$ in sub-collection \{V9=4\} based on 200 simulated $U[0,1]$ based features. }
\label{Reliachec4}
\end{figure}

Nonetheless, for the reliability check on CE calculations, we use the same simulation plan and calculate CEs: $H[Y|V0]$ as shown in the right of Figure~\ref{Reliachec4}. Due to having only one uncensored data point out of 17 in this sub-collection \{V9=4\}, we see that only V12 has a p-value at the borderline of being significant. As for the 2-feature setting, all feature-pairs have rather big p-values according to the left of Figure~\ref{Reliachec4} of the null distributions of simulated CEs of $H[T|(V0, V12)]$. For the continuum of our discussion on interacting effects, we still briefly interpret 2-feature results as follows.

\begin{figure}[h!]
 \centering
\includegraphics[width=0.7\textwidth]{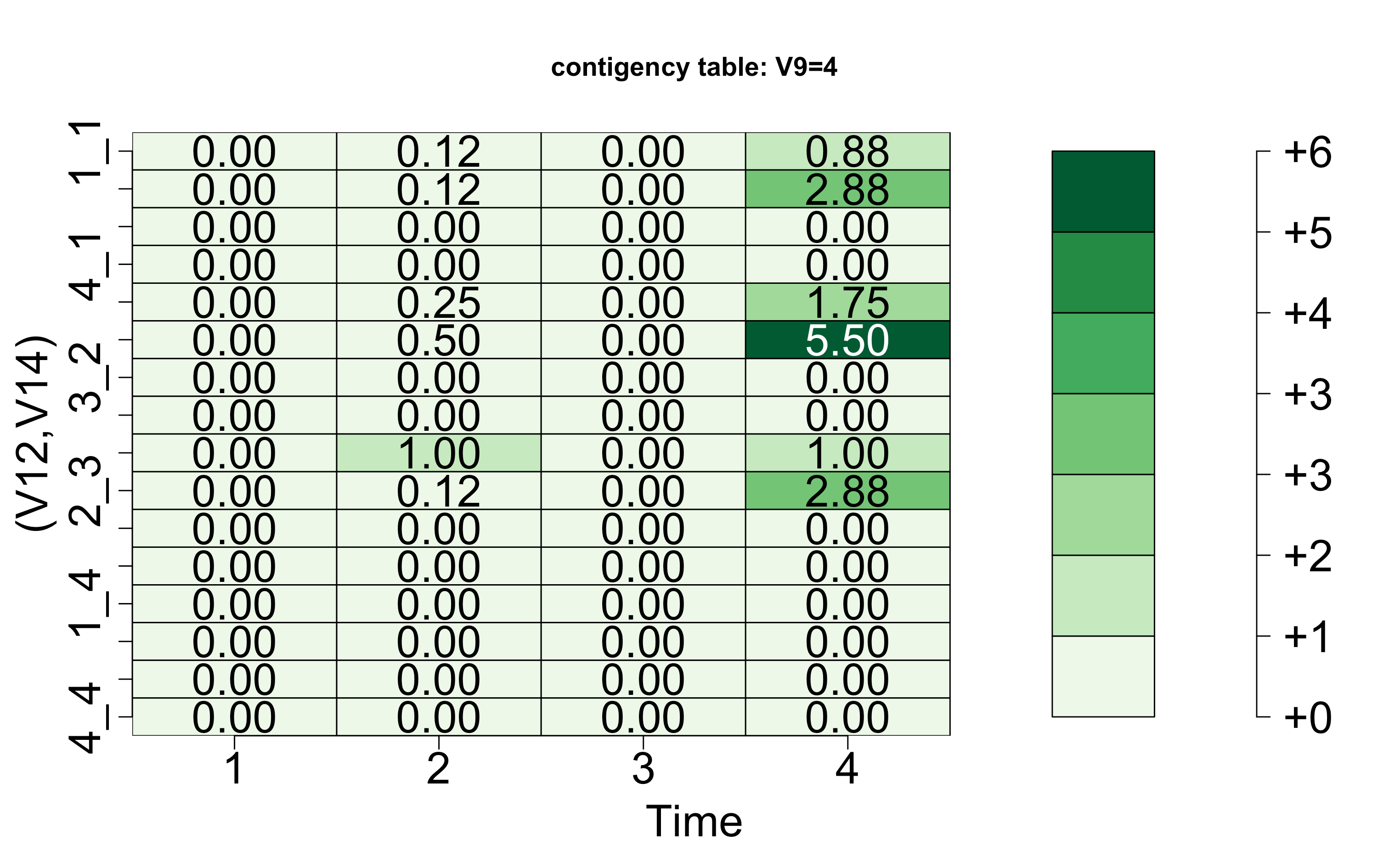}
 \caption{Contingency table of $C[(V12, V14)-vs-T]$ within sub-collection $V9=4$. }
 \label{ConV9eq4toV12V14}
 \end{figure}

\begin{figure}[h!]
 \centering
\includegraphics[width=0.7\textwidth]{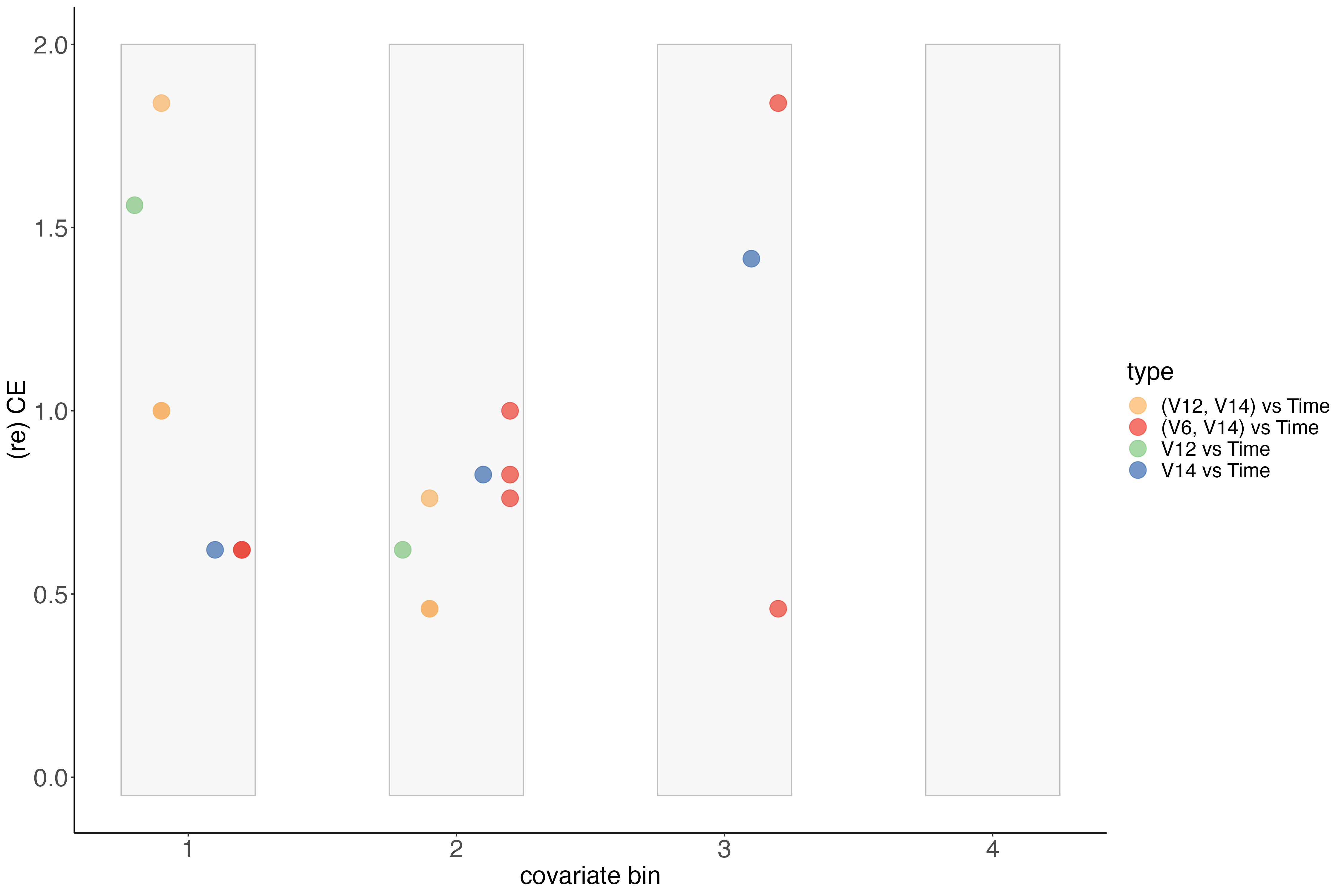}
 \caption{Heterogeneity of CE-expansions ($V9=4$): V12-vs-T (green dots); V14-vs-T (blue dots); (V12, V14)-vs-T (orange dots); (V6, V14)-vs-T (red dots). The CE of $T$ in sub-collection \{V9=4\} is equal to 0.3767.}
 \label{legendV9eq4}
 \end{figure}

As for the 2-feature results, the feature-pair \{V12, V14\} achieves the ecological effect, so they can be concurrently present as order-1 major factors. To see the result of such concurrent presence of features V12 and V14, based on Figure~\ref{ConV9eq4toV12V14}, their contingency table of $C[(V12, V14)-vs-T]$ show 4 rows having relatively low CEs out of 6 non-zero rows. More detailed CEs results are presented in the CE-expansion plots shown in Figure~\ref{legendV9eq4}. Based on these results, we see several univariate- and bivariate categories are qualified for short code-IDs.

Further, we also see feature-pairs: \{V11, V12\} and \{V11, V14\}, achieve interacting effects. Thus, if we are to choose a collection of major factors within this sub-collection, the collection of 2-order major factors \{\{V11, V12\}, \{V11, V14\}\} is one reasonable choice. This collection is rather distinct from the three collections chosen in the previous three sub-collection.

\begin{table}[h!]
\centering
\begin{tabular}{cccc|ccc}\hline
1-feature &CE&	SCE-drop &$p$-value(PH) & 2-feature& CE&	SCE-drop\\ \hline
V12&0.3380&0.0387&1.0000&V12\_V14&0.3104&0.0276\\
V14&0.3582&0.0185&1.0000&V6\_V14&0.3131&0.0451\\
V6&	0.3596&0.0171&1.0000&V1\_V14&0.3135&0.0446\\
V1&0.3600&0.0167&1.0000&V12\_V13&0.3254&0.0125\\
V3&0.3660&0.0107&1.0000&V6\_V12&0.3261&0.0119\\
V5&0.3677&0.0090&1.0000&V11\_V12&0.3261&	0.0119\\
V10&0.3704&0.0063&1.0000&V11\_V14&	0.3301&0.0280\\
V2&0.3715&0.0052&1.0000&V3\_V6&0.3323&0.0273\\
V13&0.3725&0.0042&1.0000&V3\_V12&0.3326&0.0053\\
V4&0.3749&0.0018&1.0000&V1\_V12&0.3326&0.0053\\
V11&0.3753&0.0014&1.0000&V2\_V12&0.3333&0.0047\\
V7&	0.3767&0.0000&NA&V3\_V13&0.3354&0.0305\\
V8&0.3767&0.0000&NA&V10\_V14&0.3357&0.0225\\
V15&0.3767&0.0000&NA&V10\_V12&	0.3361&0.0018\\
V16&0.3767&0.0000&1.0000&V12\_V15&	0.3362&0.0017\\\hline
\end{tabular}%
\caption{\{V9=4\}:Top 15 ranked conditional entropies (CE) and successive CE-drop under 1-feature and 2-feature settings.  $n=17$ with 1 uncensored data point.}
\label{CEdropV9eq4}
\end{table}

\paragraph{Overall results of ADNI data analysis from V9 perspective.}
Among the sub-collections: from \{V9=1\} to \{V9=4\}, we see their global distinctions as indicated from the CEDA results derived from using the censoring status as a response variable in Section~\ref{Sec:adni_analysis}. The feature V9 is shown to achieve the highest CE-drop as having the largest mutual information with the response variable $\delta_i$. With V9 as one of the natural perspectives for exploring heterogeneity in ADNI data, the first evident heterogeneous pattern is seen from the significant variations of four conditional entropies of $T$: \{0.6545, 1.3183, 0.5380, 0.3767\} across the four sub-collections. This evidence reveal structural distinctions among the four compositions of subjects belonging to the four sub-collections. That is, the V9 indeed provides significant amounts of information in sub-collections \{V9=1\}, \{V9=3\} and \{V9=4\}, but not in sub-collection \{V9=2\}. Such varying results further point to further needed investigations of finding which features or feature-sets can assist V9 to further reduce the uncertainty of $T$.

The second global heterogeneous pattern is the varying compositions of the top 5 ranked individual features and the four sets of feature-pairs having interacting effects. In particular, the 1st ranked individual features across the four sub-collections are all different. Features V7 and V6 respectively appear as individual 1st ranked features in \{V9=1\} and \{V9=3\}. Both features also appear in the top three ranked features across \{V9=1\} to \{V9=3\}. That is, V7 or V6 indeed can individually provide some essential amounts of extra information beyond V9, but not both together because this feature-pair doesn't achieve the ecological effect. V4 is ranked 1st in \{V9=2\}, while V12 is ranked 1st in \{V9=4\}. In contrast, both features are ranked rather low in \{V9=1\} and \{V9=3\}.

It is interesting to note also that the V11 plays a different role across the four sub-collections. It plays the role of candidate of order-1 major factor in \{V9=1\} and \{V9=2\}, while uniquely plays the supporting role of having an interacting effect with order-1 major factors in \{V9=3\} and \{V9=4\}. To a great extent, V10 also plays more or less the same roles as V11 across these four sub-collections. It is worth mentioning that V3 and V7 reveal their interacting effect in \{V9=1\}, not in the rest of the three sub-collections.

As for V8, it is a known important feature variable in AD. However, one significant and consistent pattern coming out of our CEDA results is: ``V8 is never an obvious order-1 major factor''. This observation seemingly means that V8 indeed doesn't provide essential amounts of information about $T$ beyond V9. Is this implication solid? We would see some clues in the next subsection.

Such heterogeneous patterns of global and sub-collection scales are evidently authentic because no man-made assumptions or structures are involved. After recognizing the fact that heterogeneity is inherent in this data set, its implied consequence is also recognized as collected manifestations, each of which rests on all those categories receiving very short code-IDs via one specific perspective. Therefore, the ideal scenario is that all subjects receive a spectrum of short code-IDs derived from many distinct perspectives. This scenario is the topic studied in the Part-II of this paper.

At the end of this section, we conclude that no homogeneous modeling structures could be suitable for this ADNI data set. Its inadequacy demonstrated through the Cox's PH hazard regression model on two scales: overall and the four sub-collections is clear and intuitive. Such intuition is  surely not new. It is understood that human aging-related diseases are too complex to be captured by any homogeneity-based modeling structures. Here we further extend our intuition to express that complexity embraced into individual person's disease dynamics can be revealed and presented via the above heterogeneity's manifestation to a great extent as depicted in our ideal scenario.

\subsection{Heterogeneity w.r.t V8 (ADAS13-bl)}
\begin{figure}[h!]
 \centering
\includegraphics[width=0.7\textwidth]{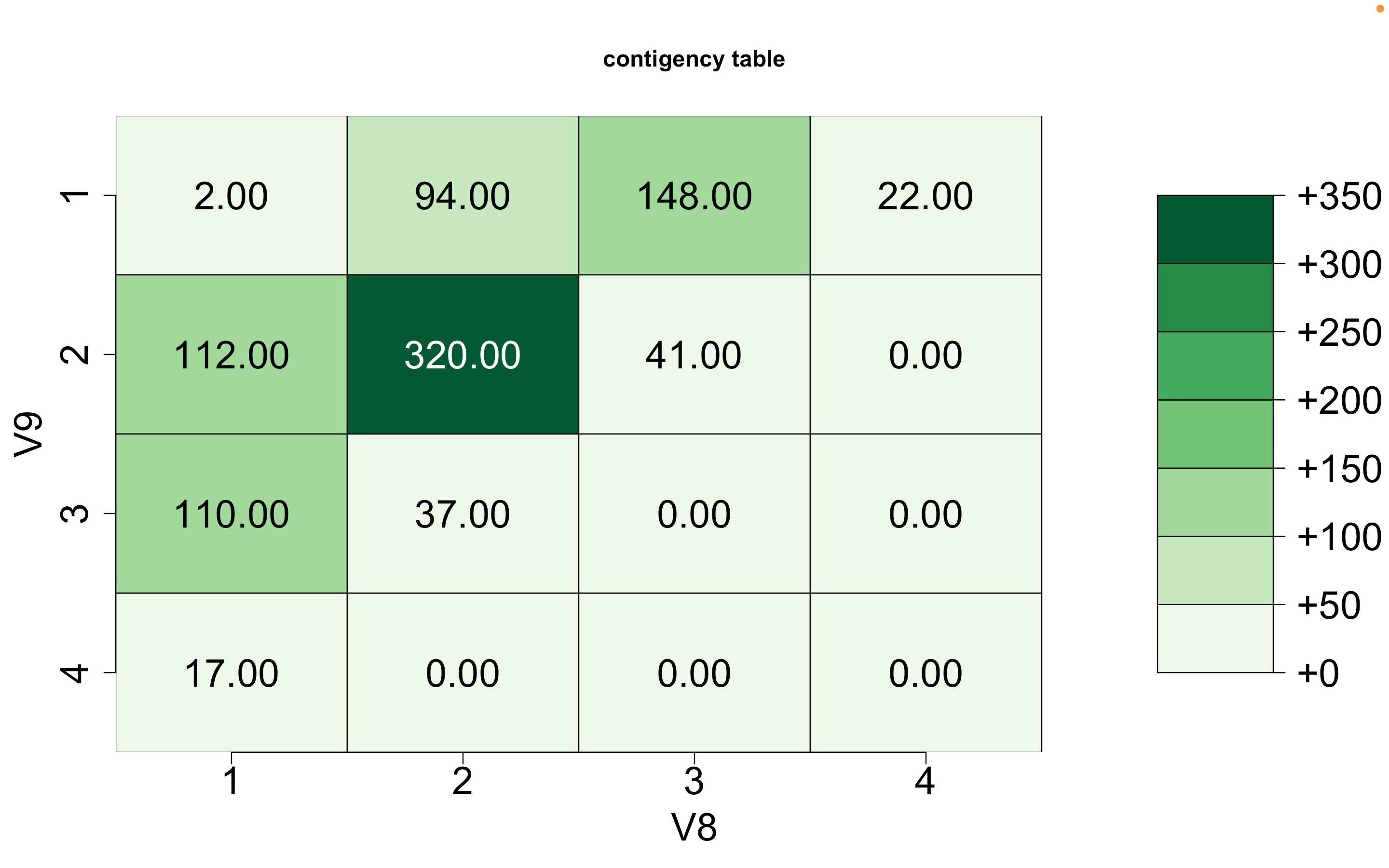}
 \caption{Contingency table of V9-vs-V8.}
 \label{contingV9V8}
 \end{figure}

\paragraph{}We choose to look into heterogeneity through the V9 perspective in the previous section. It is natural to look through V8 as well because V8 ranked 2nd in reducing the uncertainty of $\delta_i$ in Section~\ref{Sec:delta_i}. On the other hand, the associative relation between V8 and V9 is indeed complicated from their contingency table $C[V9-vs-V8]$ as shown in Figure~\ref{contingV9V8}. All cells under the diagonal of $4\times 4$ matrix lattice are zeros, while all entry counts on the diagonal are much smaller than entry counts right above the diagonal, which decreases sharply along the other diagonal direction. Such patterns signal multiple non-linear constraints embraced in this contingency table going beyond the simple observation: V9's ordinal categories are arranged in the reverse order of V8's ordinal categories. So it becomes curious to ask whether V8 generates heterogeneity that mirrors heterogeneity pertaining to V9 in the previous subsection.

The heterogeneity pertaining to V8 would be explored in this subsection with respect to four sub-collections:\{V8=1\}, \{V8=2\}, \{V8=3\} and \{V8=4\}. It is striking to note that the sub-collection \{V8=4\} consists of subjects whose survival time from MCI to AD all fall in the category of $T=1$. This fact is consistent with the fact that higher scores of V8(ADAS13.bl) strongly indicate the likelihood of progressing into a more severe disease \cite{Kueper2018}. Hence, the CE of $T$ is zero, and so are all conditional CEs with respect to all covariate features. This is one striking aspect of heterogeneity that we haven't seen within V9's sub-collections. Therefore, it is worthwhile exploring further the rest of the three sub-collections.

\paragraph{\{V8=1\}:} According to the computed CEs reported in Table~\ref{CEdropV8eq1} for the 1-feature setting of sub-collection \{V8=1\}, V9 is the most dominant covariate feature by achieving a CE-drop that is twice of the CE-drop of the 2nd ranked feature V11. And it is unusual to see V15 ranked 7th, which is lowly ranked across the four sub-collections with respect to V9.

As for the 2-feature setting, feature-pair \{V9, V10\} is ranked 1st.  We know that V9 and V10 are marginally independent. They indeed become conditional dependent given $T$ by achieving the ecological effect. In fact, all 10 feature-pairs of the top 5 ranked features are on the top 15 list and they all achieve ecological effects. Thus, they are natural candidates for order-1 major factors. We also observe the feature-pair: \{V3, V6\}, achieving the interacting effect.

Hence, if we are to choose a potential collection of major factors, then this collection should be \{V7, V9, V10, V11, V4, \{V3, V6\}\}, within this sub-collection \{V8=1\}.

\begin{table}[h!]
\centering
\begin{tabular}{ccc|ccc}\hline
1-feature& CE&	SCE-drop  & 2-feature&CE&	SCE-drop\\ \hline
V9&0.9456&0.0414&V9\_V10&	0.9126&0.0329\\
V11&0.9662&0.0207&V6\_V9&0.9222&0.0234\\
V7&0.9669&0.0201&V4\_V9&0.9231& 0.0224\\
V4&	0.9677&0.0192&V9\_V11&	0.9244&0.0211\\
V6&	0.9689&0.0181&V11\_V15&0.9255&0.0406\\
V10&0.9706&0.0164&V7\_V11&0.9277&0.0384\\
V15&0.9715&0.0155&V7\_V9&0.9282&0.0174\\
V1&	0.9733&0.0137&V1\_V11&	0.9290&0.0372\\
V16&0.9772&0.0097&V1\_V9&0.9313&0.0142\\
V13&0.9783&0.0087&V4\_V11&0.9314&0.0348\\
V12&0.9786&0.0083&V9\_V12&0.9315&0.0141\\
V14&0.9811&0.0059&V6\_V10&0.9317&0.0372\\
V3&	0.9815&0.0054&V3\_V6&0.9318&0.0371\\
V2&0.9841&0.0029&V9\_V15&0.9337&0.0118\\
V5&	0.9848&0.0021&V3\_V9&0.9344&0.0112\\\hline
\end{tabular}%
\caption{\{V8=1\}:Top 15 ranked conditional entropies (CE) and successive CE-drop under 1-feature and 2-feature settings. }
\label{CEdropV8eq1}
\end{table}

\paragraph{\{V8=2\}:} Within the sub-collection \{V8=2\}, the entropy of $T$ here is much higher than the entropy of $T$ in sub-collection \{V8=1\}. Based on CEs of 1-feature setting in Table~\ref{CEdropV8eq2}, the feature V9 is still a dominant factor by having a CE-drop almost twice as large as the CE-drop of the 2nd ranked V7. Even though all involving features are less associated than they are marginal, the number of feature-pairs achieving the ecological effect is much smaller than that in sub-collection \{V8=1\}. We only find feature-pairs \{V9, V10\}, \{V4, V9\} and \{V4, V10\} achieve the ecological effects. Thus, if we need to conclude with a potential collection of major factors, then this collection is \{V9, V4, V10\}.

\begin{table}[h!]
\centering
\begin{tabular}{ccc|ccc}\hline
1-feature &CE&	SCE-drop  & 2-feature &CE&	SCE-drop\\ \hline
V9&	1.1884&0.0702&V7\_V9&1.1537&0.0346\\
V7&1.2171&0.0415&V4\_V9&1.1592&0.0291\\
V6&1.2260&0.0327&V6\_V9&1.1606&0.0278\\
V4&	1.2364&0.0223&V9\_V10&	1.0681&0.0202\\
V11&1.2373&0.0213&V9\_V11&1.1715&0.0168\\
V5&1.2428&0.0158&V9\_V15&1.1725&0.0159\\
V10&1.2451&	0.0135&V9\_V12&1.1744&	0.0139\\
V12&1.2470&0.0117&V9\_V14&1.1746&0.0137\\
V13&1.2472&0.0114&V9\_V16&1.1765&0.0118\\
V14&1.2506&0.0080&V1\_V9&1.1781&0.0103\\
V16&1.2514&0.0072&V9\_V13&1.1804&0.0080\\
V1&1.2517&0.0069&V3\_V9&1.1841.&0.0042\\
V15&1.2527&0.0059&V2\_V9&1.1845&0.0039\\
V3&	1.2553&0.0033&V7\_V11&1.1851&0.0320\\
V2&	1.2581&0.0005&V4\_V7&1.1903&0.0267\\\hline
\end{tabular}%
\caption{\{V8=2\}:Top 15 ranked conditional entropies (CE) and successive CE-drop under 1-feature and 2-feature settings. }
\label{CEdropV8eq2}
\end{table}

\paragraph{\{V8=3\}:} Based on Table~\ref{CEdropV8eq3} for the sub-collection \{V8=3\}, V11 is ranked the 1st. That is, the 2nd-ranked V9 is no longer a dominant feature in terms of CE-drop. The most striking observation is that 14 out of the top 15 feature-pairs are having ecological effects, except \{V9, V11\}.  This phenomenal pattern is almost entirely opposite of that of the top 15 feature-pairs observed in the sub-collection \{V8=2\}. On top of this specific observation, there is only one out of the 15 pairs having interacting effects:\{V1, V11\}. We conclude that if we are to choose a collection of major factors, this collection is: \{\{V1, V11\}, V7, V6, V4, V14, V3\}, which is very distinct to the selected collection in the sub-collections \{V8=1\} and \{V8=2\}. This is an evident perspective of heterogeneity contained in this data set.

\begin{table}[h!]
\centering
\begin{tabular}{ccc|ccc}\hline
1-feature& CE&	SCE-drop  & 2-feature& CE&	SCE-drop\\ \hline
V11&0.7895&0.0461&V7\_V11&0.7363&0.0531\\
V9&0.7926&0.0429&V9\_V11&0.7482&0.0412\\
V7&	0.7931&0.0424&V7\_V9&0.7483&0.0442\\
V6&	0.8041&0.0315&V6\_V11&	0.7505&0.0389\\
V16&0.8099&0.0256&V4\_V11&0.7508&0.0386\\
V4&	0.8192&0.0163&V1\_V11&0.7519&0.0375\\
V14&0.8202&0.0153&V7\_V16&0.7542&0.0389\\
V3&0.8226&0.0129&V11\_V14&0.7545&0.0349\\
V15&0.8229&0.0126&V1\_V7&0.7553&0.0378\\
V1&0.8249&0.0106&V11\_V12&0.7554&0.0339\\
V10&0.8251&0.0104&V7\_V14&0.7560&0.0371\\
V2&0.8256&0.0099&V4\_V7&0.7562&0.0369\\
V13&0.8297&0.0058&V3\_V11&0.7573&0.0332\\
V12&0.8324&0.0031&V6\_V7&	0.7588&0.0342\\
V5&0.8331&0.0024&V3\_V7&0.7599&0.0332\\\hline
\end{tabular}%
\caption{\{V8=3\}:Top 15 ranked conditional entropies (CE) and successive CE-drop under 1-feature and 2-feature settings. }
\label{CEdropV8eq3}
\end{table}

\paragraph{Overall results of ADNI data analysis from V8 perspective.}
Across the four sub-collections, the four CEs of $T$ vary significantly: \{0.9869, 1.2586, 0.8355, 0.0000\}. This is the first evidence of heterogeneity. The feature V9 is the top-ranked in sub-collections: \{V8=1\} and \{V8=2\}, but ranked 2nd in \{V8=3\}. This is the second piece of evidence. The three collections of feature-pairs that achieve the ecological effect or interacting effect in the sense of becoming conditional dependents are very distinct. This is the third piece of evidence. Thus, we conclude that V8 is also a legitimate perspective for heterogeneity in this ADNI data set.

These pieces of evidence of heterogeneity echo the suggestion provided in the review paper \cite{Kueper2018}: the original ADAS-Cog is not an optimal outcome measure for pre-dementia studies. It needed modification. It went on to suggest that the most beneficial modification of ADAS-Cog is tests of memory. As it turns out to be V9 in this study. Since, across the first three sub-collections, V9 consistently reveals to offer extra information beyond V8.

Here we reiterate that the particular valuable piece of information offered by V8 is the zero CE in \{V8=4\}. That is, the 22 subjects' status of being in \{V8=4\} can precisely point to one and only one category $\# 1$ of $T$. From the same perspective of having very low CEs, the two pieces of information of being \{V9=3\} and \{V9=4\}, respectively, are also critical for the 147 and 17 subjects in the two sub-collections. In sharp contrast, this is not the case for the 189 subjects in sub-collection \{V8=3\}. These pieces of information are parts of the collective heterogeneity embedded within the information content of this ADNI data.

In summary, information pieces are indeed scattered among various perspectives of heterogeneity. This is the reality at least in this ADNI data set. These valuable fragmenting pieces of information need to be collected and systemized. Therefore, no matter whether the data analyzing goal is focused on an understanding of the complex system under study or just for predictive decision-making, such systemizing information pieces is a critical task. For instance, the critical issue facing the 189 subjects in sub-collection \{V8=3\} is how to more precisely describe these subjects by incorporating which feature variables? That is the chief purpose of suggesting a collection of major factors within sub-collections, with which improvements in describing subjects' characteristics can be potentially derived.

\section{Conclusions and discussions}
\paragraph{}
In the previous two subsections, we have demonstrated two perspectives of heterogeneity embedded within this data set. There are many other perspectives to be explored to build the data's information content more fully. A version of the data's full information content can be described in the ideal scenario as mentioned in the conclusion of the subsection devoted to heterogeneity from the V9 perspective. In this ideal scenario, each subject is attached to a spectrum of code-IDs from explored different perspectives of heterogeneity. These Code-IDs would tell what are relevant pieces of information pertaining to this subject: some pieces say which features or feature-sets would shed clear lights on which category of this subject's $T$ would fall into with short coding lengths and very low uncertainty, while in contrast, some pieces say which features or feature-sets can only support potential categories of $T$ with high uncertainty. Both kinds of pieces of information and the spectrum of code-IDs would help facilitate a better understanding of AD as a complex disease. At this stage, making such a manifestation of heterogeneity-based full information content is studied and deferred to Part-II of this study.

In this paper as Part-I, when combining results of heterogeneity derived from the V9 and V8 perspectives, we clearly see that V9 provides extra information beyond V8, while V8 doesn't provide extra information beyond V9. However, the results of heterogeneity derived from the four sub-collections via the V8 perspective surely offer to distinguish pieces of information beyond the results of heterogeneity from the four sub-collections via the V9 perspective. Such a discrepancy between information and entropy indeed points to a very important direction for extracting information from data: all relevant perspectives of heterogeneity are worth exploring. Different relevant perspectives of heterogeneity pertaining to different features or feature-sets mount to offer distinct pieces of information with distinct implications. Therefore, the immediate and critical issue for building the ideal scenario is how to exhaustively search for all potentially relevant perspectives of heterogeneity. Explicit resolutions to this issue would need further computational developments, which are also referred to the Part-II. Further, another critical issue arises: after knowing what important pattern information should be harvested from each perspective of heterogeneity, we need to address how to effectively display and systemize valuable, but fragmented information pieces to build a scientific understanding of the AD's prognosis dynamics.

In this paper, we adopt the CEDA paradigm to analyze this ADNI data set of time needed from MCI to an event of AD diagnosis. We first propose a formal testing protocol for checking whether the censoring mechanism is independent of the targeted time-to-event. This contingency-table-based approach is a rather straightforward application of the Re-distribution-to-the-right algorithm. However, to our knowledge, this simple approach has not yet been reported in Survival Analysis literature. After confirming the non-informative censoring mechanism, we compare CEDA results with Cox's PH results on two scales: global and sub-collection. Due to heavy censoring rates, all PH results on both scales deem to be unreliable. Therefore, heterogeneity becomes an urgent and critical issue in Survival analysis. And the immediate impact of the presence of heterogeneity in time-to-event data is that any modeling construct with a global structure, such as the linearity-based structural PH model, has slim chances of producing scientifically valid results.

Further, as we really acknowledge that a complex disease, such as AD, indeed retains very little chance to be fitted well by any global model due to existential heterogeneity, the whole disease progressing process doesn't likely sustain man-made additive effects from different features and feature-sets as typically assumed in statistical modeling. Furthermore, any interacting effects due to any sets of conditional dependent features of various order can not be regressed into any fixed formats preferred just for implementing mathematical or statistical operations. Thus, it is a conservative, robust and realistic way of approaching data analysis by simply admitting that data analysts have acquired no knowledge regarding complex formats of effects of features or feature-sets on global as well as local scales. This way of data analysis is the philosophic basis of the CEDA paradigm.

\section*{Acknowledgement}
\paragraph{} Data collection and sharing for this project was funded by the Alzheimer's Disease Neuroimaging Initiative (ADNI) (National Institutes of Health Grant U01 AG024904) and DOD ADNI (Department of Defense award number W81XWH-12-2-0012). ADNI is funded by the National Institute on Aging, the National Institute of Biomedical Imaging and Bioengineering, and through generous contributions from the following: AbbVie, Alzheimer's Association; Alzheimer's Drug Discovery Foundation; Araclon Biotech; BioClinica, Inc.; Biogen; Bristol-Myers Squibb Company; CereSpir, Inc.; Cogstate; Eisai Inc.; Elan Pharmaceuticals, Inc.; Eli Lilly and Company; EuroImmun; F. Hoffmann-La Roche Ltd and its affiliated company Genentech, Inc.; Fujirebio; GE Healthcare; IXICO Ltd.; Janssen Alzheimer Immunotherapy Research \& Development, LLC.; Johnson \& Johnson Pharmaceutical Research \& Development LLC.; Lumosity; Lundbeck; Merck \& Co., Inc.; Meso Scale Diagnostics, LLC.; NeuroRx Research; Neurotrack Technologies; Novartis Pharmaceuticals Corporation; Pfizer Inc.; Piramal Imaging; Servier; Takeda Pharmaceutical Company; and Transition Therapeutics. The Canadian Institutes of Health Research is providing funds to support ADNI clinical sites in Canada. Private sector contributions are facilitated by the Foundation for the National Institutes of Health (\hyperlink{www.fnih.org}{www.fnih.org}). The grantee organization is the Northern California Institute for Research and Education, and the study is coordinated by the Alzheimer's Therapeutic Research Institute at the University of Southern California. ADNI data are disseminated by the Laboratory for Neuro Imaging at the University of Southern California.

\bibliographystyle{vancouver}
\bibliography{ref}

\end{document}